\documentclass[a4paper,11pt]{article}
\usepackage{jcappub}
\usepackage{amssymb}
\usepackage{graphicx}
\usepackage{amsmath}
\usepackage{hyperref}
\usepackage{multirow}

\title{\boldmath  The Hubble tension: A decade review}

\author[2]{Rong-Gen Cai,}
\emailAdd{cairg@itp.ac.cn}

\author[1,3]{Shao-Jiang Wang}
\emailAdd{schwang@itp.ac.cn (Corresponding author)}

\affiliation[1]{Institute of Theoretical Physics, Chinese Academy of Sciences (CAS), Beijing 100190, China}
\affiliation[2]{Institute of Fundamental Physics and Quantum Technology \& School of Physical Science and Technology, Ningbo University, Ningbo, 315211, China}
\affiliation[3]{Asia Pacific Center for Theoretical Physics (APCTP), Pohang 37673, Korea}

\abstract{Ever since the new millennium, precision cosmology has forged the $\Lambda$-cold-dark-matter ($\Lambda$CDM) model as the standard model of concordant cosmology, withstanding various tests except for an ever-enlarging discrepancy between early-Universe observations and late-Universe measurements on the current Hubble expansion rate of our observable Universe. This Hubble-constant tension has likely become a real crisis for modern cosmology, with the discrepancy persisting regardless of whether the early-Universe observations depend on \textit{Planck} CMB or not, and the late-Universe measurements depend on distance ladders at all. If the Hubble tension originates from a different early Universe, its resolutions pertain to shrinking the sound horizon by altering either early expansion or recombination histories, but at the same time necessitating modifications to both primordial and late Universe altogether. Alternatively, if the Hubble tension arises from a different late Universe, its resolutions operate by changing the absolute magnitude of supernovae either intrinsically or effectively, both of which have been strongly constrained by the inverse distance ladders with the cosmic distance duality relation. The remaining options seem to turn to our local Universe, but a local Hubble bubble or cosmic void solution has long been ruled out as a significant contribution to the Hubble tension. In view of this dilemma, we review in this paper alternative resolutions involving interacting dark energy models, either combining early-time and late-time modifications or operating at the transition from inhomogeneity to homogeneity scales.}

\begin{document}
\maketitle
\flushbottom

\section{Introduction}\label{sec:introduction}

The Hubble constant was born as a local distance-scale measure almost a century ago. Although the terminology ``Hubble-constant tension'' or simply the ``Hubble tension'' first appeared in Ref.~\cite{DiValentino:2016hlg} and Ref.~\cite{Karwal:2016vyq}, there was the \textit{first Hubble tension} before current debates. After recognition that spiral nebulae are external galaxies, the Cepheid period-luminosity relation, Slipher's redshift measurements, and Hubble's original velocity-distance relation led to early estimates of enormously high values for the Hubble constant, around $500\,\mathrm{km\,s^{-1}Mpc^{-1}}$, implying an unacceptably young universe, which was already in conflict with geological estimates of the age of the Earth.  For more historical review, see the recent summaries~\cite{Tully:2023bmr,Cervantes-Cota:2023wet}.


When the hot Big Bang framework was firmly established by 1960s after the discovery of the cosmic microwave background (CMB) and the support from Big Bang nucleosynthesis (BBN), the value of the Hubble constant $H_0$ thereby became not merely a local distance-scale measure but a quantity tied to the cosmic age, radiation/matter content, and global dynamics of the Universe. The following decades were dominated by the well-known ``$H_0=50$ or $H_0=100$'' controversy, which is referred to here as the \textit{second Hubble tension}. Although neither extreme survived as the modern consensus, this debate was scientifically productive: it exposed the vulnerability of distance measurements to population effects, Malmquist bias, peculiar velocities, sample selection, zero-point calibration, and so on. It also drove the development of more robust secondary distance indicators. The outcome of this second Hubble tension was widely considered the first modern measurement of the Hubble constant, $H_0=72\pm8\,\mathrm{km\,s^{-1}Mpc^{-1}}$, from the Hubble Space Telescope (HST) Key Project led by Wendy Freedman~\cite{HST:2000azd} around the new millennium.

In the era of precision cosmology~\cite{Moresco:2022phi} after the new millennium, the combination of CMB, Type Ia supernovae (SNe Ia), and baryon acoustic oscillation (BAO) has established the $\Lambda$-cold-dark-matter ($\Lambda$CDM) model~\cite{Planck:2013pxb} as the standard model for concordant cosmology. Back to the era of Wilkinson Microwave Anisotropy Probe (WMAP), the rough agreement on the Hubble constant between the WMAP constraint $H_0=70.5\pm1.6\,\mathrm{km\,s^{-1}Mpc^{-1}}$~\cite{WMAP:2012nax} and distance ladder measurement $H_0=73.8\pm2.4\,\mathrm{km\,s^{-1}Mpc^{-1}}$~\cite{Riess:2011yx} did not raise any suspicion. It was the \textit{Planck} constraints on the Hubble constant, $H_0=67.3\pm1.2\,\mathrm{km\,s^{-1}Mpc^{-1}}$ (\textit{Planck} 2013~\cite{Planck:2013pxb}) and $H_0=67.8\pm0.9\,\mathrm{km\,s^{-1}Mpc^{-1}}$ (\textit{Planck} 2015~\cite{Planck:2015fie}), that raises  $2\sim3\sigma$ tensions with distance ladder measurement $H_0=73.24\pm1.74\,\mathrm{km\,s^{-1}Mpc^{-1}}$~\cite{Riess:2016jrr}. Ever since then, the \textit{third Hubble tension} stepped forward with ever-enlarging discrepancies~\cite{Riess:2018byc,Riess:2018uxu,Riess:2019cxk,Riess:2020fzl}, reaching an unprecedented $5\sigma$ threshold between local direct measurement  $H_0=73.04\pm1.04\,\mathrm{km\,s^{-1}Mpc^{-1}}$ (SH0ES)~\cite{Riess:2021jrx} and global fitting constraint $H_0=67.27\pm0.60\,\mathrm{km\,s^{-1}Mpc^{-1}}$ (\textit{Planck} 2018)~\cite{Planck:2018vyg}. This $5\sigma$ Hubble crisis is even more severe for a community consensus measurement  $H_0=73.50\pm0.81\,\mathrm{km\,s^{-1}Mpc^{-1}}$ (H0DN) from the Local Distance Network~\cite{H0DN:2025lyy}, resulting in $7.1\sigma$ from flat $\Lambda$CDM with CMB and $5.0\sigma$ with BBN+BAO. If any single systematics could be responsible for this Hubble crisis~\cite{Cai:2023sli}, it must at least contain some genuine new physics that disguises itself as ``systematics''. Therefore, the Hubble tension~\cite{Bernal:2016gxb,Verde:2019ivm,Knox:2019rjx,Riess:2020sih,Freedman:2021ahq}, among other tensions~\cite{Perivolaropoulos:2021jda,DiValentino:2022fjm}, might eventually point to some new physics beyond the $\Lambda$CDM~\cite{Bull:2015stt}.

Hubble resolutions~\cite{DiValentino:2020zio,DiValentino:2021izs,Schoneberg:2021qvd,Shah:2021onj,Abdalla:2022yfr,Poulin:2023lkg,Hu:2023jqc,Vagnozzi:2023nrq,CosmoVerseNetwork:2025alb} remain highly constrained. If solely contributing the Hubble tension to the sound-horizon ($r_s$) tension~\cite{Bernal:2016gxb,Aylor:2018drw,Verde:2019ivm,Knox:2019rjx,Riess:2020sih,Jiang:2024xnu}, then early-Universe resolutions~\cite{Krishnan:2020obg,Jedamzik:2020zmd,Hill:2020osr,Lin:2021sfs,Vagnozzi:2021gjh,Philcox:2022sgj,Kamionkowski:2022pkx,Khalife:2023qbu,Pedrotti:2024kpn,Qu:2024lpx,Poulin:2025nfb,SPT-3G:2025vyw,Toda:2025kcq,Wang:2025djw,Yin:2026gss,Bella:2026zuk} typically seek to reduce the sound horizon by modifying the pre-recombination expansion history or recombination physics, but successful models must also simultaneously modify the inflationary Universe~\cite{Braglia:2020bym,Takahashi:2021bti,Ye:2022efx,Braglia:2022phb,Lin:2022gbl,Fu:2023tfo,Fu:2025ciy} with an extremely scale-invariant primordial spectrum~\cite{Poulin:2018cxd,Niedermann:2019olb,Niedermann:2020dwg,Ye:2020btb,Ye:2021nej,Jiang:2022uyg,Jiang:2022qlj,Peng:2023bik} and the late Universe~\cite{Vattis:2019efj,Pandey:2019plg,Clark:2020miy} with matter-perturbation linear-amplitude ($S_8$) tension~\cite{DiValentino:2020vvd,Perivolaropoulos:2021jda,Abdalla:2022yfr,Poulin:2022sgp} and growth-index ($\gamma$) tension~\cite{Nguyen:2023fip} solved. 
If solely contributing the Hubble tension to the absolute-magnitude ($M_B$) tension~\cite{Poulin:2024ken}, then late-Universe resolutions~\cite{Benevento:2020fev,Camarena:2021jlr,Efstathiou:2021ocp,Cai:2021weh,Cai:2022dkh,Keeley:2022ojz,Huang:2024erq,Huang:2024gfw,Pedrotti:2025ccw,Ling:2025lmw,Wang:2026kor,Tiwari:2026pzk} instead attempt to alter the distance-redshift relation effectively or the calibrated absolute magnitude of SNe Ia intrinsically, yet these scenarios are strongly limited by inverse distance ladders~\cite{Cuesta:2014asa,Heavens:2014rja,Aubourg:2014yra,Verde:2016ccp,Alam:2016hwk,Verde:2016wmz,Macaulay:2018fxi,Feeney:2018mkj,Lemos:2018smw,Vagnozzi:2019ezj,eBOSS:2020yzd,Ling:2025lmw}, $G_\mathrm{eff}$ constraints~\cite{Banik:2024yzi,Perivolaropoulos:2025gzo}, and cosmic distance duality relation~\cite{CDDR,Bassett:2003vu,More:2008uq,Holanda:2010ay,Nair:2011dp,Meng:2011nt,Holanda:2011hh,Wu:2015ixa,Rana:2017sfr,Ruan:2018dls,Zheng:2020fth,Favale:2024sdq,Qi:2024acx,Wang:2024rxm,Jesus:2024nrl,Alfano:2025gie,Yang:2025qdg,Teixeira:2025czm,Afroz:2025iwo,Kanodia:2025jqh,Li:2025htp,Zheng:2025cgq,Alfano:2025fyq,Barua:2025dxe,Kumar:2026kbo,Tiwari:2026pzk}. Purely local explanations within general relativity, such as a cosmic void~\cite{Keenan:2013mfa,Hoscheit:2018nfl}, are not large and deep enough to resolve the full discrepancy~\cite{Wojtak:2013gda,Odderskov:2014hqa,Wu:2017fpr,Kenworthy:2019qwq}. The remaining viable possibilities appear to require either correlated early-late modifications~\cite{Vagnozzi:2023nrq}, or new physics associated with the transition from local inhomogeneity to large-scale homogeneity~\cite{Yu:2022wvg,Giani:2023aor,Huang:2024erq,Huang:2024gfw,Huang:2025som,Wang:2026kor}, both of which might be modeled by interacting dark energy scenario (e.g., Refs.~\cite{Karwal:2021vpk,Cai:2021wgv}).

In this short review, we provide concise summaries of the observational status in Sec.~\ref{sec:Observation}, the theoretical dilemma in Sec.~\ref{sec:Resolution}, and alternative resolutions in Sec.~\ref{sec:Alternative}. The observational status is given with a focus on the most recent progress to establish the Hubble tension as a pressing crisis of standard cosmology. The theoretical dilemma is organized with a series of arguments forming ``No-Go Theorems'' for both early-Universe and late-Universe solutions alone. Therefore, the alternative resolutions seem to favor either combined early-late modifications or local-scale new physics disguised as ``systematics''. Eventually, we might need some interacting dark energy-dark matter models to rebuild cosmological concordances.

\section{From Hubble tension to Hubble crisis}\label{sec:Observation}

In this section, we review both early-Universe constraints (Sec.~\ref{subsec:EarlyObservation}) and late-Universe measurements (Sec.~\ref{subsec:LateObservation}) of the Hubble constant. In view of these results, the Hubble tension turns into a pressing crisis of standard cosmology, not only for the ever-enlarging discrepancies between early and late Universe, but also for the irrelevance to any specific observations. Therefore, the Hubble-constant crisis is manifested not only in the $5\sigma$ discrepancy between the \textit{Planck} and SH0ES determinations of the Hubble constant, but also in the \textit{systematic tendency}~\cite{Riess:2022oxy,Verde:2023lmm} that direct late-Universe measurements of the Hubble constant are higher than global-fitting values inferred from the early Universe by the $\Lambda$CDM model.

\subsection{Early-Universe observations}\label{subsec:EarlyObservation}

\begin{figure}
\centering
\includegraphics[width=\textwidth]{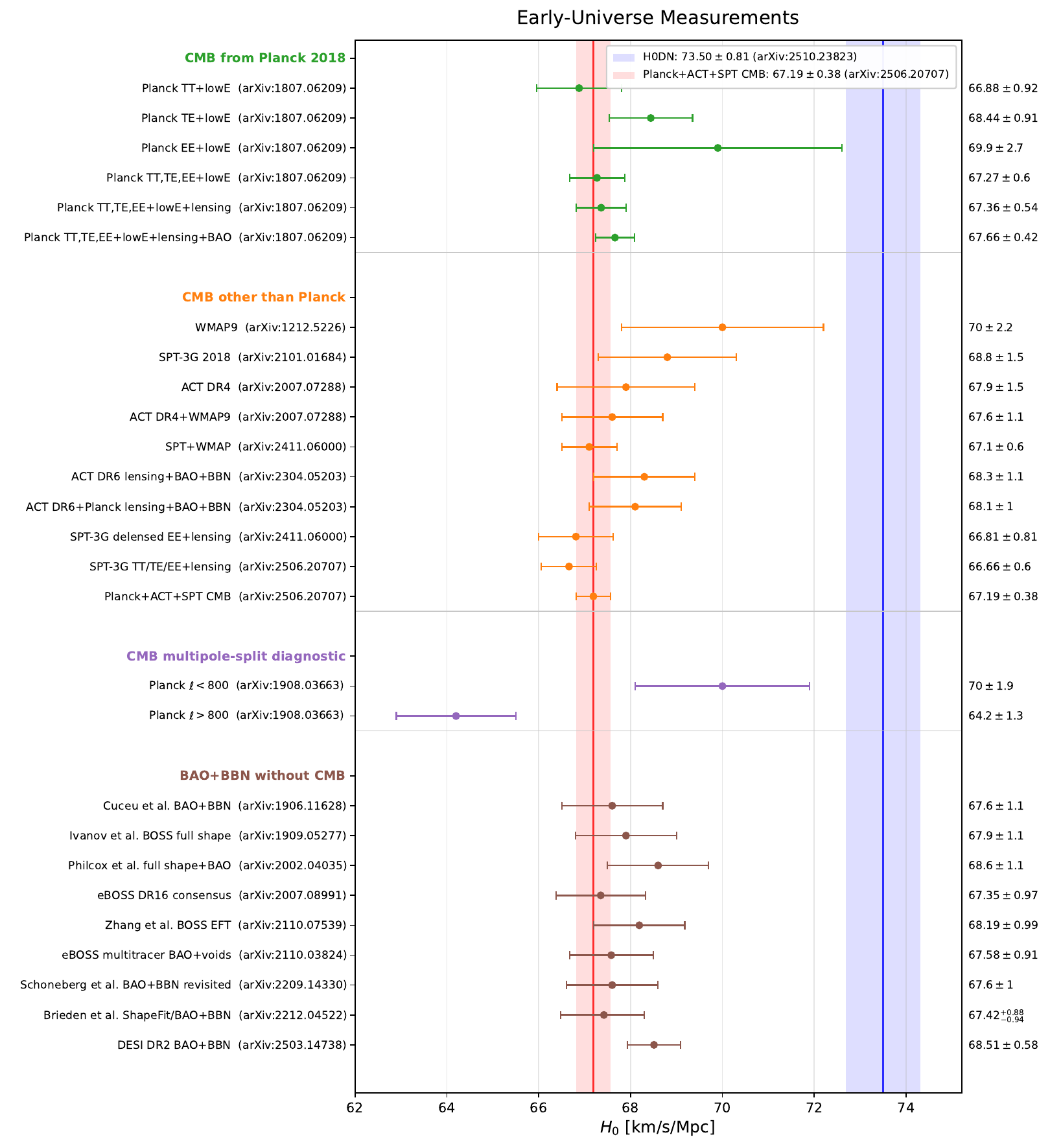}\\
\caption{Early-Universe constraints on the Hubble constant.}\label{fig:EarlyH0}
\end{figure}

Although observations of the early Universe are carried out in the present late Universe, the data we obtain in fact encode information about the early Universe. However, this information cannot be used directly to measure the present Hubble expansion rate, namely the Hubble constant. A specific cosmological model is therefore needed to extrapolate the early-Universe information to the late Universe. Thus, the resulting value of the Hubble constant is usually referred to as the global-fitting value for given early-Universe datasets. In this section, we will see that globally fitting the $\Lambda$CDM model to CMB from \textit{Planck} (Sec.~\ref{subsubsec:CMBwiPlanck}) or not (Sec.~\ref{subsubsec:CMBwoPlanck}) or even without CMB (BBN+BAO in Sec.~\ref{subsubsec:woCMB}) all lead to systematically lower values of the Hubble constant (Fig.~\ref{fig:EarlyH0}).

\subsubsection{CMB from \textit{Planck} }\label{subsubsec:CMBwiPlanck}

Within the observationally verified history described by the known standard model of particle physics, together with the standard model of cosmology, neutrinos began to decouple about one second after the birth of the early Universe, and electron-positron annihilation soon followed. Three minutes later, as the Universe cooled, light elements began to form; this is the BBN process. After about sixty thousand years, the energy densities of radiation and matter became comparable. Subsequently, the inverse process of hydrogen formation from electrons and protons with photon emission became unable to compensate for the electrons lost in the forward process. As a result, when the age of the Universe reached about 380,000 years, Thomson scattering between electrons and photons could no longer be maintained, and photons decoupled from the background plasma and began to free-stream, forming the last-scattering surface. This is the CMB radiation. These CMB photons then passed through the gravitational potential wells of matter structures, and finally, a fraction of them arrived near the Earth and were observed by us.

The determination of the Hubble constant from CMB data is a global-fitting statistical analysis, but the process can still be understood approximately and analytically. The CMB data record the temperature $T(\hat{n})$ of CMB photons arriving from each direction. Theory predicts that this temperature spectrum is a blackbody spectrum, from which an isotropic background temperature $T_0=2.7255\pm0.0006$ K (Kelvin) can be fitted~\cite{Ivanov:2020mfr}, and the temperature fluctuation in that direction is then $\Theta(\hat{n})\equiv\delta T(\hat{n})/T_0$. The observed standard deviation of the CMB temperature fluctuations is only of order $10^{-5}$. This observational fact led to the idea that there was an almost exponential inflationary epoch in the extremely early Universe, during which microscopic quantum fluctuations were rapidly stretched outside the Hubble horizon and became classical perturbations. After inflation ended, the Hubble horizon began to grow again and allowed classical perturbations of different scales to re-enter the horizon at different times, inducing perturbations in the baryon-photon fluid and transmitting them to the last-scattering surface as temperature fluctuations of the decoupled photons. It is usually assumed that the primordial quantum perturbations produced during inflation form a nearly scale-invariant, adiabatically evolving Gaussian random scalar field, and hence their power spectrum is phenomenologically parameterized as $P_\mathcal{R}(k)=A_s(k/k_\mathrm{ref})^{n_s-1}$, where $A_s$ is the amplitude, $n_s$ is the scalar spectral index, and $k_\mathrm{ref}$ is an observational reference scale. For CMB temperature fluctuations, the most sensitive pivot scale is $k_*=0.05\,\mathrm{Mpc}^{-1}$.

The two-point correlation of temperature fluctuations in different directions, $C(\theta)\equiv\langle\Theta(\hat{n})\Theta(\hat{n}')\rangle$, can be expanded as $C(\theta)=\sum\limits_\ell\frac{2\ell+1}{4\pi}C_\ell P_\ell(\cos\theta)$ with spherical-harmonic expansion $\Theta(\hat{n})=\sum\limits_{\ell=0}^\infty\sum\limits_{m=-\ell}^\ell\Theta_{\ell m}Y_{\ell m}(\hat{n})$ with correlation for the expansion coefficient of form $\langle\Theta_{\ell m}\Theta^*_{\ell'm'}\rangle=C_\ell\delta_{\ell\ell'}\delta_{mm'}$, where $C_\ell=2\pi\int_{-1}^1\mathrm{d}\cos\theta\,C(\theta)P_\ell(\cos\theta)$ is the angular power spectrum that is actually measured. A very prominent feature of this angular power spectrum is the acoustic peaks. In particular, the position of the first acoustic peak,
\begin{equation}
\theta_*\equiv \frac{r_s(z_*)}{D_M(z_*)}=\int_{z_*}^\infty\frac{c_s(z)\mathrm{d}z}{H(z)}\bigg/\int_0^{z_*}\frac{c\,\mathrm{d}z}{H(z)},
\end{equation}
is measured with the highest precision (as precise as the combination $\Omega_\mathrm{m}h^3$). Here, the redshift of the last-scattering surface $z_*$ can be defined by the condition that the Thomson-scattering optical depth $\tau$, neglecting reionization, reaches unity. The comoving angular-diameter distance $D_M(z_*)$ from us to the last-scattering surface is mainly determined by the late expansion history, $E(z)^2\equiv H(z)^2/H_0^2=\Omega_\Lambda+\Omega_m(1+z)^3+\Omega_r(1+z)^4$. By contrast, the comoving sound horizon $r_s(z_*)$ on the last-scattering surface is jointly determined by $z_*$, which depends on the recombination history, and by the Hubble parameter $H(z)$ and sound speed $c_s(z)=1/\sqrt{3(1+R(z))}$, which depend on the early expansion history. The baryon-to-photon ratio $R(z)\equiv3\rho_b(z)/(4\rho_\gamma(z))=(3/4)(\omega_b/\omega_\gamma)/(1+z)$ is determined by the physical baryon abundance $\omega_b\equiv\Omega_bh^2$ and the physical photon abundance $\omega_\gamma\equiv\Omega_\gamma h^2=2.473\times10^{-5}(T_0/2.7255)$.

The standard cosmological model usually takes the following six parameters: $\theta_*$, $\tau$, $\omega_b$, $\omega_c$, $A_s$, and $n_s$, as the base parameters in the global fit; the Hubble constant is a derived parameter. Once these six parameters are determined by globally fitting to the CMB data, the position $\theta_*$ of the first acoustic peak in the CMB angular power spectrum can be used to solve for $\omega_\Lambda\equiv\Omega_\Lambda h^2$. Finally, evaluating the Friedmann equation of the standard cosmological model, namely the $\Lambda$CDM model, $(H/100\,\mathrm{km/s/Mpc})^2=\omega_\Lambda+(\omega_b+\omega_c)(1+z)^3+\omega_r(1+z)^4$, at $z=0$ gives the value of the Hubble constant constrained by CMB+$\Lambda$CDM. The physical radiation abundance $\omega_r\equiv\Omega_r h^2$ is a piecewise function: before electron--positron annihilation it is $\omega_r=(g/2)(4/11)^{4/3}\omega_\gamma$, while after electron--positron annihilation it is $\omega_r=[1+(7/8)(4/11)^{4/3}N_\mathrm{eff}]\omega_\gamma$. Here the evolution of the effective number of relativistic degrees of freedom $g$ is determined by the standard model of particle physics, and, assuming a minimal normal neutrino mass hierarchy (see Refs.~\cite{DiValentino:2021imh,Jiang:2024viw,Lynch:2025ine,Giare:2025ath} for recent neutrino mass bound), it also fixes the neutrino-like effective number of relativistic degrees of freedom to $N_\mathrm{eff}=3.046$. This gives an approximate picture of how CMB data are used to globally fit $H_0$ in $\Lambda$CDM.

The \textit{Planck}-CMB constraints on the Hubble constant are very stable from \textit{Planck} 2013 to \textit{Planck} 2018 results~\cite{Planck:2013pxb,Planck:2015fie,Planck:2018vyg}, as shown below in Tab.~\ref{tab:Planck}. It is easy to see that adding CMB lensing, polarization, and external distance data changes the inferred $H_0$ only mildly; all combinations remain near $67\sim68\,\mathrm{km\,s^{-1}Mpc^{-1}}$. For \textit{Planck} 2013 with WMAP low-$\ell$ polarization, further adding high-$\ell$ ground-based CMB data did not significantly move the result, while adding BAO only tightens the inference while preserving the low value. For \textit{Planck} 2015 with improved temperature likelihood, calibration, foreground treatment, and polarization analysis, it favored base $\Lambda$CDM over simple one-parameter extensions, meaning that merely adding curvature, neutrino mass, effective number of neutrino-like relativistic degrees of freedom, or simple dynamical dark energy did not naturally solve the $H_0$ discrepancy without creating other tensions. For \textit{Planck} 2018 as the final legacy cosmological-parameter release, it systematically explored standard one-parameter extensions to $\Lambda$CDM and found that most simple extensions do not naturally raise $H_0$ enough. Therefore, the CMB constraints on the Hubble constant are internally stable across \textit{Planck} temperature, polarization, lensing, and BAO combinations, though there might still be some CMB anomalies~\cite{Schwarz:2015cma,Giare:2023xoc}.

\begin{table}[!h]
\begin{scriptsize}
    \centering
    \renewcommand{\arraystretch}{1.5}
    \begin{tabular}{|c|c|c|c|c|c|}
    \multicolumn{6}{c}{\textit{Planck} 2013}\\
    \hline
         \textit{Planck} & \textit{Planck} & \textit{Planck} & \textit{Planck} & \textit{Planck} & \textit{Planck} \\
          (TT) & +lensing & +WP & +WP+highL & +lensing+WP+highL & +WP+highL+BAO\\
    \hline
         $67.4\pm1.4$ & $67.9\pm1.5$ & $67.3\pm1.2$ & $67.3\pm1.2$ & $67.9\pm1.0$ & $67.80\pm0.77$ \\
    \hline
    \multicolumn{6}{c}{\textit{Planck} 2015}\\
    \hline
         TT+lowP & TT+lowP & TT+lowP & TT,TE,EE+lowP & TT,TE,EE+lowP & TT,TE,EE+lowP \\
          & +lensing & +lensing+ext &  & +lensing & +lensing+ext \\
    \hline 
         $67.31\pm0.96$ & $67.81\pm0.92$ & $67.90\pm0.55$ & $67.27\pm0.66$ & $67.51\pm0.64$ & $67.74\pm0.46$ \\
    \hline
    \multicolumn{6}{c}{\textit{Planck} 2018}\\
    \hline
         TT+lowE & TE+lowE & EE+lowE & TT,TE,EE+lowE & TT,TE,EE+lowE & TT,TE,EE+lowE \\
          &  &  &  & +lensing & +lensing+BAO \\
    \hline 
         $66.88\pm0.92$ & $68.44\pm0.91$ & $69.9\pm2.7$ & $67.27\pm0.60$ & $67.36\pm0.54$ & $67.66\pm0.42$ \\
    \hline
    \end{tabular}
    \caption{Constraints on the Hubble constant from \textit{Planck} 2013/2015/2018 with $\Lambda$CDM model.}
    \label{tab:Planck}
\end{scriptsize}
\end{table}

\subsubsection{CMB other than \textit{Planck} }\label{subsubsec:CMBwoPlanck}

The WMAP nine-year result, $H_0=70.0\pm2.2\,\mathrm{km\,s^{-1}Mpc^{-1}}$~\cite{WMAP:2012nax}, is larger than the \textit{Planck} 2018 result, $H_0=67.27\pm0.60\,\mathrm{km\,s^{-1}Mpc^{-1}}$~\cite{Planck:2018vyg}. However, other ground-based observations independent of the \textit{Planck} experiment also give results similar to \textit{Planck} , such as $H_0=68.8\pm1.5\,\mathrm{km\,s^{-1}Mpc^{-1}}$ from SPT (South Pole Telescope)-3G 2018 data~\cite{SPT-3G:2021eoc} and $H_0=67.9\pm1.5\,\mathrm{km\,s^{-1}Mpc^{-1}}$ from ACT DR4 (Atacama Cosmology Telescope Data Release 4)~\cite{ACT:2020gnv}. Although, the significantly smaller $H_0$ from \textit{Planck} than that from WMAP9 once raised concerns, ACT DR4 TT/TE/EE plus WMAP9 low/intermediate-$\ell$ TT and TE likelihood on scales that ACT does not cover well returns $H_0=67.6\pm1.1\,\mathrm{km\,s^{-1}Mpc^{-1}}$~\cite{ACT:2020gnv}, similar to SPT+WMAP constraint $H_0=67.1\pm0.6\,\mathrm{km\,s^{-1}Mpc^{-1}}$~\cite{SPT-3G:2024atg} and all consistent with \textit{Planck}.

More recently, the non-\textit{Planck} CMB observations have been greatly sharpened to yield constraints comparable to \textit{Planck}'s. Note the caveat here that some ``non-\textit{Planck}'' CMB constraints are not fully \textit{Planck}-free in practice, with the use of \textit{Planck}-derived $\tau_\mathrm{reio}$ priors or \textit{Planck}-based absolute calibration, even when the constraining power on $H_0$ comes from ACT or SPT. For example, the combination of ACT DR6 CMB lensing, galaxy BAO, and a BBN prior gives $H_0=68.3\pm1.1\,\mathrm{km\,s^{-1}Mpc^{-1}}$~\cite{ACT:2023kun}, while the combination of ACT DR6 and \textit{Planck} CMB lensing together with BAO and a BBN prior gives $H_0=68.1\pm1.0\,\mathrm{km\,s^{-1}Mpc^{-1}}$~\cite{ACT:2023kun}. Most recently, the SPT-3G delensed-EE plus lensing (includes \textit{Planck} PR4 $\tau_\mathrm{reio}$ prior and \textit{Planck}-based calibration) reported $H_0=66.81\pm0.81\,\mathrm{km\,s^{-1}Mpc^{-1}}$~\cite{SPT-3G:2024atg}; combining SPT-3G TT/TE/EE spectra with previously published SPT-3G CMB lensing results (includes \textit{Planck} PR4 $\tau_\mathrm{reio}$ prior and \textit{Planck}-based calibration) reported $H_0=66.66\pm0.60\,\mathrm{km\,s^{-1}Mpc^{-1}}$~\cite{SPT-3G:2025bzu}, which was the first time that combined ground-based (SPT+ACT) CMB primary and lensing data have reached \textit{Planck}'s constraining power on the Hubble constant. The combination of these three CMB experiments (\textit{Planck}+ACT+SPT) yields the tightest CMB constraints to date, with $H_0=67.19\pm0.38\,\mathrm{km\,s^{-1}Mpc^{-1}}$~\cite{SPT-3G:2025bzu}. 

Taken together, these results strongly disfavor the idea that the Hubble tension is merely a \textit{Planck}-specific artifact of any CMB systematics. However, a noticeable feature of CMB data~\cite{Planck:2016tof,Planck:2018vyg} is that the low-$\ell$ data prefer a relatively higher $H_0$ while the high-$\ell$ data prefer a relatively lower $H_0$ within $\Lambda$CDM model~\cite{Knox:2019rjx}. For example, a splitting of \textit{Planck} spectra at $\ell=800$ up to which the WMAP has measured would constrain $H_0=70.0\pm1.9\,\mathrm{km\,s^{-1}Mpc^{-1}}$ for $\ell<800$ and $H_0=64.2\pm1.3\,\mathrm{km\,s^{-1}Mpc^{-1}}$ for $\ell>800$. Although this feature is consistent with a statistical fluctuation (or a ``look-elsewhere'' effect), it could as well explain why \textit{Planck}/ACT/SPT constrain a relatively lower $H_0$ than WMAP, as they measure higher and higher multipoles. Usually the lensing effect would be important for higher multipoles, and \textit{Planck} 2018 measured the overall rescaling parameter $A_L$ of the lensing power spectrum to be higher than $\Lambda$CDM expectation $A_L=1$~\cite{Planck:2018vyg}. Note that an $A_L>1$ could be compensated by a negative $\Omega_K$~\cite{DiValentino:2019qzk}, which could be further compensated by a larger growth index $\gamma>0.55$ than $\Lambda$CDM expectation~\cite{Nguyen:2023fip}. Therefore, this peculiar shift in $H_0$ between low/high-$\ell$ might as well point to peculiar physics that operates differently between large/small scales or background/perturbative levels. See also possible correlation to reionization~\cite{Allali:2025wwi,Allali:2025yvp}.

\subsubsection{Without CMB at all}\label{subsubsec:woCMB}

Completely independent of CMB data, the combined BAO+BBN can also constrain the Hubble constant since these two data sets together turn BAO from a relative standard ruler into an absolute distance ladder anchored by early-Universe physics as shown below~\cite{Schoneberg:2019wmt}.

BAO observations come from large-scale-structure galaxy surveys. They record galaxy redshifts, which are converted to distances using a given fiducial model, together with angular positions and other photometric or spectroscopic data, thereby determining the positions of galaxies within the fiducial model. The spatial distribution of galaxies is not completely random. After primordial perturbations enter the horizon, they induce density perturbations in the baryon-photon fluid, and these density perturbations propagate outward from each point in space at the sound speed. Once photons decouple from the baryon-photon fluid, the baryonic component can no longer sustain the propagation of the acoustic oscillation, and the density perturbation is frozen in at a comoving scale of about $r_s\sim150$ Mpc. Later, baryons fall into the gravitational potential wells formed by dark matter and form galaxies. The two-point correlation function of galaxy positions (after reconstruction~\cite{Eisenstein:2006nk,Padmanabhan:2008dd,Padmanabhan:2012hf}) therefore exhibits a local excess at $r_s$ compared with a completely random distribution. Thus, although BAO data are obtained from observations of late-Universe galaxy distributions, the information they encode directly originates from the sound horizon left by the early Universe at the last-scattering surface, or more precisely the sound horizon at the baryon-drag epoch,
\begin{align}
r_d=\int_{z_d}^\infty\frac{c_s(z)\mathrm{d}z}{H(z)},\quad c_s(z)=\frac{c}{\sqrt{3(1+R)}},\quad R=\frac{3\rho_b}{4\rho_\gamma}.
\end{align}
However, BAO does not directly measure this sound horizon; rather, it measures some distance scales only in units of the sound horizon at the baryon drag epoch. More precisely, the two-point correlation functions of BAO observations directly measure the deviations in the directions parallel and perpendicular to the line of sight relative to the radial and transverse BAO length scales predicted by a fiducial model,
\begin{align}
\alpha_{||}(z)&=\frac{D_H(z)/r_d}{D_H^\mathrm{fid}(z)/r_d^\mathrm{fid}},\quad \frac{D_H(z)}{r_d}=\frac{c}{H(z)r_d}=\frac{c}{H_0r_d}\frac{1}{E(z)}\\
\alpha_\perp(z)&=\frac{D_M(z)/r_d}{D_M^\mathrm{fid}(z)/r_d^\mathrm{fid}},\quad  \frac{D_M(z)}{r_d}=\int_0^{z'}\frac{c\mathrm{d}z'}{H(z')r_d}=\frac{c}{H_0r_d}\int_0^{z'}\frac{\mathrm{d}z'}{E(z')}.
\end{align}
Therefore, BAO can only constrain the product $r_dH_0$ as an approximate constant~\cite{Brieden:2022heh}, as well as the uncalibrated expansion history $E(z)$ from $\alpha_\mathrm{AP}=\alpha_{||}/\alpha_\perp$ via the Alcock-Paczy\'{n}ski effect~\cite{Alcock:1979mp}. This is why, from the perspective of early-Universe model building, obtaining a larger Hubble constant generally requires shrinking the sound horizon. To break down the degeneracy of $r_dH_0$ to infer $H_0$ without using the CMB-inferred posterior or prior on the sound horizon~\cite{Vonlanthen:2010cd,Audren:2012wb,Verde:2016wmz,Aylor:2018drw}, the cleanest way is to adopt the BBN observation or the full shape of the BAO spectrum that might require extra inputs (like $\omega_b$ from BBN)~\cite{Philcox:2020vvt,DAmico:2020ods,Brieden:2021edu,Zhang:2021yna,Philcox:2021kcw,Farren:2021grl,Brieden:2022lsd}.

BBN observations, on the other hand, come from measurements of the abundances of light elements in the early Universe, such as the primordial helium abundance $Y_\mathrm{P}^\mathrm{BBN}=4n_\mathrm{He}/n_b$ and the primordial deuterium abundance $y_\mathrm{DP}=10^5n_\mathrm{D}/n_\mathrm{H}$. These are functions of the physical baryon density $\omega_b=\Omega_bh^2$ and the effective number of relativistic neutrino-like degrees of freedom $N_\mathrm{eff}$. This $N_\mathrm{eff}$-$\omega_b$ relation can be obtained from public BBN codes such as \texttt{PArthENoPE}~\cite{Pisanti:2007hk} and \texttt{PRIMAT}~\cite{Pitrou:2018cgg} under the standard model of particle physics, with possible neutrino extensions. Once the observed $Y_\mathrm{P}^\mathrm{BBN}$ and $y_\mathrm{DP}$ determine an allowed confidence region in the $N_\mathrm{eff}$-$\omega_b$ plane, the overlap between this region and the model relation gives the confidence intervals for $N_\mathrm{eff}$ and $\omega_b$. For the minimal assumption with $N_\mathrm{eff}=3.046$, BBN constrains the physical baryon density $\omega_b$. This then determines the baryon-to-photon ratio $R=3\rho_b/(4\rho_\gamma)$ in the sound speed $c_s(z)$, and hence the drag-epoch sound horizon. Thus, BBN supplies an absolute calibration of the BAO ruler without using primary CMB anisotropy spectra themselves. This is why BBN+BAO is often called an inverse distance ladder: instead of starting from local calibrators and climbing outward, one calibrates a high-redshift standard ruler and propagates the distance scale downward to infer $H_0$. 

Some of recent BAO+BBN results are: $H_0=67.6\pm1.1\,\mathrm{km\,s^{-1}Mpc^{-1}}$~\cite{Cuceu:2019for}, $H_0=67.9\pm1.1\,\mathrm{km\,s^{-1}Mpc^{-1}}$~\cite{Ivanov:2019pdj}, $H_0=68.6\pm1.1\,\mathrm{km\,s^{-1}Mpc^{-1}}$~\cite{Philcox:2020vvt}, $H_0=67.35\pm0.97\,\mathrm{km\,s^{-1}Mpc^{-1}}$ \cite{eBOSS:2020yzd}, $H_0=68.19\pm0.99\,\mathrm{km\,s^{-1}Mpc^{-1}}$~\cite{Zhang:2021yna}, $H_0=67.58\pm0.91\,\mathrm{km\,s^{-1}Mpc^{-1}}$ \cite{eBOSS:2021pff}, $H_0=67.6\pm1.0\,\mathrm{km\,s^{-1}Mpc^{-1}}$~\cite{Schoneberg:2022ggi}, $H_0=67.42_{-0.94}^{+0.88}\,\mathrm{km\,s^{-1}Mpc^{-1}}$~\cite{Brieden:2022heh}, and $H_0=68.51\pm0.58\,\mathrm{km\,s^{-1}Mpc^{-1}}$ \cite{DESI:2025zgx}. Again, these results strongly disfavor the idea that the Hubble tension is merely a Planck-specific artifact of early-Universe observations. Nevertheless, the constraining power of BAO+BBN is now mainly limited by BBN systematic uncertainties~\cite{Schoneberg:2024ifp}, as the derived baryon abundance depends nontrivially on the assumed deuterium-burning rates and on which BBN code/rate treatment is adopted, for example, \texttt{PRyMordial}~\cite{Burns:2023sgx}.


\begin{figure}
\centering
\includegraphics[width=0.75\textwidth]{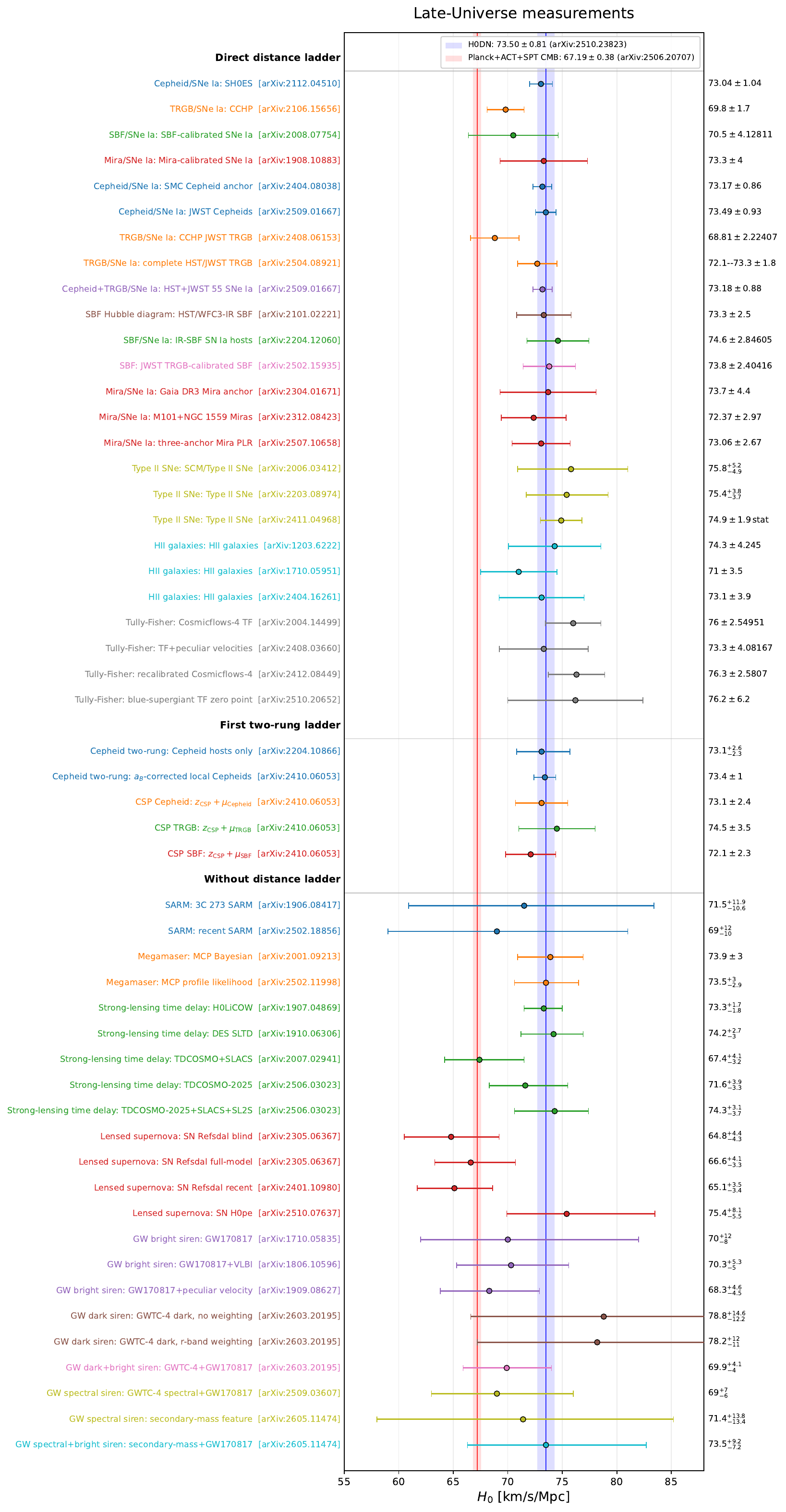}\\
\caption{Late-Universe measurements of the Hubble constant.}\label{fig:LateH0}
\end{figure}

\subsection{Late-Universe measurements}\label{subsec:LateObservation}

Unlike the early-Universe observations described above, observations of the late Universe seem able to measure directly the present Hubble expansion rate, namely the Hubble constant itself. However, the growth of matter perturbations in the late Universe underwent a nonlinear evolutionary stage in complex astrophysical environments. Measurements in the local Universe are therefore affected by more complicated systematic errors, making it difficult to extract the true contribution to the global background expansion from the late local Universe of ours. In this section, we will see that various late-Universe measurements, whether they depend on the full three-rung distance ladder (Sec.~\ref{subsubsec:fullrung}) or the first two-rung distance ladder (Sec.~\ref{subsubsec:tworung}) or even without distance ladder at all (Sec.~\ref{subsubsec:noladder}), appear to form another camp that averages into a systematically higher Hubble-constant value (Fig.~\ref{fig:LateH0}).

\subsubsection{With distance ladders}\label{subsubsec:fullrung}

The key to measuring the Hubble constant in the late local Universe is to determine the distance--redshift relation through distance measurements,
\begin{align}
d_L(z)=\frac{D_L(z)}{c/H_0}\equiv(1+z)\int_0^z\frac{\mathrm{d}z'}{E(z')},
\end{align}
where $E(z)\equiv H(z)/H_0$ depends on the specific input cosmological model parameters. The redshift is usually easy to measure, but distance measurements are otherwise difficult. Different distance-measurement methods, however, are applicable over different ranges. It is therefore necessary to calibrate different distance anchors together to form a direct distance ladder~\cite{Efstathiou:2020wxn}. The lowest rung uses geometric distance measurements~\cite{Follin:2017ljs,Riess:2022mme}, such as trigonometric parallax, masers, and detached eclipsing binaries, to calibrate distance indicators at intermediate distances, such as Cepheids, the tip of the red-giant branch (TRGB)~\cite{Jang:2017dxn}, surface-brightness fluctuations (SBF), and Mira variable stars. On the second rung, these calibrated intermediate-distance indicators can be used as calibrators for more distant indicators, such as SNe Ia, SBF, Type II supernovae, HII galaxies, and the baryonic Tully-Fisher relation. Finally, on the third rung, these calibrated distant indicators can be used to measure the Hubble constant in the Hubble flow of our late Universe.

Currently, the most precise direct distance ladder is built from Cepheid-calibrated SNe Ia by the SH0ES team led by Adam Riess. SNe Ia can be used as standard candles because they are remnants of explosions of carbon-oxygen white dwarfs in binary systems. In such a system, a white dwarf accretes matter from its companion, such as a main-sequence star, subgiant, red giant, or helium star, until it reaches the Chandrasekhar limit, about $1.44$ solar masses. It then reaches the ignition temperature, reignites carbon fusion, and undergoes an explosion. Thus the absolute luminosities at the maximum of their light curves are almost identical (after proper standardization), denoted by $M$. According to the definition of the distance modulus $\mu\equiv m-M$, the apparent magnitude of a SNe Ia is
\begin{align}
m\equiv M+5\lg\frac{D_L}{10\,\mathrm{pc}}
=5\lg d_L(z)+M+5\lg\frac{c}{H_0\,\mathrm{Mpc}}+25
\equiv 5\lg d_L(z)-5a,
\end{align}
where the intercept $M+5\lg(c/H_0/\mathrm{Mpc})+25\equiv -5a$ of this apparent magnitude-logarithmic distance relation can be precisely constrained by the uncalibrated third-rung SNe Ia model-independently. Here, the model-independency is guaranteed by selecting the third-rung SNe Ia in the Hubble flow between $0.023\lesssim z\lesssim0.15$ so that the dimensionless luminosity distance $d_L(z)=z[1+(1/2)(1-q_0)z+\cdots]$ is dominated by the leading redshift term $z$, as the second term is much smaller than the first $(1/2)(1-q_0)z\lesssim0.1$ for $z\lesssim0.13$ and $q_0\sim-0.5$ from \textit{Planck}-$\Lambda$CDM expectation. The lower Hubble-flow redshift limit $z\gtrsim0.023$ is chosen at the transition scale from inhomogeneity to homogeneity to yield a meaningful definition of the Hubble constant to be measured. Therefore, as long as the absolute magnitude $M$ of SNe Ia can be determined alternatively through the first and second rungs of the distance ladder, the Hubble constant can be directly inferred from the well-constrained intercept $-5a$.

In specific: For the first rung, the geometric distance measure $D_\mathrm{geometry}^\mathrm{1st}$ of the corresponding apparent magnitude of the Cepheid $m_\mathrm{Cepheid}^\mathrm{1st}$ determines the absolute magnitude $M_\mathrm{Cepheid}^\mathrm{1st}$ of the first-rung Cepheid, which is assumed to be the same as the second-rung Cepheid $M_\mathrm{Cepheid}^\mathrm{2nd}=M_\mathrm{Cepheid}^\mathrm{1st}$. Next, for the second-rung Cepheid, the absolute magnitude $M_\mathrm{Cepheid}^\mathrm{2nd}$ together with its apparent magnitude $ m_\mathrm{Cepheid}^\mathrm{2nd}$ determines the distance modulus $\mu_\mathrm{Cepheid}^\mathrm{2nd}$ of second-rung Cepheid, which is assumed to be the same as the second-rung SNe Ia $\mu_\mathrm{SNIa}^\mathrm{2nd}=\mu_\mathrm{Cepheid}^\mathrm{2nd}$ in the same host galaxy of Cepheid. Then, this second-rung SNe Ia distance modulus $\mu_\mathrm{SNIa}^\mathrm{2nd}$ together with its apparent magnitude $ m_\mathrm{SNIa}^\mathrm{2nd}$ determines the absolute magnitude $M_\mathrm{SNIa}^\mathrm{2nd}$, which is assumed to be the same as the third-rung SNe Ia $M_\mathrm{SNIa}^\mathrm{3rd}=M_\mathrm{SNIa}^\mathrm{2nd}$. Finally, the highly constrained intercept $-5a^\mathrm{3rd}_\mathrm{SNIa}\equiv M_\mathrm{SNIa}^\mathrm{3rd}+5\lg(c/H_0/\mathrm{Mpc})+25$ from Hubble-flow SNe Ia, together with the previously measured $M_\mathrm{SNIa}^\mathrm{3rd}$, can directly infer the Hubble constant $H_0$. There are three main systematics in this direct distance ladder: 
\begin{enumerate}
    \item the calibration systematics $M_\mathrm{Cepheid}^\mathrm{1st}=M_\mathrm{Cepheid}^\mathrm{2nd}$ with no evolution of Cepheid from the first to second rungs (see, however, Refs.~\cite{Mortsell:2021nzg,Mortsell:2021tcx,Wojtak:2022bct}), which is limited by small samples;
    \item the calibration systematics $M_\mathrm{SNIa}^\mathrm{2nd}=M_\mathrm{SNIa}^\mathrm{3rd}$ with no evolution of SNe Ia from the second to third rungs~\cite{Murakami:2023xuy}, whose standardization is dominated by host-mass corrections; 
    \item the statistical systematics in constraining the intercept $-5a_\mathrm{SNIa}^\mathrm{3rd}\approx m_\mathrm{SNIa}^\mathrm{3rd}-5\lg z_\mathrm{SNIa}^\mathrm{3rd}$, which can be significantly pinned down due to the large sample.
\end{enumerate}
With SH0ES calibrations of SNe Ia absolute magnitude, most late-time non-transition models converge to $73$--$74$ $\mathrm{km\,s^{-1}Mpc^{-1}}$~\cite{Dhawan:2020xmp}. Therefore, what really matters is the calibration of the SNe Ia absolute magnitude. For example, calibrating SNe Ia with Cepheids, TRGB, SBF, and Miras yields $H_0=73.04\pm1.04\,\mathrm{km\,s^{-1}Mpc^{-1}}$~\cite{Riess:2021jrx}, $H_0=69.8\pm1.7\,\mathrm{km\,s^{-1}Mpc^{-1}}$~\cite{Freedman:2021ahq}, $H_0=70.50\pm2.37\mathrm{(stat)}\pm3.38\mathrm{(sys)}\,\mathrm{km\,s^{-1}Mpc^{-1}}$~\cite{Khetan:2020hmh}, and $H_0=73.3\pm4.0\,\mathrm{km\,s^{-1}Mpc^{-1}}$~\cite{Huang:2019yhh}, respectively. Hence, the $H_0$ tension is also referred to as the $M_B$ tension.

Recently, significant progress has been made for this direct distance ladder with different calibrators. For the \textbf{Cepheid-calibrated ladder}, a new geometric distance anchor is provided by 88 Cepheid variables in the Small Magellanic Cloud from HST to yield a joint measurement $H_0=73.17\pm0.86\,\mathrm{km\,s^{-1}Mpc^{-1}}$~\cite{Breuval:2024lsv}, and James Webb Space Telescope (JWST) Cepheid observations in the perfect host with no background contamination provides the most rigorous test of Cepheid distances with measurement $H_0=73.49\pm0.93\,\mathrm{km\,s^{-1}Mpc^{-1}}$~\cite{Riess:2025chq}, since JWST analysis has rejected the unrecognized crowding in HST Cepheid photometry and distance-dependent bias or offset at $7s\sim8\sigma$~\cite{Riess:2023bfx,Riess:2024ohe,Riess:2024vfa}. For the \textbf{TRGB-calibrated ladder}, the Chicago-Carnegie Hubble Program (CCHP) reported $H_0=68.81\pm1.79\mathrm{(stat)}\pm1.32\mathrm{(sys)}\,\mathrm{km\,s^{-1}Mpc^{-1}}$~\cite{Freedman:2024eph} with inclusion of JWST TRGB, while the SH0ES team reported $H_0=72.1-73.3\pm1.8\,\mathrm{km\,s^{-1}Mpc^{-1}}$~\cite{Li:2025lfp} for the \textit{complete} HST/JWST TRGB sample with less sample-selection artifact. The latest combination of HST+JWST Cepheid and 35 TRGB calibrations from HST+JWST spanning 55 SNe Ia concludes $H_0=73.18\pm0.88\,\mathrm{km\,s^{-1}Mpc^{-1}}$~\cite{Riess:2025chq}, yielding almost $6\sigma$ tension.

For the \textbf{SBF-calibrated ladder}, a direct HST/WFC3-IR SBF Hubble diagram (in replacing SNe Ia) for 63 early-type galaxies out to 100 Mpc, with Cepheid/TRGB zero-point calibration, measured $H_0=73.3\pm0.7\mathrm{(stat)}\pm2.4\mathrm{(sys)}\,\mathrm{km\,s^{-1}Mpc^{-1}}$~\cite{Blakeslee:2021rqi}; IR-SBF distances to SN Ia hosts used to test/calibrate the SN Ia rung measured $H_0=74.6\pm0.9\mathrm{(stat)}\pm2.7\mathrm{(sys)}\,\mathrm{km\,s^{-1}Mpc^{-1}}$~\cite{Garnavich:2022hef}; JWST TRGB distances tied to the NGC 4258 megamaser calibrate the HST SBF zero point, independent of Cepheids and SNe Ia, and measured $H_0=73.8\pm0.7\mathrm{(stat)}\pm2.3\mathrm{(sys)}\,\mathrm{km\,s^{-1}Mpc^{-1}}$~\cite{Jensen:2025aai}. For the \textbf{Mira-calibrated ladder}, Gaia DR3 Milky-Way O-rich Mira PLR as an additional anchor for the Mira variables in the Type Ia host galaxy NGC 1559 measured $H_0=73.7\pm4.4\,\mathrm{km\,s^{-1}Mpc^{-1}}$~\cite{Sanders:2023jkl}; using M101 and NGC 1559 as calibrator hosts reported $H_0=72.37\pm2.97\,\mathrm{km\,s^{-1}Mpc^{-1}}$~\cite{Huang:2023frr}; employing a three-anchor solution (LMC, NGC 4258, and 41 O-rich Miras in 18 globular clusters) to determine distances to the same M101 and NGC 1559 as calibrator hosts measured $H_0=73.06\pm2.67\,\mathrm{km\,s^{-1}Mpc^{-1}}$~\cite{Bhardwaj:2025kbw}. Both of the above SBF- and Mira-calibrated ladders are consistent with the Cepheid-based SH0ES measurement.

In addition to replacing the calibrators, it is then intriguing to replace the distance indicators, for example, SBF (Hubble diagram mentioned above), Type II SNe, HII galaxies, and the baryonic Tully-Fisher relation. 
\textbf{Type II SN}~\cite{Rodriguez:2018wsx} can measure $H_0$~\cite{deJaeger:2023vkm} as their hydrogen-rich, core-collapse ejecta enter a photospheric/plateau phase whose luminosity can be standardized or geometrically inferred. Physically, the plateau is set by hydrogen recombination in an expanding red-supergiant envelope; brighter SNe II maintain larger photospheric radii and higher expansion velocities, giving a luminosity-velocity correlation. Recent measurements include $H_0=75.8_{-4.9}^{+5.2}\,\mathrm{km\,s^{-1}Mpc^{-1}}$~\cite{deJaeger:2020zpb}, $H_0=75.4_{-3.7}^{+3.8}\,\mathrm{km\,s^{-1}Mpc^{-1}}$~\cite{deJaeger:2022lit}, and $H_0=74.9\pm1.9\mathrm{(stat)}$~\cite{Vogl:2024bum}. 
\textbf{HII galaxies} can measure $H_0$ as the compact starburst galaxies and giant extragalactic HII regions can be standardizable candles through the empirical luminosity-velocity-dispersion relation $\log L(\mathrm{H}\beta)=\alpha\log\sigma+\kappa+\cdots$, where the Balmer-line luminosity $L(\mathrm{H}\beta)$ is set by the ionizing photon output of young massive clusters, and the ionized-gas velocity dispersion $\sigma$ scales with the mass of the young star-forming system. After correcting the observed Balmer flux for extinction, underlying absorption, aperture effects, age/equivalent-width effects, metallicity/size where relevant, and peculiar velocities at low redshift, one infers $D_L=[L(\mathrm{H}\beta)/4\pi F(\mathrm{H}\beta)]^{1/2}$, and hence a Hubble diagram. Locally, the zero point is calibrated with giant HII regions in galaxies whose distances are known from Cepheids, TRGB, or NGC 4258-like geometric anchors; the slope is constrained by nearby HII galaxies. Recent measurements include $H_0=74.3\pm3.1\mathrm{(stat)}\pm2.9\mathrm{(sys)}\,\mathrm{km\,s^{-1}Mpc^{-1}}$~\cite{Chavez:2012km}, $H_0=71.0\pm2.8\mathrm{(stat)}\pm2.1\mathrm{(sys)}\,\mathrm{km\,s^{-1}Mpc^{-1}}$~\cite{FernandezArenas:2017dux}, and $H_0=73.1\pm3.9\,\mathrm{km\,s^{-1}Mpc^{-1}}$~\cite{Chavez:2024twa}.

\textbf{The Tully-Fisher relation} measures $H_0$ by using spiral galaxies as distance indicators. A rotationally supported disk galaxy with a larger total gravitational mass rotates faster and is also more luminous, or more baryon-rich. Observationally, the inclination-corrected HI 21 cm linewidth gives a distance-independent proxy for $2V_\mathrm{rot}$, while optical/infrared photometry gives an apparent magnitude. A calibrated relation of the form, $M=a+b(\log W-2.5)+\cdots$, then predicts the absolute magnitude $M$, hence the distance modulus $\mu=m-M$, and therefore the distance $D$. Comparing the cosmological recession velocity (corrected for peculiar motion) with this distance gives $H_0\sim v_\mathrm{cos}/D$. In the baryonic Tully-Fisher version, luminosity is replaced by total baryonic mass $M_b=M_\star+M_\mathrm{gas}$, motivated by the empirical relation $M_b\propto V_\mathrm{rot}^4$. The method is therefore a galaxy-based distance ladder, mostly independent of SNe Ia standardization (see also fundamental plane relation~\cite{Scolnic:2024hbh}), although its absolute zero point still requires calibrators such as Cepheids, TRGB, SNe Ia, masers, or other primary distance indicators. The recent Cosmicflows-4 calibration of optical and infrared Tully-Fisher relations measured $H_0=76.0\pm1.1\mathrm{(stat)}\pm2.3\mathrm{(sys)}\,\mathrm{km\,s^{-1}Mpc^{-1}}$~\cite{Kourkchi:2020iyz}. Recently, an improved Tully-Fisher estimate of $H_0=73.3\pm2.1\mathrm{(stat)}\pm3.5\mathrm{(sys)}\,\mathrm{km\,s^{-1}Mpc^{-1}}$~\cite{Boubel:2024cqw} is obtained by fitting the Tully-Fisher relation and the peculiar-velocity field simultaneously, including selection effects, and using the full Cosmicflows-4 $i$-band and W1-band samples, which was recalibrated/reanalyzed in Ref.~\cite{Scolnic:2024oth} to yield $H_0=76.3\pm2.1\mathrm{(stat)}\pm1.5\mathrm{(sys)}\,\mathrm{km\,s^{-1}Mpc^{-1}}$. Most recently, using blue-supergiant distances from the flux-weighted gravity-luminosity relation to calibrate the Tully-Fisher zero point, Ref.~\cite{Kudritzki:2025hly} found $H_0=76.2\pm6.2\,\mathrm{km\,s^{-1}Mpc^{-1}}$, with the large error driven by the still-small blue-supergiant calibrator sample.

\subsubsection{First two-rung ladders}\label{subsubsec:tworung}

It is possible to construct a two-rung Cepheid distance ladder by removing the usual third rung of SNe Ia and measuring $H_0$ directly from geometrically calibrated Cepheid distances and the redshifts of their host galaxies~\cite{Kenworthy:2022jdh}. The key point is that once Cepheid absolute luminosities are anchored by geometric distances in the Milky Way, NGC 4258, and the LMC, the Cepheid distance modulus gives each host-galaxy distance, while the corrected host-galaxy redshift gives its recession velocity; therefore $H_0$ is simply the relative scale between Cepheid distances and redshifts in the low-redshift Hubble law. This provides an important test of whether SNe Ia calibration or population systematics could be responsible for the Hubble tension. The price is that Cepheid hosts lie at much lower redshift, $z\sim0.006$, so peculiar velocities are no longer negligible and can be $\sim20\%$ of the observed redshift. 

To avoid SNe systematics, Ref.~\cite{Kenworthy:2022jdh} uses 35 SH0ES Cepheid hosts with homogeneous HST photometry, geometrical Cepheid anchors, Pantheon+ group-corrected CMB-frame redshifts, two external peculiar-velocity reconstructions of the nearby density field, a full covariance model for correlated large-scale-structure velocities, and a hierarchical Bayesian model with latent true distances. It also explicitly models selection effects, distinguishing distance-limited and redshift-limited samples, because Malmquist-type biases become important at such low redshift. The main result is a fiducial two-rung constraint $H_0=73.1_{-2.3}^{+2.6}\,\mathrm{km\,s^{-1}Mpc^{-1}}$, with different selection assumptions giving central values between about $H_0=71.8-77.0\,\mathrm{km\,s^{-1}Mpc^{-1}}$; using peculiar-velocity reconstructions reduces the fractional uncertainty from roughly $6\%$ to roughly $3\%$. Therefore, it is not that the two-rung ladder is more precise than the standard three-rung ladder, but that even after removing SNe Ia entirely, the inferred $H_0$ remains high and in $2.6\sigma$ tension with Planck, making unknown SN Ia systematics an unlikely explanation of the Hubble tension (see also~\cite{Stiskalek:2025ktq}). 

With a similar conclusion as Ref.~\cite{Kenworthy:2022jdh}, we argue in Ref.~\cite{Huang:2024gfw} that the Hubble tension can be sharpened from a broad \textit{Planck}-SH0ES discrepancy into a more specific conflict between Planck-CMB inference and the first two rungs of the local distance ladder. The key observation is that the third-rung PantheonPlus SNe Ia show a stable intercept $-5a_B$ calibrated by either \textit{Planck} or SH0ES, while an apparent $a_B$ anomaly is identified for local Cepheid-hosted second-rung SNe Ia. If this is caused by a redshift-side systematic, most plausibly peculiar-velocity-induced bias at $z\lesssim0.01$, then rebuilding the $a_B$ consistency for the second-rung SNe Ia with corrected local redshifts and using Cepheid distance moduli alone can infer $H_0=73.4\pm1.0\,\mathrm{km\,s^{-1}Mpc^{-1}}$, without relying on Hubble-flow SNe Ia. The key difference is that Ref.~\cite{Kenworthy:2022jdh} used explicit density/velocity-field reconstructions and selection modeling, while Ref.~\cite{Huang:2024gfw} used $a_B$ consistency as a simpler empirical redshift-correction criterion and obtains a tighter two-rung constraint. The Carnegie Supernova Project (CSP) cross-check with Cepheid, TRGB, and SBF calibrators further supports our method, and arrived at consistent measurements with SH0ES,
\begin{align}
H_0=\begin{cases}
73.1\pm2.4\;\mathrm{km/s/Mpc}, & z_\mathrm{CSP}+\mu_\mathrm{Cepheid},\\
74.5\pm3.5\;\mathrm{km/s/Mpc}, &   z_\mathrm{CSP}+\mu_\mathrm{TRGB},\\
72.1\pm2.3\;\mathrm{km/s/Mpc}, & z_\mathrm{CSP}+\mu_\mathrm{SBF}.
\end{cases}
\end{align}
In particular, for the TRGB case, the result is obtained by removing four TRGB calibrators with large deviations from the average intercept of the whole sample. This confirms the sample selection effect in the TRGB-calibrated ladder that either completing~\cite{Li:2025lfp} or properly removing~\cite{Huang:2024gfw} TRGB samples could all agree with SH0ES~\cite{Kenworthy:2022jdh} instead of CCHP~\cite{Freedman:2019jwv}.

\subsubsection{Without distance ladder}\label{subsubsec:noladder}

The construction of a multi-rung distance ladder requires calibration between two luminosity-distance indicators on neighboring rungs in anchor galaxies. Consequently, calibration errors are inevitably propagated from rung to rung and introduce considerable observational and systematic errors at the final rung. Over the past two decades, the SH0ES team has made extraordinary efforts to reduce the total calibration error of all rungs of the distance ladder to below $1\%$. Nevertheless, a method that avoids the construction of a distance ladder and realizes direct distance measurement at large distances would significantly reduce observational and systematic errors in measuring $H_0$ in late/local Universe. Such methods include spectroastrometry and reverberation mapping, Megamasers, the Tully-Fisher relation, strong-lensing time delays, and gravitational-wave standard sirens, as well as FRB~\cite{Wu:2021jyk,James:2022dcx}.

The spectroastrometry and reverberation mapping (SARM)~\cite{Wang:2019gaq} of broad-line regions (BLRs) constitutes a novel means to probe the geometric distance of active galactic nuclei (AGNs), which has recently become practically feasible owing to successful interferometric observations with the Very Large Telescope Interferometer/GRAVITY. The BLR is a compact distribution of photoionized gas clouds orbiting the central supermassive black hole. When the ionizing continuum varies, the broad emission lines respond after a light-travel-time delay; reverberation mapping therefore gives a physical/linear BLR scale, roughly $R_\mathrm{BLR}\sim c\tau$. Meanwhile, spectroastrometry with VLTI/GRAVITY measures the wavelength-dependent photocenter shift across a broad line, i.e. the angular structure and kinematics of the BLR on tens-of-microarcsecond scales. Combining the two gives a geometric angular-diameter distance, $D_A\sim \Delta R/\Delta\theta$, which can then be converted to $H_0$ through the redshift-distance relation $D_A(z;H_0,\Omega_m,\Omega_\Lambda)$. This is why the method is often described as an AGN/quasar ``parallax'' or ``geometric distance'' method, although it is not annual parallax. The first such measurement~\cite{Wang:2019gaq} used the quasar 3C 273 and reported $D_A(\mathrm{3C\, 273})=551.5_{-78.7}^{+97.3}$ Mpc, leading, for a Planck-like flat $\Lambda$CDM background, to $H_0=71.5_{-10.6}^{+11.9}\,\mathrm{km\,s^{-1}Mpc^{-1}}$. Most recently, Ref.~\cite{SARM:2025ffz} reported $H_0=69_{-10}^{+12}\,\mathrm{km\,s^{-1}Mpc^{-1}}$. In the future, GRAVITY+ and a larger AGN sample could make SARM a competitive, independent geometric probe of $H_0$.

The Megamaser method is one of the cleanest geometric routes to $H_0$ directly as a single-rung distance ladder. It uses 22 GHz water maser emission from sub-parsec, nearly edge-on circumnuclear disks around supermassive black holes. VLBI maps the angular positions and line-of-sight velocities of individual maser spots; long-term single-dish monitoring measures their centripetal accelerations. In the simplest Keplerian picture, high-velocity masers on the disk midline determine $v_\mathrm{rot}(\theta)$, while systemic masers on the near side give $a_\mathrm{los}\simeq v_\mathrm{rot}^2/r$. Comparing the physical radius $r$ inferred from dynamics with the angular radius $\theta$ gives an angular-diameter distance, schematically, $D_A\sim r/\theta\sim v_\mathrm{rot}^2/(a_\mathrm{los}\theta)$. For galaxies sufficiently in the Hubble flow, this geometric distance is combined with recession velocity to infer $H_0$. The Megamaser Cosmology Project (MCP) used geometric distances to six maser galaxies well into the Hubble flow to measure $H_0=73.9\pm3.0\,\mathrm{km\,s^{-1}Mpc^{-1}}$~\cite{Pesce:2020xfe}, and its frequentist profile-likelihood re-analysis of the same MCP Megamaser sample, which recovered essentially the same $H_0=73.5_{-2.9}^{+3.0}\,\mathrm{km\,s^{-1}Mpc^{-1}}$~\cite{Barua:2025zqo} as the original Bayesian result.

Strong-lensing time delay (SLTD)~\cite{Birrer:2022chj,Treu:2022aqp,Treu:2023mih} measures the Hubble constant by measuring the arrival-time differences between different lensed images in a strong gravitational-lensing system. The source in such a system is usually a quasar or even a supernova, while the lens is a foreground galaxy. When light from the source passes near the lens, the deflection by the gravitational potential well produces multiple images of the same source when the observed light rays are traced backward. These multiple images are generally arranged asymmetrically, so different light paths have different path lengths and hence different travel times, known as the geometric time delay. General relativistic effects also introduce a Shapiro time delay caused by the change in the effective propagation speed of light. The total time delay is therefore $\Delta t=\frac{1}{c}D_{\Delta t}\Delta\phi$, where the time-delay distance
\begin{align}
D_{\Delta t}=\frac{(1+z_d)D_A(z_d)D_A(z_s)}{D_A(z_s)-\frac{1+z_d}{1+z_s}D_A(z_d)}\propto\frac{1}{H_0}
\end{align}
depends on the angular-diameter distances from the observer to the lens and source and from the lens to the source, while the Fermat potential $\Delta\phi$ of the lens galaxy is obtained by estimating the radial mass-profile function of the galaxy. The main theoretical/systematic issue is the mass-sheet degeneracy~\cite{Birrer:2020jyr,Denzel:2020zuq}: a transformation of the lens surface density can leave image positions and shapes almost invariant while rescaling the inferred time-delay distance, and hence $H_0$. For example, the radial mass-profile function was initially assumed to be a power law. Under this assumption, the H0LiCOW ($H_0$ Lenses in COSMOGRAIL's Wellspring) team measured $H_0=73.3_{-1.8}^{+1.7}\,\mathrm{km\,s^{-1}Mpc^{-1}}$~\cite{Wong:2019kwg}, while the DES (Dark Energy Survey) team measured $H_0=74.2_{-3.0}^{+2.7}\,\mathrm{km\,s^{-1}Mpc^{-1}}$~\cite{DES:2019fny}. However, when the TDCOSMO (Time-Delay Cosmography) team used the SLACS (Sloan Lens Advanced Camera for Surveys) galaxy sample, which does not show strong-lensing phenomena, to infer the actual radial mass-profile function of lens galaxies, the resulting Hubble constant decreased to $H_0=67.4_{-3.2}^{+4.1}\,\mathrm{km\,s^{-1}Mpc^{-1}}$~\cite{Birrer:2020tax}. The most recent TDCOSMO milestone cosmological analysis gives a result around $H_0=71.6_{-3.3}^{+3.9}\,\mathrm{km\,s^{-1}Mpc^{-1}}$~\cite{TDCOSMO:2025dmr} for the TDCOSMO-2025 time-delay sample in flat $\Lambda$CDM when combined with Pantheon+ supernova constraints on $\Omega_m$. Including non-time-delay lenses from SLACS and SL2S with kinematic information improves the constraint to $H_0=74.3_{-3.7}^{+3.1}\,\mathrm{km\,s^{-1}Mpc^{-1}}$~\cite{TDCOSMO:2025dmr}. \textbf{Strongly lensed supernovae} are now becoming a new complementary route to $H_0$~\cite{Vega-Ferrero:2017yqr,Grillo:2018ume,Pierel:2019pnr,Kelly:2023wzm}. For SN Refsdal, cluster-lens time-delay modeling gave $H_0=64.8_{-4.3}^{+4.4}\,\mathrm{km\,s^{-1}Mpc^{-1}}$ in a blind analysis, with a full-model combined value near $H_0=66.6_{-3.3}^{+4.1}\,\mathrm{km\,s^{-1}Mpc^{-1}}$~\cite{Kelly:2023mgv}, both of which agree well with the most recent constraint $H_0=65.1_{-3.4}^{+3.5}\,\mathrm{km\,s^{-1}Mpc^{-1}}$~\cite{Grillo:2024rhi}. For the JWST-discovered multiply imaged SNe Ia (SN H0pe), lens models yield a higher value, approximately $H_0=75.4_{-5.5}^{+8.1}\,\mathrm{km\,s^{-1}Mpc^{-1}}$~\cite{Agrawal:2025tuv}. The special advantage of lensed SNe Ia is that their standardizable luminosity can constrain magnification and help break lens-model degeneracies that are harder for quasars.

Gravitational-wave (GW) standard sirens~\cite{Schutz:1986gp,Krolak:1987ofj,Sathyaprakash:2009xs} (see Ref.~\cite{Jin:2025dvf} for a recent review) realize direct distance measurement at cosmological distances by using the GW waveform emitted during the inspiral phase of compact binary systems,
\begin{align}
h_\mathrm{GW}\propto\frac{\cos\iota}{D_L(z)}\left(\frac{G\mathcal{M}_c(z)}{c^2}\right)^{5/3}\left(\frac{\pi f}{c}\right)^{2/3}
\sin[\Phi(f)],
\end{align}
where $M_c(z)$ is the redshifted chirp mass $M_c=(m_1m_2)^{3/5}/(m_1+m_2)^{1/5}\propto f^{-11/5}\dot{f}^{3/5}$, $D_L(z)$ is the luminosity distance, $\iota$ is the inclination angle of the binary orbital plane relative to our line of sight, and $\Phi(f)$ is the phase. Note that $D_L$ is degenerated with $\iota$. By template-matching the measured inspiral GW signal with the waveform above, the luminosity distance $D_L$ can be measured directly. 
A \textbf{bright siren} is a GW event with a unique electromagnetic counterpart and host galaxy, as in a binary containing a neutron star, and if its electromagnetic signal can be observed, namely if the orbital inclination is smaller than about $20$ degrees, then its redshift $z$ is known. The Hubble constant can then be measured directly from the Hubble--Lema\^{i}tre law $cz=H_0D_L$ at low redshift. For example, the binary neutron-star merger GW170817~\cite{TheLIGOScientific:2017qsa} associated with GRB 170817A/AT2017gfo in NGC 4993 gave measurements of $H_0=70.0_{-8.0}^{+12.0}\,\mathrm{km\,s^{-1}Mpc^{-1}}$~\cite{Abbott:2017xzu}, $H_0=70.3_{-5.0}^{+5.3}\,\mathrm{km\,s^{-1}Mpc^{-1}}$~\cite{Hotokezaka:2018dfi}, and $H_0=68.3_{-4.5}^{+4.6}\,\mathrm{km\,s^{-1}Mpc^{-1}}$~\cite{Mukherjee:2019qmm}. 
A \textbf{dark siren} has no unique electromagnetic counterpart. The GW event gives a 3D localization volume: sky position plus luminosity-distance posterior. One then cross-matches this volume with a galaxy catalog~\cite{Oguri:2016dgk,Mukherjee:2020hyn,Borhanian:2020vyr}. Each galaxy in the localization region is treated as a possible host with a measured or photometric redshift, weighted by sky localization, distance consistency, catalog completeness, and sometimes by luminosity or stellar mass. A single dark siren gives a broad, often multimodal $H_0$ posterior; many events can combine statistically~\cite{DES:2020nay}. For instance, by networking the LISA space gravitational-wave detector of ESA/NASA with China's Taiji/TianQin detector~\cite{Ruan:2019tje,Ruan:2020smc,Cai:2023ywp,Jin:2023sfc}, the dark-siren method is expected, with five years of network observations, to improve the precision of the Hubble-parameter constraint to below $1\%$~\cite{Wang:2020dkc}. The most recent dark siren constraints from the fourth Gravitational-Wave Transient Catalog (GWTC-4) reported $H_0=78.8_{-12.2}^{+14.6}\,\mathrm{km\,s^{-1}Mpc^{-1}}$ without luminosity weighting and $H_0=78.2_{-11.0}^{+12.0}\,\mathrm{km\,s^{-1}Mpc^{-1}}$ applying $r$-band luminosity weighting, which, after combined with the bright siren GW170817, led to the final constraint $H_0=69.9_{-4.0}^{+4.1}\,\mathrm{km\,s^{-1}Mpc^{-1}}$~\cite{Alfradique:2026dsq} (see, e.g., Ref.~\cite{Palmese:2021mjm,LIGOScientific:2021aug,Mukherjee:2022afz} for previous measurements). 
A \textbf{spectral siren}~\cite{Stiskalek:2025ibp,Mastrogiovanni:2023emh,Pierra:2023deu,Farah:2024xub,Mali:2024wpq} is a physically distinct method from dark sirens. If the intrinsic source-frame compact-object mass distribution has identifiable features, e.g. neutron-star mass scale, black-hole mass gaps, pair-instability features, or population peaks, then changing $H_0$ changes the inferred source-frame mass distribution. A hierarchical fit to cosmology plus population hyperparameters can therefore infer $H_0$ without a unique host galaxy. The most recent GWTC-4 spectral-siren analyses reported $H_0=69_{-6}^{+7}\,\mathrm{km\,s^{-1}Mpc^{-1}}$~\cite{MaganaHernandez:2025cnu} when combined with GW170817, while secondary-mass-feature analysis~\cite{Li:2026amt} reports $H_0=71.4_{-13.4}^{+13.8}\,\mathrm{km\,s^{-1}Mpc^{-1}}$ from spectral sirens alone, and $H_0=73.5_{-7.2}^{+9.2}\,\mathrm{km\,s^{-1}Mpc^{-1}}$ when combined with the bright siren GW170817. 
A \textbf{tidal siren} uses neutron-star tidal effects to break the mass-redshift degeneracy internally~\cite{Chatterjee:2021xrm}. In a binary neutron-star waveform, the point-particle inspiral accurately measures redshifted masses, while the late inspiral also contains finite-size tidal corrections governed by dimensionless tidal deformabilities, schematically related to neutron-star compactness and EoS. If the neutron-star EoS, or sufficiently tight universal relations, are known, the tidal phase information constrains the source-frame masses. Comparing source-frame masses with detector-frame masses gives $1+z$, and hence $H_0$ from $D_L(z)$. All these siren methods measure $H_0$ with their central values compatible with both Planck and local-distance-ladder values, but the uncertainties remain too large to adjudicate the Hubble tension. \textbf{Stochastic siren}~\cite{Cousins:2025bas} uses GW background from binary black hole mergers.

\section{From Hubble crisis to Hubble dilemma}\label{sec:Resolution}

Although the constraints on the Hubble constant from early- and late-Universe observations are not identical, one trend cannot be ignored: direct measurements of the Hubble constant in the late Universe are systematically higher than global-fitting values from the early Universe. Because the observational and systematic errors of different probes are very different, it is difficult to imagine a common observational systematic error that would produce such a systematic deviation. If there is, it could better be some theoretical ``systematics'' (new physics) that disguises itself as unrecognized observational systematics.

The simplest new-physics constructions are simple and direct extensions of the standard cosmological model, such as introducing a small spatial curvature $\Omega_K$, introducing late-Universe dynamical dark energy through the Chevallier-Polarski-Linder (CPL) parametrization $w=w_0+w_a(1-a)$~\cite{Chevallier:2000qy,Linder:2002et}, introducing a small amount of new neutrino-like relativistic degrees of freedom before BBN $N_\mathrm{eff}$, and combinations of the above. However, many studies, for example Refs.~\cite{Guo:2018ans,DiValentino:2019dzu,Okamatsu:2021jil}, have shown that simple extensions of the standard cosmological model merely enlarge the uncertainties of model parameters and are still insufficient to fully resolve the Hubble-constant crisis. 

\begin{figure}[!htbp]
    \centering
    \includegraphics[width=\linewidth]{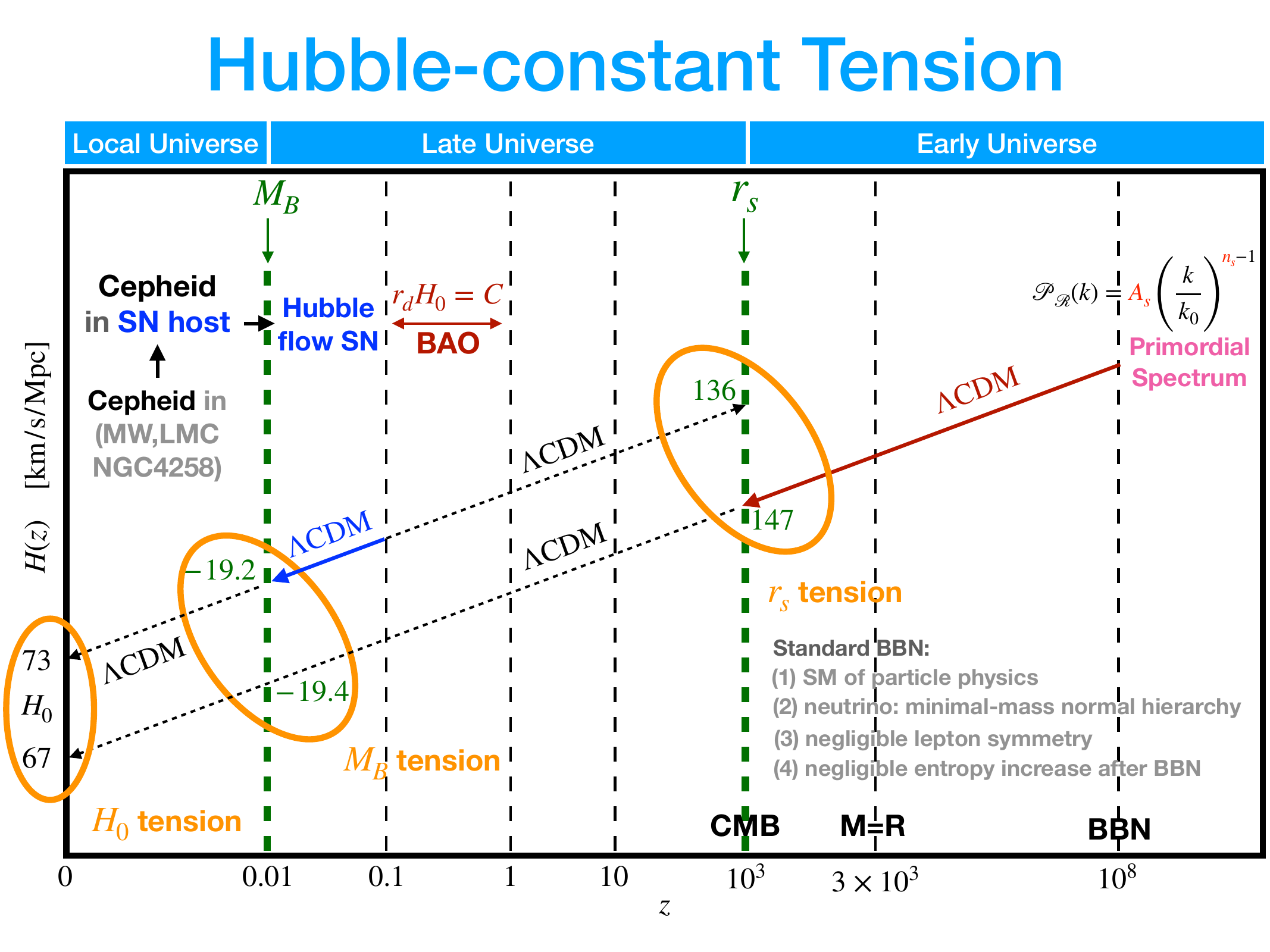}
    \caption{A schematic illustration of the Hubble expansion rate $H(z)$ with respect to redshift. The extrapolations to the present day along $\Lambda$CDM model (diagonal dotted lines) between early-Universe CMB constraint (diagonal red line) and late-Universe Cepheid-SNe Ia measurement (diagonal blue line) leads to the Hubble tension,  which can be transformed into either sound-horizon ($r_s$) tension or absolute-magnitude ($M_B$) tension, depending on the specific causes when tracing the Hubble tension.}
    \label{fig:H0Tension}
\end{figure}

It is therefore necessary to introduce highly nontrivial modifications of the standard cosmological model, such as new energy-density components, new interaction forms, new modified-gravity effects, or even attempts that change the time evolution of fundamental physical constants or challenge the cosmological principle itself. Since the Hubble tension can be roughly regarded as a contradiction between current observations from the early Universe and those of the late Universe (see Fig.~\ref{fig:H0Tension}), its model constructions can also be roughly divided into early-Universe solutions (Sec.~\ref{subsec:EarlyResolution}) and late-Universe modifications (Sec.~\ref{subsec:LateResolution}).

\subsection{Early-Universe solutions}\label{subsec:EarlyResolution}

If we trace the cause of the Hubble tension to the early Universe alone, then we expect not to change the late Universe, that is, leaving the dimensionless comoving angular-diameter distance to the last scattering surface $z_*$, $d_M(z_*)=H_0D_M(z_*)=\int_0^{z_*}\mathrm{d}z/E(z)$, unchanged. For these early-Universe solutions to at least satisfy CMB and BAO constraints, we expect the angular scale of the sound horizon, $\theta_*=r_s(z_*)/D_M(z_*)=H_0r_s/d_M$, is unchanged, which is also consistent with a constant BAO degeneracy $H_0r_s=\mathrm{const.}$. Therefore, increasing $H_0$ while keeping $\theta_*$, $H_0r_s$, and $d_M(z_*)$ unchanged would require reducing the sound horizon $r_s$.

\begin{figure}[!htbp]
    \centering
    \includegraphics[width=\linewidth]{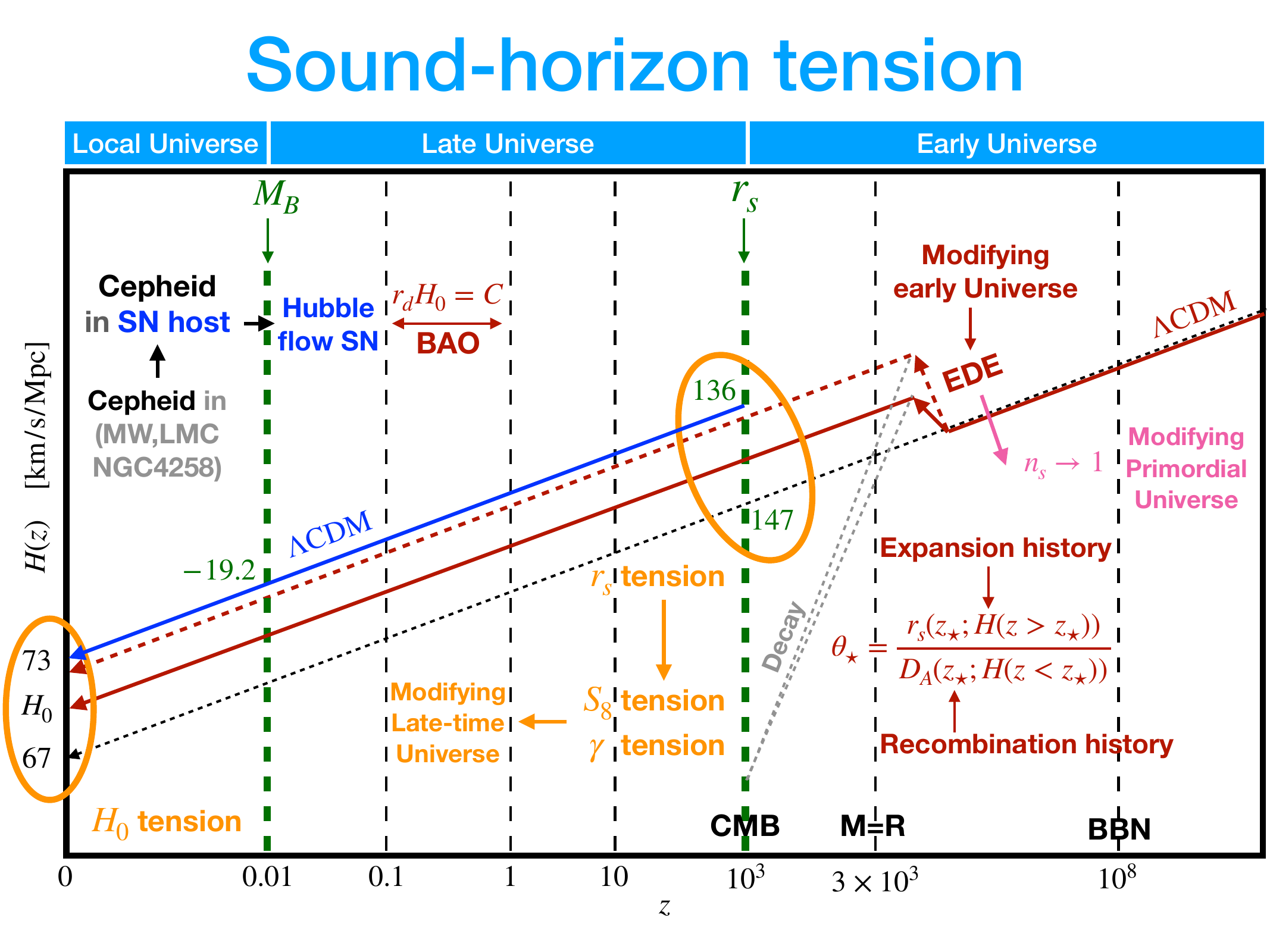}
    \caption{A schematic illustration of the Hubble-constant ($H_0$) tension as the sound-horizon ($r_s$) tension and its early-Universe solutions. Any original intention to modify the early Universe alone would necessarily end up with modifying both the primordial Universe and the late Universe, to produce an extremely scale-invariant primordial scalar power spectrum and to evade linear amplitude and growth rate tensions in matter perturbations, respectively.}
    \label{fig:rsTension}
\end{figure}

There are two ways to reduce the sound horizon to solve this sound-horizon ($r_s$) tension (Fig.~\ref{fig:rsTension} and Sec.~\ref{subsubsec:rstension}): one is to shorten the duration of sound-wave propagation, and the other is to directly reduce the sound speed itself. The duration of sound-wave propagation can be shortened by modifying the recombination history during photon decoupling, thereby making recombination occur earlier and reducing the sound horizon. The sound speed can be reduced by modifying the pre-decoupling expansion history, thereby changing the relative amounts of radiation and baryonic matter in the baryon-to-photon ratio. However, such early-Universe modifications would necessarily encounter a series of ``Early-time no-go theorems''~\cite{Krishnan:2020obg,Jedamzik:2020zmd,Lin:2021sfs,Vagnozzi:2021gjh,Philcox:2022sgj,Vagnozzi:2023nrq,Pedrotti:2024kpn}  (Sec.~\ref{subsubsec:earlynogo}) as well as requiring an extremely scale-invariant primordial scalar spectrum (Sec.~\ref{subsubsec:HZns}). Therefore, any original intention to solve the Hubble tension from solely modifying the early Universe alone would necessarily require extra modifications to both the primordial Universe and late Universe.

\subsubsection{The sound-horizon tension}\label{subsubsec:rstension}

(1) The recombination history can be modified by primordial magnetic field or non-standard recombination history, which is highly constrained by multiple observations~\cite{Jedamzik:2020zmd,Lucca:2023cdl,Lynch:2024gmp,Lynch:2024hzh,Mirpoorian:2024fka}.

\textbf{Primordial magnetic field} (PMF)~\cite{Jedamzik:2023csc} is motivated by present astronomical and cosmological observations, such as those of galaxies, galaxy clusters, and voids, that frequently encounter magnetic-field environments~\cite{Vachaspati:2020blt}. Their origin remains unknown, but it is generally thought that they may have been produced in the early Universe, for example during the electroweak phase transition or inflation~\cite{Di:2020kbw,Yang:2021uid,Hosking:2022umv}. Such primordial magnetic fields~\cite{Jedamzik:2011cu,Zucca:2016iur,Sutton:2017jgr,Jedamzik:2018itu} would induce small-scale inhomogeneities and force baryons to move along magnetic fields toward regions with lower magnetic energy density, thereby accelerating recombination, reducing the sound horizon, and raising the Hubble constant~\cite{Jedamzik:2020krr}. 
More recent PMF-specific recombination calculations, including~\cite{Jedamzik:2023rfd,Jedamzik:2025cax,Schiff:2025vjf}, move beyond simple phenomenological clumping by emphasizing radiative-transfer and MHD effects; especially Ref.~\cite{Jedamzik:2025cax} argues that updated MHD/Lyman-$\alpha$ modeling can yield mild-to-moderate hints for pG-level PMFs and may improve fits to Planck+Dark Energy Spectroscopic Instrument
(DESI) data, whereas Ref.~\cite{Schiff:2025vjf} stresses that simplified clumping prescriptions may be insufficient and that a physically consistent PMF recombination must couple baryon dynamics, magnetic stresses, and Lyman-$\alpha$ transport. 

However, independent phenomenological tests are more constraining~\cite{Thiele:2021okz,Rashkovetskyi:2021rwg,Galli:2021mxk,SPT-3G:2022hvq,AtacamaCosmologyTelescope:2025nti}: ACT DR4 test to small-scale baryon inhomogeneities found no evidence for the clumping needed to solve the tension~\cite{Thiele:2021okz}; Planck high-$\ell$ damping-tail data and BAO consistency strongly restrict the required parameter space~\cite{Rashkovetskyi:2021rwg};  Planck, ACT, and SPT data alone do not require magnetically assisted recombination~\cite{Galli:2021mxk} although adding a SH0ES prior can produce a weak preference; including related clumping/PMF extensions in SPT-3G cosmology found consistency with $\Lambda$CDM~\cite{SPT-3G:2022hvq}; ACT DR6 finds no evidence for PMFs or modified recombination among extended models~\cite{AtacamaCosmologyTelescope:2025nti}. 
Overall, the PMF scenario remains physically motivated because it attacks the Hubble tension through a concrete recombination mechanism rather than an ad hoc shift in $H_0$, but detailed PMF/MHD modeling still leaves room for a viable signal, while high-resolution CMB damping-tail and modified-recombination constraints from Planck, ACT, and SPT substantially limit the simplest baryon-clumping implementations.

\textbf{Non-standard recombination history}~\cite{Chluba:2023xqj} can be realized by varying fundamental constant (VFC) like fine-structure constant $\alpha_\mathrm{EM}$ and electron mass $m_e$, which changes hydrogen binding energies, recombination rates, the visibility function, and the Thomson scattering rate, thereby shifting the sound horizon $r_s$, the damping scale, and the CMB-inferred $H_0$. However, $\alpha_\mathrm{EM}$ variation is much more tightly constrained because it also changes the Thomson cross section and Silk damping, so this route is mainly a varying $m_e$ rather than a pure varying-$\alpha_\mathrm{EM}$ solution. This mechanism was already foreshadowed by the nonstandard-recombination analysis~\cite{Chiang:2018xpn,Hart:2019dxi,Liu:2019awo} and the Planck constraints on VFC like $\alpha_\mathrm{EM}$ and $m_e$~\cite{Planck:2014ylh,Hart:2017ndk,Smith:2018rnu}. Explicit proposals that increasing $m_e$ can alleviate the Hubble tension include Refs.~\cite{Sekiguchi:2020teg,Hoshiya:2022ady,Seto:2022xgx,Seto:2024cgo,Toda:2024ncp,Toda:2024uff,Toda:2025dzd}, the last of which (Ref.~\cite{Toda:2025dzd}), using Planck/PR4-like CMB, SNe, BAO, and DESI BAO DR2, finds that DESI mildly favors a larger $m_e$, that varying $m_e$ reduces the Hubble tension more efficiently than varying $\alpha_\mathrm{EM}$, and that simultaneously varying $m_e$ and $\alpha_\mathrm{EM}$ does not add much improvement because of degeneracies and damping-scale constraints. Model-building efforts such as hyperlight coupled scalars~\cite{Baryakhtar:2024rky} and the axio-dilaton~\cite{Smith:2025uaq} aim to embed $m_e$ variations in scalar-field dynamics, though at the risk of conflict with solar-system/fifth-force constraints. Recent review~\cite{Schoneberg:2024ynd} on $m_e$ effect on reducing the Hubble tension has been shown to depend on model/data choices.

However, the above non-standard recombination history scenarios also receive strong observational constraints. Ref.~\cite{Smith:2022iax} demonstrate that varying-$m_e$ models do not raise $H_0$ as efficiently as extra pre-recombination energy-density models in sound-horizon-free tests, while the addition of late-time dynamical dark energy~\cite{Smith:2025icl} may weaken the $m_e$-driven $H_0$ gain. Other combined constraints~\cite{Lee:2022gzh,Khalife:2023qbu,Lynch:2024hzh,Toda:2024ncp,Toda:2025kcq,Wang:2025dzn} also weaken the model preference when including the new data. Crucially, any post-recombination shift in constants (see Refs.~\cite{Uzan:2002vq,Uzan:2010pm,Uzan:2024ded,Martins:2017yxk} for reviews) is tightly bounded by BBN and quasar/atomic-clock constraints~\cite{Cooke:2017cwo,Aver:2015iza,Seto:2023yal,Murphy:2016yqp,Evans:2014yva,Murphy:2017xaz,Songaila:2014fza}. Overall, varying $m_e$ is one of the more viable early-Universe recombination-based avenues to raise the CMB-inferred $H_0$, but it is not yet a clean solution; varying $\alpha_\mathrm{EM}$ alone is strongly disfavored as a full solution because the same change that shifts recombination also distorts the damping tail, and combined $m_e+\alpha_\mathrm{EM}$ variations mostly introduce degeneracies rather than a decisive improvement. See also Ref.~\cite{Mirpoorian:2025rfp} for possible explanation to DESI result.

\noindent(2) The early expansion history can be modified by injecting new energy components to raise the pre-recombination expansion rate, such as dark radiation and early dark energy. 

\textbf{Dark radiation}~\cite{Gariazzo:2023hch} before BBN is highly constrained, since BBN already strongly constrains the effective number of neutrino-like relativistic degrees of freedom before BBN; hence dark radiation as a solution to the Hubble tension must be introduced only after BBN in order not to spoil the BBN constraint on the earlier $N_\mathrm{eff}$ unless with lepton asymmetry~\cite{Gelmini:2020ekg,Matsumoto:2022tlr}. If the introduced dark radiation is a free-streaming radiation component like photons, then it will wash out small-scale radiation perturbations and hence change the Silk damping scale of the CMB power spectrum at small scales. In fact, introducing free-streaming dark radiation cannot keep both the acoustic peaks and the Silk damping scale unchanged. One can therefore only introduce non-free-streaming dark radiation~\cite{Ghosh:2019tab}, such as strongly self-interacting neutrinos~\cite{Kreisch:2019yzn,Das:2020xke,He:2023oke} without spoiling BBN~\cite{Blinov:2020hmc}, but this leads to CMB polarization features inconsistent with CMB data~\cite{RoyChoudhury:2020dmd}. Other models include interacting stepped dark radiation~\cite{Aloni:2021eaq,Schoneberg:2023rnx}, pseudoscalar sterile-neutrino self-interactions~\cite{Corona:2021qxl}, Wess Zumino dark radiation~\cite{Schoneberg:2022grr}, and recoupled sterile-neutrino-pseudoscalar fluid~\cite{Sharma:2026ngx}. However, both models from Ref.~\cite{Schoneberg:2023rnx,Sharma:2026ngx} are disfavored by ACT DR4/DR6 as a solution to the Hubble tension. This is because the ACT DR6 extended-model analysis~\cite{AtacamaCosmologyTelescope:2025nti} finds no evidence for new free-streaming light species, $N_\mathrm{eff}=2.86\pm0.13$, tightened to $N_\mathrm{eff}=2.89\pm0.11$ with external BBN, no evidence for neutrino self-interactions, and $N<0.134$ for self-interacting dark radiation, concluding that dark radiation models introduced to raise $H_0$ or lower CMB-inferred clustering are not favored by the data. Other constraints include Refs.~\cite{Cyr-Racine:2013jua,Blinov:2019gcj,Lu:2023uhc,Camarena:2024daj,Bagherian:2024obh,Poudou:2025qcx} on the self-interacting dark-radiations and Refs~\cite{Forastieri:2015paa,Forastieri:2017oma,Chu:2018gxk} on the secret sterile-neutrino interactions.

\textbf{Early dark energy}~\cite{Karwal:2016vyq,Kamionkowski:2022pkx,Poulin:2023lkg} is essentially also a form of dark radiation, or at least diluted faster than radiation with Hubble expansion to better solve the Hubble tension. The simplest example is an axion field~\cite{Poulin:2018cxd}. By adjusting the shape of the axion potential so that the axion mass is much smaller than the Hubble parameter at that time, the axion field is frozen by Hubble friction at a certain point on the potential during most of the period before the matter-radiation equality, acting as an effective cosmological constant, namely early dark energy (EDE). As the Universe expands, when the Hubble parameter falls to a value comparable to the axion mass, the axion rolls down its potential and begins to oscillate and decay. By choosing a suitable potential shape, the energy density corresponding to this oscillation can be made to decay at a rate comparable to or even faster than radiation (anti-de Sitter EDE~\cite{Ye:2020btb,Jiang:2021bab,Jiang:2022uyg,Wang:2025dtk}, Rock `n' roll realizations~\cite{Agrawal:2019lmo}, axion-dilaton destabilization~\cite{Alexander:2019rsc}, acoustic EDE~\cite{Lin:2019qug,Lin:2020jcb}, oscillating scalar field~\cite{Smith:2019ihp,Gonzalez:2020fdy,Berghaus:2019cls}, EDE from massive neutrino~\cite{Sakstein:2019fmf}, rotating axion EDE~\cite{Co:2024oek}, multifield EDE~\cite{Bella:2026zuk}). This in turn allows one to set a larger initial early-dark-energy density and thereby substantially change the early expansion history to shrink the sound horizon even smaller, and hence a larger $H_0$. Data analyses supporting this model~\cite{Chudaykin:2020acu,Fujita:2020ecn,Smith:2020rxx,Chudaykin:2020igl,Poulin:2021bjr,Herold:2021ksg,Simon:2022adh,Smith:2022hwi,Gomez-Valent:2022bku,Herold:2022iib,Poulin:2025nfb} indicate that early dark energy must reach about ten percent of the total energy slightly before matter-radiation equality (around $z\sim5000$) and then decay away faster than radiation. On the other hand, \textbf{new early dark energy} (NEDE)~\cite{Niedermann:2023ssr}, instead of a slowly rolling or oscillating axion-like scalar, realizes the EDE injection through a dark-sector vacuum phase transition shortly before recombination~\cite{Niedermann:2019olb,Niedermann:2020dwg}, which was further developed into hot/cold types~\cite{Niedermann:2021ijp,Niedermann:2021vgd,Cruz:2023cxy,Cruz:2023lmn,Garny:2025kqj} with corresponding observational constraints~\cite{Poulin:2021bjr,Cruz:2022oqk,Garny:2024ums,Chatrchyan:2024xjj}.

However, the simplest one-field EDE solution is under strong pressure from combinations of Planck full-shape CMB spectra, CMB lensing, BAO, SNe Ia, BOSS/eBOSS full-shape galaxy clustering, weak lensing, and especially the small-scale CMB damping tail~\cite{Hill:2020osr,Ivanov:2020ril,DAmico:2020ods,Murgia:2020ryi,Jedamzik:2020zmd,Seto:2021xua,Hill:2021yec,LaPosta:2021pgm,Reeves:2022aoi,Murai:2022zur,Simon:2023hlp,Smith:2023oop,Efstathiou:2023fbn,Gsponer:2023wpm,Eskilt:2023nxm,Goldstein:2023gnw,Qu:2024lpx} (see also the mini-review~\cite{McDonough:2023qcu}), where the small-scale-CMB/DESI analyses are particularly important because SPT-3G and Planck NPIPE tend not to reproduce the stronger ACT DR4 preference, and DESI BAO plus ACT/Planck lensing analyses report no evidence for axion-like EDE in their baseline combinations. Although the EDE scenarios might be data/prior-dependent against observations~\cite{Poulin:2025nfb}, they remain among the most concrete and developed early-Universe Hubble-tension mechanisms~\cite{McDonough:2021pdg,McDonough:2022pku}. It is worth noting that EDE as a solution to the Hubble tension can significantly suppress the recent DESI evidence for dynamical dark energy~\cite{Wang:2024dka,Pang:2025lvh,Wang:2025djw,Adi:2025hyj,Gonzalez-Fuentes:2026rgu} and allow for a canonical evolving scalar field or cosmological constant. This might as well be the reason to further explore alternative forms of EDE, but also highlights the importance of re-examining the nature of DE within the broader context of cosmological tensions.

\subsubsection{Early-time no-go theorems}\label{subsubsec:earlynogo}

Most early-Universe modifications, in particular, the EDE scenario, suffer from three common issues. First, there is a fine-tuning problem~\cite{Pedrotti:2026dwj}: for EDE to reach a fraction of about a few percent properly before matter-radiation equality, the initial value of the EDE field must be carefully tuned. Second, there is a coincidence problem~\cite{Lin:2022phm}: the epoch at which EDE accumulates and then rapidly decays must occur appropriately before matter-radiation equality. Third, there is the $S_8$ problem: modifying the recombination history to advance the recombination will shorten the time for matter perturbations to grow, while modifying the early expansion history to accelerate the expansion will also suppress the growth of early matter perturbations; thus, the amount of matter must be increased simultaneously to compensate for this effect at late time. But this increased matter content enhances matter perturbations at linear scales at late times, namely $S_8$ tension, and thus conflicts with late-Universe large-scale-structure constraints on matter perturbations. In fact, the third issue applies to almost all early-Universe modification models~\cite{Jedamzik:2020zmd,Pogosian:2020ded}; they are either inconsistent with observed galaxy-clustering properties or with galaxy weak-lensing constraints. 

This picture was recently further confirmed in Ref.~\cite{Pedrotti:2024kpn} without any CMB data (or even CMB distance priors) at all to isolate constraints coming as model-independently as possible from the late-time background expansion. They found conclusively that, once the late-time matter fraction $\Omega_m$ is calibrated by BAO and/or uncalibrated SNe Ia, while the baryon density $\omega_b$, any increase in $H_0$ must be accompanied by an increase in the physical cold-dark-matter density $\omega_c$, which, under a nearly standard primordial scalar spectrum, typically also raises $S_8\equiv\sigma_8(\Omega_m/0.3)^{1/2}$. Although a higher $\omega_c$ can help compensate the excess early-ISW power that often appears when the sound horizon is reduced, the accompanying higher $S_8$ creates pressure from weak-lensing and structure-growth data. Therefore, the familiar $H_0-S_8$ correlation seen in many early-time solutions is not merely a model-by-model accident, but follows from the background relation; consequently, successful early-time new physics must not only reduce $r_d$ (unless can be measured model-independently~\cite{Giare:2024syw}), but must do so along a parameter direction that can tolerate a larger $\omega_c$ while not catastrophically worsening $S_8$, suggesting that a fully successful scenario may need early-plus-late new physics with late-time freedom in dark energy or other mechanisms to relax $\Omega_m$ and/or $S_8$ tension (see Sec.~\ref{subsec:Early-Late}).

One should bear in mind that the $S_8$ tension is much less clean as a target for new physics (see also Refs.~\cite{Sanchez:2020vvb,Forconi:2025cwp} for alternative $S_{12}$). Planck 2018 $\Lambda$CDM~\cite{Planck:2018vyg} gives the high-$S_8$ CMB anchor, roughly $S_8\simeq0.83$. The older picture was driven by Kilo-Degree Survey (KiDS)-1000, DES Year 3 (Y3), and Hyper Suprime-Cam (HSC) Y3, which all gave lower-than-Planck growth amplitudes at varying significance. KiDS-1000 found $S_8=0.759_{-0.021}^{+0.024}$~\cite{KiDS:2020suj} in flat $\Lambda$CDM and described this as a strong ($3\sigma$) KiDS-Planck tension; DES Year 3 (Y3), using the full $3\times2$pt combination of cosmic shear, galaxy clustering, and galaxy-galaxy lensing, found $S_8=0.776\pm0.017$~\cite{DES:2021wwk} and hence supported the ``low growth'' tendency but did not by itself claim a decisive crisis; HSC Y3 cosmic shear found $S_8=0.769_{-0.034}^{+0.031}$~\cite{Li:2023tui}, again low relative to Planck, with an advertised $\sim2\sigma$ tension. It seems that there is an empirical pattern that several independent weak-lensing surveys preferred $S_8\simeq0.76-0.78$, producing a persistent but not uniformly high-significance tension with Planck $\Lambda$CDM. Similar $S_8$ tension also appears in RSD measurement~\cite{Nunes:2021ipq}. The major recent change is the complete KiDS-Legacy cosmic-shear analysis, $S_8=0.815_{-0.021}^{+0.016}$~\cite{Wright:2025xka}, in agreement ($0.73\sigma$) with Planck-CMB and BOSS CMASS measurements~\cite{Xu:2024cix}, which is mainly attributable to updated redshift calibration methodology/sample changes and statistical noise from the newly added survey area. However, the latest DES Y6 cosmic shear found somewhere in between~\cite{DES:2026mkc}: $S_8=0.798_{-0.015}^{+0.014}$ for a non-linear intrinsic-alignment model and $S_8=0.783_{-0.015}^{+0.019}$ for a tidal alignment and tidal torque model, retaining a $2.0-2.3\sigma$ discrepancy with Planck-CMB.

In addition to the above $S_8$ tension, there is growing interest in the dubbed $\gamma$ tension~\cite{Nguyen:2023fip} recently: if one keeps the Planck-calibrated $\Lambda$CDM expansion history but allows the linear growth rate to be written as $f(a)=\mathrm{d}\ln D/\mathrm{d}\ln a\simeq\Omega_m(a)^\gamma$, then the low-redshift growth data prefer a larger value, $\gamma=0.633_{-0.024}^{+0.025}$, i.e. a slower late-time growth in nearly $4\sigma$ tension with general relativity (GR)+$\Lambda$CDM expectation $\gamma\simeq6/11\simeq0.545$, indicating a redshift-dependent suppression rather than a pure normalization shift in $S_8$. This motivated constructive directions such as Horndeski/modified-gravity scans~\cite{Wen:2023bcj}, phenomenological freedom in the growth-index sector~\cite{Sakr:2023bms}, a varying $G_N$~\cite{Cortes:2025wni}, and unified-tension scenarios such as $\Lambda_s$CDM \cite{Escamilla:2025imi,Ambelu:2026mkd}, where a late sign-switching cosmological constant can reduce the $\gamma$, $S_8$, and $H_0$ tensions but is not uniformly favored by model comparison. 

However, the evidence for $\gamma$ tension is not yet decisive: Planck-only growth-index deviations are entangled with the $A_L$ anomaly and implementation choices, while ACT/SPT and CMB-lensing combinations tend to restore consistency with $\gamma\simeq0.55$~\cite{Specogna:2023nkq}; ordinary smooth or clustering CPL dark energy has little ability to generate $\gamma>0.6$, even after using DESI DR2 BAO background constraints~\cite{Cortes:2024yon}; and the powerful ACT+SPT+Planck CMB-lensing combination reports a high-precision growth amplitude consistent with $\Lambda$CDM rather than a large growth suppression~\cite{ACT:2025qjh}. Therefore, the $\gamma$ tension remains an important diagnostic and possible anomaly, especially for low-redshift RSD/weak-lensing growth, but the interpretation as new physics is constrained by CMB-lensing consistency. Future Stage-IV analyses, such as those forecast~\cite{Tsedrik:2024cdi}, will be decisive only if they jointly model cosmic shear, RSD, CMB lensing, nonlinear and screening effects, baryonic feedback, intrinsic alignments, and scale/redshift-dependent growth. If the $\gamma$ tension turns out to be real, its relation to $S_8$ and $H_0$ tensions would be crucial for the new physics.

\subsubsection{The Harrison-Zeldovich spectrum}\label{subsubsec:HZns}

Most early-Universe modifications, in particular, the EDE scenario, tend to push the inferred primordial scalar tilt upward, approximately by $\delta n_s\simeq0.4\delta H_0/H_0$~\cite{Ye:2021nej}, implying that a complete shift to $H_0\simeq73\,\mathrm{km\,s^{-1}Mpc^{-1}}$ tends to drive $n_s$ close to a scale-invariant Harrison-Zeldovich value, $n_s=1$. For example, the original EDE model for an axionlike potential with the most preferable $n=3$ case prefers $n_s=0.9812(0.9880)\pm0.0080$~\cite{Poulin:2018cxd}, while the NEDE model~\cite{Niedermann:2019olb} increases $n_s=0.9889(0.9912)_{-0.0066}^{+0.0067}$~\cite{Niedermann:2020dwg}, and the AdS EDE uplifts $n_s=0.9976(0.9974)_{-0.0045}^{+0.0046}$~\cite{Ye:2020btb}. This observational trend can be made more concrete by combining ACT DR4, SPT-3G, and Planck data~\cite{Jiang:2022uyg} or without Planck-CMB data~\cite{Peng:2023bik,Peng:2025tqt}, which calls for a return to the Harrison-Zeldovich spectrum~\cite{Jiang:2022qlj} if the early-Universe solution is indeed adopted to address the Hubble tension. Another associated feature~\cite{Jiang:2023bsz,Wang:2024tjd} is that EDE does not merely move $n_s$ upward; it also changes the allowed $n_s$--$r$ contour, typically pushing $n_s$ toward unity while tightening the upper bound on the tensor-to-scalar ratio $r$, so inflation models that look viable in $\Lambda$CDM can move relative to the allowed region once EDE is included. Note that this preference for $n_s=1$ should not be simply interpreted as a ruled-out $\Lambda$CDM limit~\cite{Li:2025nnk} but as a diagnostic of whether the early Universe is being mis-modeled~\cite{Giare:2024akf}. Intriguingly, the current ACT result tends to increase $n_s$ as well~\cite{AtacamaCosmologyTelescope:2025blo,AtacamaCosmologyTelescope:2025nti}.

In fact, this preference for a Harrison-Zeldovich spectrum can be well-understood as an observational degeneracy: reducing the sound horizon $r_s$ and raising $H_0$ changes both the fits to the CMB peak and damping-tail, but a larger $n_s$ helps restore small-scale scalar power and maintain a good CMB fit. If a scale-invariant Harrison-Zeldovich spectrum is indeed inferred from resolving the Hubble tension with EDE models, the standard canonical single-field slow-roll inflationary model with plateau potential should be modified accordingly~\cite{Braglia:2020bym}; otherwise, the standard prediction $n_s\simeq 1-\mathcal{O}(1)/N_*\to1$ would require a much longer $e$-folding number than what we actually needed for solving the horizon problem. The most direct modification involves multifield configurations as proposed recently in, for example, axion curvaton~\cite{Takahashi:2021bti} and hybrid waterfall~\cite{Ye:2022efx,Braglia:2022phb} models (see also~\cite{Lin:2022gbl} for D-term inflation in a braneworld scenario or bounce inflation~\cite{Li:2024rgq,Zhang:2026cux}). Since no evidence has been reported yet for multifield inflation models, it would be more appealing to reproduce the Harrison-Zeldovich spectrum within the single-field inflationary scenario without enlarging the $e$-folding number, for example, a non-minimal derivative coupling~\cite{Fu:2023tfo,Fu:2025ciy}, or a large step of inflaton potential~\cite{Yuan:2026xcg}. Nevertheless, if ultraviolet (UV) completions to inflation out of the swampland could bound $n_s$ from above, then this could fundamentally limit the ability of EDE as a solution to the Hubble tension.

\subsection{Late-Universe solutions}\label{subsec:LateResolution}

\begin{figure}[!htbp]
    \centering
    \includegraphics[width=\linewidth]{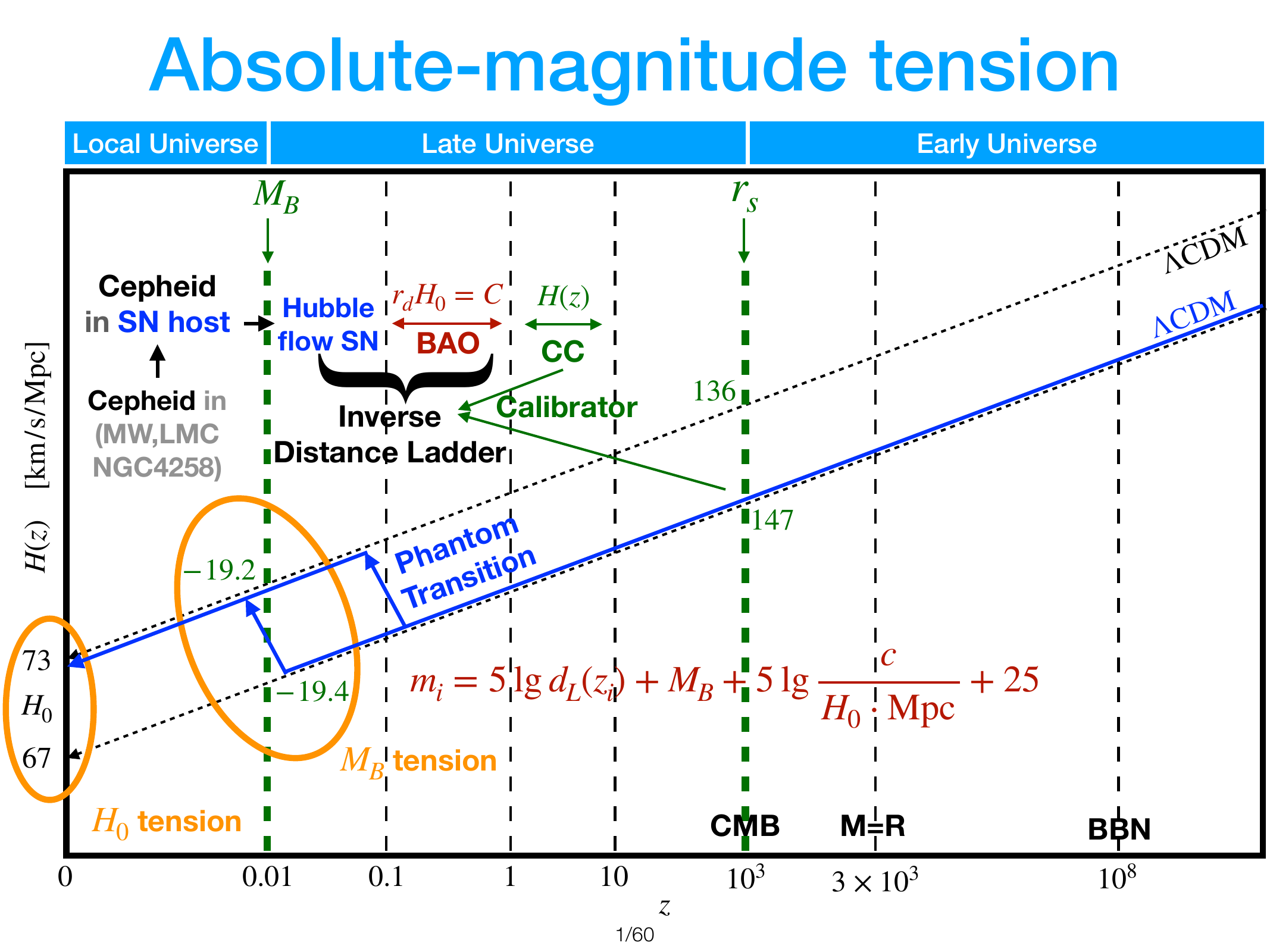}
    \caption{A schematic illustration of the Hubble-constant ($H_0$) tension as the absolute-magnitude ($M_B$) tension and its late-Universe solutions, which can be realized by either intrinsically or effectively changing the absolute magnitude of SNe Ia.}
    \label{fig:MBTension}
\end{figure}

If we trace the cause of the Hubble tension to the late Universe alone~\cite{Perivolaropoulos:2024yxv}, then we expect not to change the early Universe, that is, for a fixed angular scale of the sound horizon $\theta_*=r_s(z_*)/D_M(z_*)=H_0r_s/d_M$ from observations, $r_s$ is fixed by the unchanged early Universe, then a larger $H_0$ requires a larger dimensionless comoving angular-diameter distance $d_M(z_*)=H_0D_M(z_*)=\int_0^{z_*}\mathrm{d}z/E(z)$ to the last scattering surface $z_*$. There are basically two approaches to increase $d_M(z)=d_L(z)/(1+z)$ if the cosmic distance duality relation (CDDR) is assumed: one is to \textit{intrinsically} change the absolute magnitude of SNe Ia via, like varying the Newtonian constant $G_\mathrm{eff}$; the other is to \textit{effectively} change the absolute magnitude of SNe Ia via, like phantom-like dark energy transition around $z\sim0.1$ or even $z\sim0.01$, both of which are motivated to resolve the absolute-magnitude ($M_B$) tension (Fig.~\ref{fig:MBTension} and Sec.~\ref{subsubsec:MBtension}) but highly constrained by a series of ``Late-time no-go theorems''~\cite{Benevento:2020fev,Camarena:2021jlr,Efstathiou:2021ocp,Cai:2021weh,Cai:2022dkh,Keeley:2022ojz,Huang:2024erq,Huang:2024gfw,Pedrotti:2025ccw,Ling:2025lmw,Wang:2026kor,Tiwari:2026pzk} via inverse distance ladders (IDL) and also the CDDR (Sec.~\ref{subsubsec:latenogo}). In particular, we highlight the newly identified cosmological intercept tension~\cite{Huang:2024erq,Huang:2024gfw,Ling:2025lmw,Wang:2026kor} as an alternative view of the Hubble tension but independent of either early or late calibrations (Sec.~\ref{subsubsec:intercept}). This would eventually narrow down the late-Universe solutions to either inhomogeneous modifications (spatial/scale-dependent dark energy models to evade both IDL and CDDR constraints) or an effective change of the absolute magnitude of SNe Ia just across the homogeneity scales.

\subsubsection{The absolute-magnitude tension}\label{subsubsec:MBtension}

When going back in time with increasing redshift, the magnitude-distance/redshift relation,
\begin{align}
m=5\lg d_L(z)+M+5\lg\frac{c}{H_0\mathrm{Mpc}}+25,
\end{align}
holds for the observed $m$ and $z$, calibrated $M$, and inferred $H_0$. For the inferred $H_0$ to decrease from $73\,\mathrm{km\,s^{-1}Mpc^{-1}}$ to $67\,\mathrm{km\,s^{-1}Mpc^{-1}}$, it can only be compensated by either decreasing $M$ or $d_L(z)$ when going to the higher redshifts. The former choice is what we call an \textit{intrinsic} change in $M$, where $M$ is actually smaller (namely brighter) at higher redshifts (from $-19.2$ at lower redshifts to $-19.4$ at higher redshifts); the latter choice is what we call an \textit{effective} change in $M$, which is actually contributed by a reduced $d_L(z)$ (namely enhanced $E(z)$) at higher redshifts. Recall that,
\begin{align}
E(z)^2=\Omega_r(1+z)^4+\Omega_m(1+z)^3+\Omega_\mathrm{DE}\exp\left[3\int_0^z\frac{1+w_\mathrm{DE}(z')}{1+z'}\mathrm{d}z'\right],
\end{align}
an enhanced $E(z)$ at higher redshifts corresponds to a transition of the dark-energy equation of state (EoS) around $z_t$ from $w_\mathrm{DE}(z\lesssim z_t)<-1$ (phantom) to $w_\mathrm{DE}(z\gtrsim z_t)>-1$ (quintessence), that is, a phantom-like dark-energy transition: a would-be decreasing $\Omega_\mathrm{DE}(t)$ with increasing time turns into an increasing $\Omega_\mathrm{DE}(t)$ with increasing time (see earlier~\cite{Escamilla:2023oce}).

\textbf{The intrinsic change} of absolute magnitude of SNe Ia can be realized by a varying Newton constant $G_\mathrm{eff}$. With $M$ and $L$ the effective absolute magnitude and peak luminosity of higher-$z$ SNe affected by $G_\mathrm{eff}$, and $M_0$ and $L_0$ the local-calibrated and measured values corresponding to current $G_N$, the observed SN Hubble diagram gives
\begin{align}
M-M_0=-\frac52\lg\left(\frac{H_0^\mathrm{SH0ES}}{H_0^\mathrm{Planck}}\right)^2=-\frac52\lg\frac{L}{L_0}=-\frac52\lg\left(\frac{G_\mathrm{eff}}{G_N}\right)^{-3p/2},
\end{align}
assuming the SN Ia peak-luminosity scaling $L\propto M_\mathrm{Ch}^p$ in the Chandrasekhar mass scaling $M_\mathrm{Ch}\propto G^{-3/2}$ with Newton constant. Since $H_0^\mathrm{SH0ES}>H_0^\mathrm{Planck}$, then $M-M_0<0$, that is, the Hubble-flow SNe must be intrinsically brighter than the local calibrator SNe. Numerically, this requires $L/L_0\simeq1.18$, i.e., about $18\%$ luminosity increase or $\Delta M\simeq-0.18$ mag. However, various choices of $p$-value in the literature affect the trend in $G$ and corresponding constraints. 

The original $G$-step model~\cite{Marra:2021fvf} around $z\simeq0.01$ (about $\sim44$ Mpc away and $\sim144$ Myr ago) used $p=1$, with $\mu_G=G_\mathrm{eff}/G_N=1$ locally but $\mu_G=(H_0^\mathrm{SH0ES}/H_0^\mathrm{Planck})^{-4/(3p)}\simeq0.9$ beyond transition, would raise the Chandrasekhar mass, suppress the SN Ia absolute magnitude to be brighter by roughly the required $\sim0.2$ mag, and could simultaneously reduce growth, hence addressing both the $H_0$ and $S_8$ tensions~\cite{Liu:2024vlt}. The latter proposal~\cite{Ruchika:2023ugh} adopted an inverse scaling $p=-0.97$ from the semi-analytic SN Ia light-curve analysis~\cite{Wright:2017rsu}, which would correspond to a $12\%$ increase in $G$ in the far past and at higher redshift. Their own fit to calibrator SNe instead gives $p=-1.68\pm0.68$~\cite{Ruchika:2023ugh}, and their preferred solution has $G$ about $4\%$ larger at earlier times, with a transition at about 22.4 Mpc, which was further refined in Ref.~\cite{Ruchika:2024ymt} significantly different from the original $G$-step model~\cite{Marra:2021fvf}. Here, this $4\%$ increase in $G$ in the far past is smaller than the pure SN-only estimate (about $7\%$) because their transition also affects Cepheid calibration distances, not only the intrinsic luminosity of Hubble-flow SNe. Once Cepheids are affected, the net shift in $H_0$ is not simply $H_0^\mathrm{SH0ES}/H_0^\mathrm{Planck}=\sqrt{L/L_0}$ as there is also a Cepheid period-luminosity (PL) calibration bias, $H_0^\mathrm{SH0ES}/H_0^\mathrm{Planck}=\mathrm{Cepheid\,PL\,bias}\times\sqrt{L/L_0}$.

Challenges to a sharp change in $G$ as a solution to the Hubble tension were arisen in Ref.~\cite{Banik:2024yzi} that the required $\mathcal{O}(4\%)-\mathcal{O}(9\%)$ sharp $G$-step is not merely a harmless calibration trick: such a transition should affect stellar luminosities and lifetimes, distance indicators, cosmic chronometers, paleoclimate/geophysical observables, and other local-to-intermediate-redshift astrophysics. In specific, they argue that such a step has severe side effects: Sun-like stellar luminosities scale approximately as $L_*\propto G^{5.6}$, the Solar luminosity would have changed enough to affect Earth's climate, the length of the year would jump by about $10\%$, stellar evolution and helioseismic ages would be distorted, and old-star/Universe-age consistency would be damaged. The later rebuttal and reassessment~\cite{Perivolaropoulos:2025gzo} disputes the strength of these objections, arguing that modern stellar modeling gives a weaker $L_*\propto G^4$ scaling, that Earth's ancient fluid-like behavior changes the day/year argument, that paleoclimate and distance-indicator uncertainties allow larger $\Delta G/G$, and that the discrete nature of the transition avoids some cosmic-chronometer objections.

Therefore, the varying-$G$ model remains as a clever, testable late/local-calibration loophole rather than an established solution to the Hubble tension: its ultra-locality ($\lesssim50$ Mpc) avoids directly changing the CMB, BAO, and SN Hubble-diagram but reinterprets the distance-ladder calibration. The model-building difficulty is that it requires a very sharp, recent, and selectively coupled change in the effective gravitational coupling relevant for stellar/SN physics, while evading Solar-System, binary-pulsar, stellar-evolution, Cepheid, TRGB/Mira/SBF, geological, and fifth-force constraints~\cite{Hogas:2025mii}. In specific, the phenomenological $G_\mathrm{eff}$ used in the distance ladder is not automatically the same object as the gravitational coupling in the Friedmann equation, the Poisson equation for growth, lensing, local Cavendish experiments, Cepheid hydrostatic equilibrium, or SN Ia explosion physics. Thus, it is observationally more surgical but theoretically more contrived. Fully compelling scalar-tensor, screened modified-gravity, or phase-transition realizations are desirable in future works. 

\textbf{The effective change} of absolute magnitude of SNe Ia can be realized by a phantom-like dark energy transition~\cite{Mortonson:2009qq,Keeley:2019esp,DiValentino:2020naf,Alestas:2021xes,Alestas:2020zol} around some redshift $z_t$, which is also feasible to solve the $S_8$ tension. In specific, this modified $H(z)$ after recombination, typically through phantom dark-energy EoS $w(z\lesssim z_t)<-1$, includes a sharp $w=-1$ crossing (quintom model~\cite{Feng:2004ad,Feng:2004ff,Guo:2004fq} or phantom divide~\cite{Hu:2004kh}), an AdS-to-dS/sign-switching dark-energy models~\cite{DiGennaro:2022ykp}, phenomenological/generalized emergent dark energy (PEDE/GEDE)-like models, or interacting-dark-sector models whose effective EoS is phantom (see Sec.~\ref{subsubsec:IDEDM} and Sec.~\ref{subsubsec:EffAppEoS} for more discussions). In concrete terms, (1) the PEDE model~\cite{Li:2019yem} (see also Ref.~\cite{Sakharov:2026vwc}), where dark energy is negligible in the past and emerges at late time with phantom-like behavior, gives a substantially improved fit when a local-$H_0$ prior is imposed; this is generalized as GEDE in Ref.~\cite{Yang:2021eud}, where a deformation parameter interpolates between $\Lambda$CDM and PEDE and Planck+R19 or Planck+BAO+R19 can yield higher $H_0$, although the evidence becomes inconclusive once SN or BAO combinations are used more conservatively. (2) A related but more radical class is graduated/sign-switching dark energy~\cite{Akarsu:2019hmw}, whose effective dark-energy density can become negative in the finite past and then switch to the present positive value; this inspired $\Lambda_s$CDM/sign-switching cosmological constant models~\cite{Akarsu:2023mfb,Akarsu:2021fol}, AdS-to-dS or dS-to-dS transitions~\cite{Akarsu:2025gwi,Akarsu:2025dmj}, and later ``omnipotent''~\cite{Adil:2023exv} or AdS-dS transition parametrizations. For this kind of model to keep CMB angular scales and BAO rulers consistent while increasing $H_0$, one often needs $H(z)$ below $\Lambda$CDM at intermediate redshift and above it very near today, which naturally maps onto a phantom or sign-changing effective dark-energy component. (3) Interacting dark energy can realize the same behavior without a fundamental ghost by making the effective $w_\mathrm{DE}$ phantom (see, e.g.,  Refs.~\cite{Yang:2018euj,Yang:2018uae,Pan:2019gop,Kumar:2021eev,Nunes:2021zzi,Gariazzo:2021qtg,Nunes:2022bhn} and recent reviews~\cite{vanderWesthuizen:2025vcb,vanderWesthuizen:2025mnw,vanderWesthuizen:2025rip}).

However, the late-time phantom transitions are tightly constrained by the inverse distance ladder and by the fact that SH0ES calibrates the SN absolute magnitude $M_B$, not a model-independent global $H_0$ prior~\cite{Benevento:2020fev,Camarena:2021jlr}. The key message here is that the SH0ES result is not just an abstract Gaussian prior on $H_0$ but a Cepheid-calibrated SN Ia absolute-magnitude calibration applied to the same low-redshift SN Hubble diagram that BAO and inverse-distance-ladder analyses also use. Therefore, if one tests a late-time phantom transition, PEDE/GEDE, hockey-stick dark energy, local void, or other $z\ll1$ deformation simply by adding a Gaussian $H_0$ prior to CMB+BAO+SN, one can artificially make the model look successful~\cite{Benevento:2020fev}, because the true disagreement is between two calibrations of the SN Ia absolute magnitude $M_B$ and sound horizon $r_d$, not merely between two labels called $H_0$. In fact, the $M_B$ calibrated by the local distance ladder is incompatible with the $M_B$ required by the inverse-distance-ladder SN+BAO+CMB fits, and a sudden very-low-redshift phantom transition does not remove this $M_B$ disagreement~\cite{Camarena:2021jlr}. To see this, recall that our original expectation for a phantom-like dark-energy transition around the Hubble-flow (HF) regime is based on the distance-ladder balance between the modified $d_L(z)$ model and inferred $H_0$ value without intrinsically changing $M_B$,
\begin{align}
5\lg d_L(z\lesssim z_t)+M_B+5\lg\frac{c}{H_0^\mathrm{SH0ES}\,\mathrm{Mpc}}=5\lg d_L(z\gtrsim z_t)+M_B+5\lg\frac{c}{H_0^\mathrm{Planck}\,\mathrm{Mpc}}.
\end{align}
However, if the inverse-distance-ladder (IDL=SN+BAO+CMB) constraint $(M_B,H_0)^\mathrm{IDL}\simeq(-19.4,H_0^\mathrm{Planck})$ is considered in addition to the distance-ladder (DL=Cepheid+HF-SN) constraint $(M_B,H_0)^\mathrm{DL}\simeq(-19.2,H_0^\mathrm{SH0ES})$, then such a balance is already dominated by the compensation between DL/IDL-inferred $M_B$ and $H_0$ values, 
\begin{align}
5\lg d_L(z\lesssim z_t)+M_B^\mathrm{DL}+5\lg\frac{c}{H_0^\mathrm{DL}\,\mathrm{Mpc}}=5\lg d_L(z\gtrsim z_t)+M_B^\mathrm{IDL}+5\lg\frac{c}{H_0^\mathrm{IDL}\,\mathrm{Mpc}},
\end{align}
so that there is little room for a modified $d_L(z\simeq z_t)$ within the Hubble-flow regime, roughly
\begin{align}
\frac{d_L(z\lesssim z_t)}{d_L(z\gtrsim z_t)}-1=\frac{H_0^\mathrm{DL}}{H_0^\mathrm{IDL}}10^{(M_B^\mathrm{IDL}-M_B^\mathrm{DL})/5}-1\simeq-0.6\%.
\end{align}
This key observation~\cite{Camarena:2023rsd} was made more concretely in Ref.~\cite{Efstathiou:2021ocp}, where an explicit phantom dark energy transition model calibrated by the inverse distance ladder would fail to solve the Hubble tension with a small SH0ES-inferred $H_0$ when fitting the Hubble-flow SNe Ia with IDL-calibrated $M_B$. This conclusion is robust against more flexible $w(z)$ parametrization~\cite{Keeley:2022ojz} that spans all cosmological distances allowed by data.

The above arguments can be reinforced by replacing the high-redshift CMB anchor with cosmic chronometers (CC)~\cite{Cai:2021weh,Cai:2022dkh} to calibrate the inverse distance ladder (SN+BAO) via a globally model-independent parametrization based on cosmic age (PAge). The ages of the oldest astrophysical objects~\cite{Cimatti:2023gil} can also put strong constraints on late-Universe modifications~\cite{Vagnozzi:2021tjv} and an upper bound on the Hubble constant that can raise~\cite{Krishnan:2021dyb}.  A fully model-independent illustration~\cite{Huang:2024erq} can be achieved by fitting the model-independent late-time model (PAge) to the model-independent inverse distance ladder (HF-SN+2D-BAO) calibrated by the model-independent high-redshift calibrator (CC). We will detail these ``late-time no-go theorems''~\cite{Benevento:2020fev,Camarena:2021jlr,Efstathiou:2021ocp,Cai:2021weh,Cai:2022dkh,Keeley:2022ojz,Huang:2024erq,Huang:2024gfw,Pedrotti:2025ccw,Ling:2025lmw,Wang:2026kor,Tiwari:2026pzk} more carefully in Sec.~\ref{subsubsec:latenogo} with biased focus on our works and a possible loose end that such a phantom dark energy transition should also be checked, not around the Hubble-flow regime, but below it~\cite{Huang:2024erq,Huang:2024gfw} around the late-to-local Universe, which will be discussed in Sec.~\ref{subsubsec:intercept} with possible implications for the recent DESI evidence of phantom transition~\cite{Huang:2025som,Ling:2025lmw,Wang:2026kor} but along the direction to worsen the Hubble tension~\cite{Turner:2026zlj}.

\subsubsection{Late-time no-go theorems}\label{subsubsec:latenogo}

If the new-physics model modifying the late Universe depends only on time, then it is a homogeneous modification, as in most dynamical dark-energy models. However, almost all homogeneous late-Universe modification models are strongly constrained by the inverse distance ladder, as we sketch above. Unlike the forward distance ladder, such as SNe Ia calibrated by Cepheids, the inverse distance ladder~\cite{Cuesta:2014asa,Heavens:2014rja,Aubourg:2014yra} combines uncalibrated SNe Ia in the Hubble flow with BAO data to form an inverse distance ladder from low redshift ($z\sim0.1$) to high redshift ($z\sim1$), and calibrates it at high redshift. This high-$z$ calibration is usually provided by the sound horizon inferred from CMB observations under the standard cosmological model as a prior for BAO~\cite{Vonlanthen:2010cd,Aylor:2018drw,Lemos:2018smw}. Since the inverse distance ladder only requires the early-Universe sound horizon calibration, it does not depend on the late-Universe model and can therefore provide model-independent constraints on late-Universe models. The Hubble-constant values obtained from these inverse-distance-ladder constraints tend to agree with early-Universe measurements~\cite{Verde:2016ccp,Alam:2016hwk,Verde:2016wmz,Macaulay:2018fxi,Feeney:2018mkj,eBOSS:2020yzd}, unless the sound-horizon prior given by the early-Universe model is changed, which in turn supports the view that the required model modification should come from the early Universe. Even if the high-redshift calibration of the inverse distance ladder is replaced by other high-redshift observations, such as strong-lensing time-delay observations~\cite{Taubenberger:2019qna,Arendse:2019hev} or gravitational-wave standard sirens~\cite{Zhang:2020uan}, the resulting Hubble-constant constraints still favor early-Universe values. Therefore, homogeneous late-Universe modifications also seem unable to fully solve the Hubble-constant problem.

The trick behind many late-time transition models is to place a sharp feature below the redshift range where most BAO/SN constraints have leverage, and hence one can change a formal local $H_0$, but one cannot simultaneously fit the calibrated SN luminosities well into the Hubble flow once the forward and inverse distance ladders are combined, and the internal inconsistency of these models would become apparent. Specifically, the Hubble constant $H_0$ and supernova absolute magnitude $M_B$ constrained by the Hubble-flow SNe Ia with the forward distance ladder conflict with the $H_0$ and $M_B$ constrained by the CMB-calibrated inverse distance ladder (HF-SNIa+BAO). However, the combined intercept $-5a_B=M_B+5\lg(c/H_0\mathrm{Mpc})+25$ is almost the same across the forward and inverse distance ladders, leaving almost no room for model modification in $5\lg d_L(z)=m_B+5a_B$. This is the essential argument for the late-time no-go formulation~\cite{Benevento:2020fev,Camarena:2021jlr,Efstathiou:2021ocp,Cai:2021weh,Cai:2022dkh,Keeley:2022ojz,Huang:2024erq,Huang:2024gfw,Pedrotti:2025ccw,Ling:2025lmw,Wang:2026kor,Tiwari:2026pzk}. Even if the $M_B$ constrained by the inverse distance ladder is used to calibrate the Hubble-flow SNe Ia in the forward distance ladder, the resulting $H_0$ remains in conflict with the $H_0$ constrained by the forward distance ladder. Therefore, regardless of what global homogeneous modification is made to the late Universe, the Hubble tension persists.

This no-go formulation can be strengthened~\cite{Cai:2021weh,Cai:2022dkh} by improving the traditional inverse distance ladder. The traditional inverse distance ladder requires a high-redshift calibrator, usually the CMB-inferred sound horizon under a given early-Universe model, thereby yielding constraints independent of late-Universe models. But it obviously still depends on the early-Universe model, especially for some combined early-late models. We instead chose cosmic-age, or cosmic chronometer (CC), measurements~\cite{Jimenez:2001gg} as the high-redshift calibrator to the inverse distance ladder. By continuously tracking the age-redshift relation of a class of passively-evolving early-time massive galaxies, CC measurements directly measure the high-redshift Hubble expansion rate at a redshift $z$,
\begin{align}
H(z)=-\frac{1}{1+z}\frac{\mathrm{d}z}{\mathrm{d}t},
\end{align}
thereby avoiding the integral of $H(z)$ over a cosmological model that appears in traditional distance measurements, and hence remaining independent of any cosmological model. To match the use of cosmic-chronometer data, we further adopted a parameterization method based on the cosmic age (PAge)~\cite{Huang:2020mub,Luo:2020ufj,Huang:2020evj,Huang:2021aku,Huang:2021tvo,Huang:2022txw,Li:2022inq,Wang:2023mir,Wang:2024nsi,Wang:2024rus,Ling:2025lmw}. The essence of PAge model is to parameterize the extensive quantity (like the cosmic age) instead of the intensive quantity (like the expansion rate) since the intensive quantity can evolve abruptly thus more difficult to parameterize than the extensive quantity. In the matter-dominated era, one has $Ht-2/3=0=H_0t_0-2/3$ for the cosmic age $t$ and its current value $t_0$~\cite{Bernal:2021yli}. Since the early radiation-dominated era is negligible and the matter-dominated era contributes most of the cosmic age, one can parameterize various late-time homogeneous dark energy models as deviations from the matter-dominated era by Taylor expansions in the cosmic age as~\cite{Huang:2020mub,Luo:2020ufj}
\begin{align}
Ht-\frac23=\left(H_0t_0-\frac23(1+\eta)\right)\left(\frac{t}{t_0}\right)+\frac23\eta\left(\frac{t}{t_0}\right)^2,
\end{align}
where a free parameter $\eta$ is introduced to characterize such a deviation from a pure matter Universe with $\eta=0$, while $H_0t_0\equiv p_\mathrm{age}$ is another free parameter in the PAge model together with $\eta$ to solve for $E(z)\equiv H(z)/H_0$ with the help of the definition $H(z)\equiv-\mathrm{d}z/\mathrm{d}t/(1+z)$. 

Higher-order expansions like the cubic term in cosmic age~\cite{Huang:2021aku} are not indispensable, as shown in Ref.~\cite{Cai:2022dkh}, since the quadratic order is precise enough to parameterize the usual late-time homogeneous models~\cite{Huang:2020mub,Luo:2020ufj} even up to a higher redshift $z\gtrsim1$~\cite{Cai:2021weh,Cai:2022dkh} where some of our CC calibrators live. For a given model to be parameterized by the PAge model, one is free to choose a moment when these two models precisely match, which is usually convenient to be fixed at our present day when the two PAge parameters can be related by 
\begin{align}
    \eta=1-\frac32p_\mathrm{age}^2(1+q_0)
\end{align}
via the current value $q_0$ of the deceleration parameter $q(t)\equiv-\ddot{a}a/\dot{a}^2$. For example, if we represent the $\Lambda$CDM model with the PAge model and match the PAge model with the $\Lambda$CDM model with $q_0=-1+\frac32\Omega_m$ at the present day, then it holds $\Omega_m=(1-\eta)/(\frac94p_\mathrm{age}^2)$. 
The advantage of the PAge model over other model parameterizations is that it is a global, economical and faithful parameterization: with only two parameters, it can describe very accurately and faithfully the evolution of various late-Universe models over almost the whole redshift range deep into the matter-dominated era. Other parameterization methods, such as Taylor expansions in redshift $z$ or in $y=1-a$, already deviate visibly from the cosmological models they intend to parameterize in the first place at intermediate to high redshifts, such as $z\sim1$. With this improved inverse distance ladder~\cite{Cai:2021weh,Cai:2022dkh,Huang:2024erq}, the cosmological constraints on $H_0$, $M_B$, and $r_d$ are almost indistinguishable for both PAge and phantom dark energy transition models with respect to the $\Lambda$CDM model. There is strong evidence, with a BIC difference greater than 10, that the new-physics models parameterized by PAge are not favored over the standard $\Lambda$CDM model. In other words, homogeneous modification models of the late Universe do not solve the Hubble-constant crisis better than $\Lambda$CDM.

Apart from the inverse distance ladder constraints, the late-Universe modifications are also strongly constrained by the possible violation of the cosmic distance duality relation (CDDR)~\cite{CDDR,Bassett:2003vu,More:2008uq,Holanda:2010ay,Nair:2011dp,Meng:2011nt,Holanda:2011hh,Wu:2015ixa,Rana:2017sfr,Ruan:2018dls,Zheng:2020fth,Favale:2024sdq,Qi:2024acx,Wang:2024rxm,Jesus:2024nrl,Alfano:2025gie,Yang:2025qdg,Teixeira:2025czm,Afroz:2025iwo,Kanodia:2025jqh,Li:2025htp,Zheng:2025cgq,Alfano:2025fyq,Barua:2025dxe,Kumar:2026kbo,Tiwari:2026pzk}, as the combination of the luminosity distance from SNe Ia and the angular diameter distance from BAO implicitly assumes the CDDR, $d_L(z)=(1+z)d_M(z)=(1+z)^2d_A(z)$, whose violation could enlarge the room for late-Universe model modifications~\cite{Teixeira:2025czm,Bansal:2026axl}. Unfortunately, after mapping the Hubble tension onto the $M_B$--$r_d$ plane independent of any cosmological model or parameterization assumptions, such violation of CDDR is strongly ruled out by current constraint~\cite{Tiwari:2026pzk}, filling the last loophole and completing the strongest no-go theorem of purely late-time solutions to the Hubble tension.

\subsubsection{The intercept tension of ladders}\label{subsubsec:intercept}

From the above experience of constraining late-Universe modifications, the intercept $-5a_B=M_B+5\lg(c/H_0/\mathrm{Mpc})+25$ in the apparent magnitude-logarithmic distance relation $m_B=5\lg d_L(z)-5a_B$ plays a key role in cleanly separating the new-physics model of $d_L$ from apparent observations of $m_B$ and $z$. For $M_B$ not intrinsically changed during cosmological evolution, the increasing $H_0$ from early-Universe to late-Universe expectations would decrease the $5\lg(c/H_0/\mathrm{10\,pc})$ term and thus $-5a_B$ is also decreased, which could be compensated by an enhanced $d_L(z)$ and thus reduced $E(z)$ to the lower redshifts, or equivalently, an enhanced $E(z)$ to the higher redshifts, which is achieved by a quintessence-to-phantom transition along the cosmic time. For the inverse distance ladder, the $M_B$ constraints vary between the forward distance ladder and the inverse distance ladder that share the \textit{same} Hubble-flow SNe Ia, but the $a_B$ values are almost the same across the forward and inverse distance ladders, and this is why there is no room for late-Universe modifications in $d_L$. However, if the forward distance ladder and inverse distance ladder target different SNe Ia samples, for example, local SNe Ia and Hubble-flow SNe Ia, then we can identify an $a_B$ transition across the homogeneity scale at $z\sim0.023$ (roughly $70\,h^{-1}$ Mpc)~\cite{Huang:2024erq} or at an order of magnitude around $z\sim0.01$~\cite{Huang:2024gfw}. Furthermore, there is seemingly another $a_B$ transition at the late-time Universe to the Hubble flow. See the recent mini-review~\cite{Wang:2026kor} for a concise summary of these intercept tensions.

\textbf{Local intercept tension}: After showing that the late Universe cannot mildly deviate from $\Lambda$CDM, we further test in Ref.~\cite{Huang:2024erq} whether the remaining new physics could reside in the local Universe, $z<0.0233$, by fitting 336 local Pantheon+ SNe Ia below the homogeneity scale with well-separated peculiar-velocity corrections~\cite{Peterson:2021hel,Carr:2021lcj}, calibrated not by SH0ES but by the fully model-independent inverse-distance-ladder (IDL) posterior on $M_B$. For $\Lambda$CDM and PAge, IDL-calibrated local SNe all give $H_0\simeq66\,\mathrm{km\,s^{-1}Mpc^{-1}}$ and $M_B\simeq-19.4$, broadly consistent with IDL constraints separately, yet their joint $H_0$-$M_B$ contours are sharply displaced with each other along the diagonal direction corresponding to the intercept $-5a_B$. This diagonal displacement defines a local-late Universe $a_B$ tension: the pure IDL inference of $a_B$ differs at the $3\sim7\sigma$ level from local-SN inferences calibrated either indirectly through the IDL $M_B$ posterior or directly through IDL posteriors of the cosmological parameters using the covariance-weighted estimator of Ref.~\cite{Efstathiou:2021ocp}. The tension persists across $\Lambda$CDM, PAge, and phantom dark energy transition parameterizations, and appears calibration-independent because an analogous separation is also seen between SH0ES-calibrated local and Hubble-flow SNe. Intriguingly, a phantom dark energy transition at ultra-low redshift $z_t\simeq0.002$ is strongly preferred over $\Lambda$CDM model by the local-SN fit, with Bayes information criterion $\Delta{\rm BIC}=-36.9$ and Bayes factor evidence $\Delta\ln\mathcal{Z}=16.4$, which roughly coincides with the $G_\mathrm{eff}$ model also at ultra-low redshift~\cite{Ruchika:2023ugh,Ruchika:2024ymt}, suggesting that the apparent local phantom dark energy transition model preference may reflect some unknown local systematics.

Having identified the $a_B$ tension as robust against the choice of model and calibration, we can isolate its possible origin more sharply than in the usual $H_0$ tension: since $-5a_B=M_B+5\lg(c/H_0/{\rm Mpc})+25$ absorbs the calibration-dependent absolute magnitude, a discrepancy in $a_B$ must arise either from the modeling of the luminosity distance $d_L(z)$ or from the measured redshifts $z$, especially through peculiar velocity correction at $z\lesssim0.01$, rather than from the calibration of $M_B$ itself. It can be explicitly checked that the volumetric or Malmquist selection effect, estimated by the shift $5\lg d_L(z_{\rm HD}+10^{-6}/z_{\rm HD})-5\lg d_L(z_{\rm HD})$~\cite{Kenworthy:2022jdh}, is far too small to explain the observed local--late intercept mismatch. The standard peculiar-velocity uncertainty $\sigma_{z_{\rm HD}}$ has already been included in the Pantheon+ covariance matrix~\cite{Peterson:2021hel,Carr:2021lcj}, but a new physical effect that mimics peculiar velocities could in principle generate an apparent shift of order $5\lg d_L(z_{\rm HD}\pm\sigma_{z_{\rm HD}})-5\lg d_L(z_{\rm HD})$, which is large enough to cover the observed $a_B$ discrepancy, indicating a new-physics origin of this local-late $a_B$ tension. Therefore, any viable resolution of the $a_B$ tension should interpolate between the local and Hubble-flow intercept bands within this peculiar-velocity-like envelope. A Gaussian-process reconstruction of the intercept evolution directly from Pantheon+ data independently reproduces the same data-driven discrepancy between the local and late Universe. 

This local-late $a_B$ tension was later confirmed by comparing calibration-insensitive intercept constraints from late-time PantheonPlus SNe with those from the very local Universe in Ref.~\cite{Huang:2024gfw} without using any BAO data. In specific, the Hubble-flow range $0.0233<z<0.15$ is taken as the cleanest representative of late-time SNe, while the local sample is further compressed to $z<0.01$, with all $z>0.01$ PantheonPlus SNe regarded as late-time SNe. The late-time sample is calibrated in two independent ways: a forward-distance-ladder calibration using SH0ES Cepheid distances and Cepheid-hosted SNe to infer an $M_B$ prior, and an inverse-distance-ladder calibration using Planck to impose an $H_0$ prior. Although these two calibrations reproduce the usual Planck-SH0ES disagreement in $H_0$ and the corresponding offset in $M_B$, they yield mutually consistent late-time values of $a_B\simeq-4.761$, close to the previous Hubble-flow constraint $a_B=-4.7612\pm0.0018$ from BAO and cosmic-chronometer inverse calibration~\cite{Huang:2024erq}, indicating that late-time PantheonPlus SNe are internally well calibrated. Repeating the calibration for local SNe with priors inherited from the Planck-calibrated and SH0ES-calibrated late-time samples partially relieves the separate $H_0$ and $M_B$ tensions, but the inferred local $a_B\simeq-4.79$ remains displaced from the late-time $a_B$ by more than $5\sigma$ independently of whether the calibration is early-time or local. Therefore, the discrepancy is not primarily an $H_0$- or $M_B$-calibration artifact, but an intercept-level local-late mismatch that must originate either from new physics or from unaccounted local systematics, with volumetric selection effects and peculiar velocities identified as potentially large redshift-bias sources that must be understood before interpreting the $a_B$ tension as evidence for new physics~\cite{Huang:2024erq}. If this local-late $a_B$ tension is not interpreted as new physics but forces the $a_B$ consistency between local and late Universe, then this simply reproduces the first two-rung measurement of the Hubble constant~\cite{Kenworthy:2022jdh} in Sec.~\ref{subsubsec:tworung}.

\textbf{Late-time intercept tension}: The same $a_B$ diagnosis can be applied to the late-time dynamical dark energy signature from recent DESI evidence~\cite{DESI:2024mwx,DESI:2024uvr,DESI:2024lzq,DESI:2025zgx,DESI:2025zpo,DESI:2025fii,DESI:2025ejh}, which seemingly prefers CPL parameters $w_0>-1$ and $w_a<0$ with phantom-to-quintessence transition around $z\simeq0.5$~\cite{DESI:2025fii,DESI:2025wyn,Gonzalez-Fuentes:2025lei}. Novelly, DESI DR2 has found a strong preference over $\Lambda$CDM at about $3\sigma$ for DESI BAO+CMB and $3-4\sigma$ after adding different SNe samples (PantheonPlus~\cite{Riess:2021jrx,Brout:2022vxf,Scolnic:2021amr,Brout:2021mpj,Peterson:2021hel,Carr:2021lcj,Popovic:2021yuo}, UNION3~\cite{Rubin:2023ovl}, and DESY5~\cite{DES:2024jxu,DES:2024hip} compilations, as well as recent updated DES-Dovekie analysis~\cite{Popovic:2025glk,DES:2025sig}), even though DESI+CMB combined with DESY5 gives a stronger preference for evolving dark energy than DESI+CMB combined with Pantheon+ or Union3, while Pantheon+ alone is broadly consistent with Planck $\Lambda$CDM. This raises many concerns~\cite{Gialamas:2024lyw,Efstathiou:2024xcq,Huang:2025som,Capozziello:2025qmh} that the DESI evidence for dynamical dark energy may be driven less by BAO physics than by the external low-$z$ supernova anchor entering the DES Year Five compilation. This is because BAO data actually measure $\Omega_m$ that later goes into SNe Ia at lower redshifts to constrain the dynamical dark energy. Ref.~\cite{Efstathiou:2024xcq} argued that the DESY5 SN sample itself is discrepant with Planck $\Lambda$CDM at about the $3\sigma$ level and that common SNe in DESY5 and Pantheon+ show a $\sim0.04$ mag low-high redshift offset, large enough to move DESY5 back toward Pantheon+ and Planck $\Lambda$CDM if treated as a systematic distance shift. An alternative view of this offset~\cite{Huang:2025som} is that the low-$z$ sample combined with DES-SN in DESY5 shows large scatter and an intercept mismatch of about $\Delta(-5a_B)\simeq0.043$ mag relative to the DES-SN sample, and either correcting this offset or removing the low-$z$ sample substantially reduces the DESI+DESY5 dynamical-DE preference to below $2\sigma$, or even below $1\sigma$ after removing the additional CMB-driven interference.

The possible origin of the above offset might be that DESY5 must join $\sim200$ nearby SNe from diverse low-redshift surveys onto a homogeneous DES high-redshift photometric sample, so relative calibration, selection functions, bias corrections, intrinsic-scatter modeling, host-galaxy corrections, color-luminosity training, and peculiar-velocity/redshift systematics can all mimic a redshift-dependent Hubble diagram tilt, which is precisely the degree of freedom absorbed by $w_0w_a$CDM. The DES Collaboration rebuttal~\cite{DES:2025tir} reproduced part of the reported offset but argued that it is largely explained by scientifically motivated DES-SN5YR updates to intrinsic-scatter and host-galaxy modeling, together with different selection functions and bias corrections, and that using PantheonPlus-like modeling only reduces the DESY5 evidence from $3.9\sigma$ to $3.3\sigma$ rather than eliminating it. Nevertheless, subsequent calibration work~\cite{Popovic:2025glk} reinforced the systematic-risk interpretation by showing that small cross-calibration filter changes can be amplified by SALT training and color-luminosity relations into distance shifts of order $d\mu/dz\simeq0.025$, with potentially larger impact in the $w_0$-$w_a$ plane. Most importantly, the updated DES-Dovekie reanalysis~\cite{DES:2025sig}, with improved photometric cross-calibration, white-dwarf cross-calibration between DES and low-$z$ surveys, SALT3 retraining, and a corrected host-galaxy color-law approximation, lowers the DESI DR2+CMB+DES supernova rejection significance of $\Lambda$CDM from $4.2\sigma$ for DES-SN5YR to $3.2\sigma$ and only weak Bayesian odds, showing that the DESI dynamical-DE hint survives but is quantitatively sensitive to precisely the low-$z$ cross-calibration ingredients under criticism.

Despite of all these criticisms on low-$z$ SNe, this phantom-to-quintessence transition is actually evident even with DESI-BAO data alone as shown in Fig.6 of Ref.~\cite{DESI:2025zgx} with $[D_H(z\simeq0.5)/r_d]_\mathrm{DESI}<[D_H(z\simeq0.5)/r_d]_\mathrm{Planck}$ but $[D_M(z\simeq0.5)/r_d]_\mathrm{DESI}\gtrsim[D_M(z\simeq0.5)/r_d]_\mathrm{Planck}$, which can only be achieved by $[D_H(z_i)/r_d]_\mathrm{DESI}>[D_H(z_i)/r_d]_\mathrm{Planck}$ for some $z_i<0.5$ since $D_M(z)=\int_0^zD_H(z')\mathrm{d}z'$ is an integral over $D_H(z'<z)$ below $z$, therefore $D_H(z)$ has to cross the Planck-$\Lambda$CDM best-fit curve in such a way that turns phantom into quintessence just below $z\simeq0.5$. This might as well be explained by non-cold dark matter~\cite{Yang:2025ume,Wang:2025zri,Kumar:2025etf,Braglia:2025gdo,Abedin:2025dis,Li:2025eqh}.

However, this transition direction is not the same as a successful SH0ES-$H_0$ solution, because the preferred $H_0$ values generally remain near or even below the Planck-$\Lambda$CDM $H_0$ value~\cite{Keeley:2025stf}. This is because the DESI-preferred direction mainly relieves a BAO-CMB distance-ladder mismatch rather than the local SH0ES-Planck Hubble tension: once the CMB fixes the acoustic scale and hence the sound horizon $r_d$, the preferred late-time expansion history tends to lower the inverse-distance-ladder value of $H_0$, thereby worsening rather than solving the discrepancy with the Cepheid-SNe distance ladder. This is why imposing a SH0ES prior to a certain EDE solution to the Hubble tension would dilute the apparent DESI evidence for evolving late-time dark energy, reducing the dynamical-dark-energy preference to roughly $1.5\sigma$, $1.4\sigma$, and $2.4\sigma$ for Pantheon+, Union3, and DESY5, respectively, and suggesting a direct statistical tension between high local $H_0$ and the DESI-favored phantom-crossing trajectory~\cite{Pang:2025lvh}. In this sense, DESI does not by itself provide a late-Universe solution of the Hubble tension; rather, it sharpens the target for viable solutions: early-time mechanisms such as EDE may raise $H_0$ by reducing $r_d$ while simultaneously suppressing the need for a strongly phantom late-time phase~\cite{Wang:2025djw}, whereas interacting dark energy~\cite{Gomez-Valent:2020mqn,Cai:2021wgv,Yu:2022wvg,Karwal:2021vpk,Pitrou:2023swx,Uzan:2023dsk,Wolf:2024stt,Ye:2024zpk,Tiwari:2024gzo,Chakraborty:2025syu,Khoury:2025txd,Wolf:2025jed,Bedroya:2025fwh,Brax:2025ahm,Chakraborty:2024xas,Wang:2024hks,Giare:2024smz,Li:2024qso,Aboubrahim:2024cyk,Li:2025owk,Sabogal:2025mkp,Tsedrik:2025cwc,Zhai:2025hfi,Shah:2025ayl,Silva:2025hxw,Pan:2025qwy,Yashiki:2025loj,Barman:2025ryg,Li:2025ula,Andriot:2025los} or scalar-tensor/modified-gravity models~\cite{Ye:2024ywg,Pan:2025psn,Cai:2025mas,Ye:2026yqk} can in principle combine a higher inferred $H_0$ with effective phantom crossing, but only at the price of adding new dark-sector or gravitational degrees of freedom beyond minimally coupled quintessence~\cite{Ye:2024zpk,Pan:2025psn,Zhang:2025dwu,Yao:2025wlx,Efstratiou:2025iqi}. Therefore, DESI might indeed favor a non-$\Lambda$CDM deformation of the low-redshift distance-redshift relation, possibly interpretable as an apparent phantom-to-quintessence transition, but the Hubble tension still requires either an early-Universe reduction of $r_d$, a nontrivial dark-sector interaction, modified gravity, or unresolved calibration and systematic effects rather than direct extensions~\cite{RoyChoudhury:2024wri,RoyChoudhury:2025dhe} or a naive CPL modeling of such a phantom crossing alone that was strongly disfavored by recent re-analysis of the Bayesian inference~\cite{Ong:2025utx,Ong:2026tta,Ong:2025ver,Ong:2025cwv}.

\section{From Hubble dilemma to Hubble resolutions}\label{sec:Alternative}

We have so far briefly reviewed the Hubble-constant tension from both observational and theoretical perspectives. On the observational side, the systematic tendency that early-Universe constraints are coherently lower than late-Universe measurements supports, to some extent, the reliability of the Hubble-constant tension. On the theoretical side, the strong constraints from various observations on both early- and late-Universe model constructions illustrate the real difficulty of solving this problem. From our limited knowledge and inevitably biased perspective, the model that currently appears most promising for simultaneously solving the Hubble tension and hopefully the $S_8$ tension as well might be: (1) correlated modifications to both early and late Universe (for example, certain kinds of interacting dark energy-dark matter models) as discussed in Sec.~\ref{subsec:Early-Late}; (2) local-scale new physics that disguises as local systematics (for example, ultra-low redshift transition in $G_\mathrm{eff}$ and its effective analogue, or even the effects of nonlinear local inhomogeneities~\cite{Umeh:2022prn,Umeh:2022hab,Umeh:2022kqs} to build a smooth cosmic distance ladder and a similar radical solution~\cite{Clifton:2024mdy}) as discussed in Sec.~\ref{subsec:HubbleDiversity}, after appreciating both ``early-time no-go theorems''~\cite{Krishnan:2020obg,Jedamzik:2020zmd,Lin:2021sfs,Vagnozzi:2021gjh,Philcox:2022sgj,Vagnozzi:2023nrq,Pedrotti:2024kpn} and ``late-time no-go theorems''~\cite{Benevento:2020fev,Camarena:2021jlr,Efstathiou:2021ocp,Cai:2021weh,Cai:2022dkh,Keeley:2022ojz,Huang:2024erq,Huang:2024gfw,Pedrotti:2025ccw,Ling:2025lmw,Wang:2026kor,Tiwari:2026pzk}.

\subsection{Correlated early-late physics}\label{subsec:Early-Late}

A useful way to classify combined early-Universe and late-Universe model-building for the Hubble tension is to separate the physical role of the two sectors: the pre-recombination sector must reduce the sound horizon $r_d$ by altering recombination history or pre-recombination expansion history so that CMB-calibrated distances permit a larger $H_0$, while the post-recombination sector must repair the residual BAO/SNe/chronometer distance ladder, the DESI-preferred $w_0$-$w_a$ direction, and often the accompanying $S_8$ problem. This logic is motivated by the well-known obstruction that purely late-time modifications have difficulty increasing $H_0$ without spoiling BAO and Hubble-flow SNe, whereas purely early-time modifications such as EDE often require correlated shifts in $\omega_{\rm cdm}$, $n_s$, and the growth amplitude that can worsen $S_8$ or leave residual distance-ladder tensions. The DESI DR1/DR2 result then adds a new model-building target: the late sector should preferably reproduce, or at least not erase, the phenomenological trend $w_0>-1$, $w_a<0$, which corresponds to a dark-energy density that is larger in the past and weaker today, often described as an effective phantom-to-quintessence transition. Therefore, a successful combined model is not simply ``EDE plus arbitrary late dark energy'', but must align three degeneracy directions at once: lower $r_d$, maintain the CMB acoustic scale $\theta_*$ and matter-radiation equality scale, and fit the low-redshift absolute distance ladder without pushing $\Omega_m$, $S_8$, or $H_0r_d$ into conflict. Naively, we can classify the correlated early-late physics into two catalogues: one is a direct combination of existing early and late resolutions (Sec.~\ref{subsubsec:early-late})~\cite{Jhaveri:2026bla}, the other is a unified framework to produce the desired early-late modifications (Sec.~\ref{subsubsec:IDEDM}). Although a direct combination is more contrived, a unified framework is nevertheless challenging. Once achieved, the current paradigm of the standard model of cosmology would be totally shifted.

\subsubsection{Combined early-late resolutions}\label{subsubsec:early-late}

Perhaps the most intriguing hint to include early new physics necessarily to reduce the sound horizon is the most recent sound-horizon-free $H_0$ measurements~\cite{Zaborowski:2025umc}: $69.2_{-1.4}^{+1.3}$, $70.3_{-1.2}^{+1.4}$, and $69.6_{-1.8}^{+1.3}$ km/s/Mpc, respectively, when including measurements from i) \textit{Planck}/ACT CMB lensing$\times$unWISE galaxies, ii) the DES Year 3 $6\times2$pt analysis, and iii) \textit{Planck}/ACT CMB lensing + the DES Year 5 supernova analysis, to the base analyses of DESI (FS+BAO) with priors from CMB($\theta_*$)+BBN($\omega_b$)+Ly$\alpha$AP($\Omega_m$). These constraints are significantly higher than their previous results~\cite{Zaborowski:2024car} (see also Ref.~\cite{Pogosian:2024ykm,Pantos:2026cxv}), though the robustness of sound horizon-free determinations~\cite{Smith:2022iax} is largely criticized by the implicit use of $\Lambda$CDM priors~\cite{Jiang:2025ylr}, which would require for new technique~\cite{Gadbail:2026gkx} that reaches a similar conclusion.

The most direct construction is axion-like EDE plus late-time dynamical dark energy~\cite{Giare:2026tyk}. In its minimal form (see also cascading DE~\cite{Rezazadeh:2022lsf}), EDE is a scalar field frozen by Hubble friction before recombination, contributing a fraction $f_{\rm EDE}\sim{\cal O}(0.1)$ near $z_c\sim 10^3$-$10^4$, and then diluting faster than matter after it starts rolling or oscillating; this reduces $r_d$ and raises the CMB-inferred $H_0$. The late sector can then be taken as $w$CDM, CPL $w(a)=w_0+w_a(1-a)$, a binned $w(z)$, or a scalar-field reconstruction, with the aim of fitting DESI BAO and SNe while not undoing the EDE-induced increase in $H_0$. DESI-era analyses show the nontrivial point: adding pre-recombination EDE can suppress the apparent need for a strongly phantom late-time phase, and in $w_0w_a$CDM+EDE even a quintessence-like late component can become consistent with CMB+ACT+SPT+DESI+Pantheon+SH0ES, with a substantially improved $\chi^2$ relative to late-only $w_0w_a$CDM~\cite{Pang:2025lvh,Wang:2025djw}. Thus this class is attractive because it assigns the Hubble-tension job to EDE and the DESI-distance-shape job to the late field, but it also teaches that the DESI phantom-crossing evidence is conditional: once $r_d$ is reduced, the late-time preference can move closer to canonical quintessence or $\Lambda$~\cite{Wang:2024dka,Pang:2025lvh,Wang:2025djw,Adi:2025hyj}.

A natural construction is to combine EDE with a late dark-sector mechanism designed mainly to cure the $S_8$ cost of EDE. The problem is that EDE fits usually prefer a larger $\omega_{\rm cdm}$ and sometimes a larger $n_s$, which raises small-scale power; this motivates adding decaying dark matter~\cite{Poulin:2016nat,Vattis:2019efj,Haridasu:2020xaa,Blinov:2020uvz,Clark:2020miy,DES:2020mpv,FrancoAbellan:2021sxk,Clark:2021hlo,Alvi:2022aam,Anchordoqui:2022gmw,Simon:2022ftd,Nygaard:2023gel,Fuss:2024dam,Bucko:2024izb,Montandon:2025xpd,Zhou:2025ikl}, interacting dark energy-dark matter~\cite{Yashiki:2025loj}, or an EDE-dark-matter drag force after recombination~\cite{Poulin:2022sgp,Simon:2024jmu}. In the EDE+decaying-DM realization, the early component raises $H_0$ by lowering $r_d$, while a fraction of dark matter decays into dark radiation at late times, reducing the late matter abundance and suppressing structure growth, thereby moving both $H_0$ and $S_8$ in the desired directions. In EDE+interacting-DEDM or EDE+DM-drag models, the interaction term modifies the continuity and perturbation equations so that the late dark sector damps growth without necessarily changing the pre-recombination acoustic physics too much~\cite{DiValentino:2019ffd}. The drawback is that these models add parameters whose preferred directions can compete: EDE wants to reduce $r_d$, while the interaction often shifts $\Omega_m$ and the angular-diameter distance in a way that limits the net $H_0$ gain, so current analyses typically report only modest simultaneous improvement rather than a complete resolution. 

A similar construction is new early dark energy (NEDE) from a phase transition, possibly combined with late dark energy. NEDE replaces the smooth axion-like roll by a dark-sector first-order phase transition at eV scales~\cite{Niedermann:2019olb,Niedermann:2020dwg}: a false-vacuum component behaves as early dark energy before recombination and then decays rapidly into a fluid after bubble nucleation, with perturbations matched across the transition hypersurface. Hot, cold, and triggered versions of NEDE alter how the trigger field contributes to the energy budget and growth, and the ``cold NEDE'' variant has been argued to reduce both $H_0$ and $S_8$ tensions without adding further ingredients if the trigger field has a small but non-negligible dark-matter-like fraction~\cite{Niedermann:2021vgd,Cruz:2023lmn}. The natural late-time extension is to let the same dark sector leave behind a residual vacuum, quintessence field, or interacting component that behaves as dark energy today; this is theoretically appealing because it converts the two-coincidence problem, one near equality and one near today, into a structured dark-sector history. The challenge is to keep the phase transition sharp enough to reduce $r_d$, avoid excessive CMB and matter-power signatures, and still reproduce the DESI-like weakening of dark energy at low redshift without inserting an independent CPL sector by hand.

A related possibility is to invoke extra radiation, neutrino physics, a dark-radiation bath, or even an additional barotropic fluid~\cite{Carloni:2025jlk,Carloni:2026yut} at early times, supplemented by late-time dark energy dynamics. Extra free-streaming radiation, self-interacting neutrinos, Majoron-like neutrino interactions, or dark matter-dark radiation scattering can reduce the inferred sound horizon or modify the CMB damping tail, while the late sector is then allowed to be $w_0w_a$CDM, interacting dark energy, or a reconstructed $w(z)$ to match DESI BAO and SNe~\cite{Blinov:2019gcj,Brinckmann:2020bcn,Das:2021guu,Khalife:2023qbu,Buen-Abad:2025bgd}. The advantage is that neutrino/dark-radiation physics naturally affects precisely the pre-recombination observables that calibrate $r_d$, and dark-radiation drag can suppress growth, potentially addressing both $H_0$ and $S_8$. The difficulty is that BBN, CMB phase shifts, damping-tail measurements, laboratory bounds on mediators, and neutrino-mass constraints strongly restrict the viable parameter space; DESI also tightens neutrino-mass limits in $\Lambda$CDM while relaxing them in dynamical-DE extensions, making this sector tightly entangled with the assumed late-time model. Consequently, neutrino/dark-radiation plus DDE models remain possible, but they are less minimal than EDE+CPL and require a UV-complete mediator sector that survives both cosmological and particle-physics constraints.

Another construction is to combine modified recombination physics with late-time dynamical dark energy. Instead of increasing the early expansion rate, one changes atomic or plasma physics so that recombination occurs earlier or more abruptly, reducing the sound horizon and allowing larger $H_0$; examples include a time-varying electron mass, varying fine-structure constant, and primordial magnetic fields that induce baryon inhomogeneities and speed recombination \cite{Jedamzik:2020krr,Jedamzik:2023csc,Seto:2024cgo,Toda:2025dzd,Toda:2025kcq}. The most explicit combined case study is varying $m_e$ plus a post-recombination sign-switching cosmological constant, where the pre-recombination sector changes the recombination scale while the late sector produces an AdS-to-dS transition~\cite{Toda:2024ncp}; this attempt failed to solve the Hubble tension because the two ingredients drive $\Omega_m$ and the relevant distance scales in conflicting directions. More recently, varying $m_e$ has been combined with CPL dark energy to test whether DESI-like phantom crossing survives when early recombination physics is allowed; the result is that $\Lambda$CDM+$m_e$ can raise $H_0$ more efficiently, while adding CPL improves the global fit and preserves a residual preference for phantom-divide crossing, but the late-time freedom can also weaken the very degeneracy-breaking that made varying $m_e$ useful for $H_0$~\cite{Smith:2025icl}. Therefore, a successful pre+post recombination model must not only add two helpful mechanisms, but ensure that their preferred $\Omega_m$-$H_0$-$r_d$ degeneracy directions are mutually compatible.

Another possibility is a phenomenological late-sector transition combined with an early sound-horizon solution. Examples include emergent dark energy, generalized emergent dark energy, sign-switching $\Lambda_s$CDM, vacuum metamorphosis, little-rip/phantom models, and other rapid low-redshift transitions; by themselves these late-time models often fail the inverse-distance-ladder and BAO/SNe consistency tests, but when combined with an early reduction of $r_d$ they can be reinterpreted as the low-redshift distance-shape sector rather than the whole Hubble-tension solution. This is the closest phenomenological language to the DESI result: one fits the pre-recombination calibration with EDE, varying $m_e$, PMFs, or dark radiation, and then fits the residual DESI/SNe trend with an effective $w(z)$ that can cross $-1$. The model-building danger is that a late transition strong enough to raise $H_0$ at fixed $r_d$ tends to distort the BAO-calibrated Hubble diagram, while a late transition mild enough to fit DESI usually lowers or leaves unchanged the inverse-distance-ladder $H_0$; hence these models are viable mostly as part of a two-step solution in which early physics handles the absolute scale and late physics handles the redshift dependence.

Another similar option is to add local or astrophysical late-time ingredients on top of early modifications, although this is less a fundamental cosmological model than a calibration-sector model. Examples include redshift evolution of the SNe Ia absolute magnitude $M_B$, a varying effective Newton constant that changes the Chandrasekhar mass and hence SNe luminosities, local voids, peculiar-velocity systematics, or environment-dependent screening that affects Cepheids and SNe differently. In a combined scenario, early physics lowers $r_d$ enough to reduce the CMB-SH0ES gap, DESI-like dynamical dark energy adjusts the nonlocal BAO/SNe distance curve, and local physics explains the residual local intercept or $M_B$ offset. This is broad enough to fit many anomalies, but it is also the least predictive unless the local sector is tied to a concrete screened scalar, varying-$G_{\rm eff}$, or astrophysical progenitor model. The conservative hierarchy is therefore: first test early+$w(z)$ and early+dark-sector-interaction models against CMB+ACT/SPT+DESI+SNe; then add local calibration physics only if the remaining discrepancy is demonstrably confined to the local distance ladder or to $a_B/M_B$-like intercept observables.

A more appealing construction is a single scalar degree of freedom, unifying early and late dark energy through attractor dynamics~\cite{Pitrou:2023swx,Uzan:2023dsk,Ye:2024zpk,Khoury:2025txd}. The goal is to avoid the usual EDE coincidence problem by arranging a saddle or tracking solution near matter-radiation equality, followed by evolution toward a late-time dark-energy attractor; in this sense the same field supplies both the brief pre-recombination energy injection and the present acceleration~\cite{Ramadan:2023ivw,Liu:2023rvo,Yashiki:2025loj}. This is conceptually the cleanest way to combine early and late modifications, because it explains why the scalar is relevant at two special epochs rather than simply adding two unrelated fluids. See Refs.~\cite{Kamenshchik:2001cp,Bilic:2001cg,Bento:2002ps,Makler:2002jv,Sandvik:2002jz,Scherrer:2004au,Zhang:2004gc,Cai:2015rns,Koutsoumbas:2017fxp,Ferreira:2018wup,Frion:2023xwq,Wang:2024rus} for typical unified dark-fluid scenarios. However, explicit attractor constructions show how difficult this is: the intermediate saddle can behave like an unwanted smooth matter component during the matter era, changing the CMB lensing and matter power spectrum, and the late attractor must satisfy both DESI distance constraints and local-gravity stability constraints. The model-building direction remains promising if the scalar potential, kinetic structure, or coupling to dark matter can shorten the matter-era saddle while preserving an early burst and late acceleration; otherwise the unification principle is elegant but observationally overconstrained.

A more involved unified construction is to seek scalar-tensor~\cite{Jusufi:2026rfd}, Horndeski~\cite{Pogosian:2021mcs,Tiwari:2023jle}, chameleon, or conformal/disformal dark-sector gravity~\cite{Adi:2020qqf}, in which the same field changes either the effective Planck mass, the matter-frame expansion rate, or the dark-sector friction at different epochs. Early modified gravity can mimic EDE by increasing the pre-recombination expansion rate or changing the effective gravitational strength relevant for recombination and perturbations, while late modified gravity can produce effective phantom crossing without introducing a fundamental ghost, because the inferred $w_{\rm eff}<-1$ can arise from a varying Planck mass or non-minimal coupling~\cite{BarrosoVarela:2024htf,BarrosoVarela:2024htf} rather than from a negative kinetic term~\cite{Karwal:2021vpk,Nojiri:2022ski,Odintsov:2023cli,Bansal:2025usn}. Chameleon EDE uses environmental dependence to trigger the early field around equality, and disformal dark-sector couplings can generate an EDE-like kinetic phase that later dilutes as $a^{-6}$ before leaving a potential-driven late dark-energy epoch. This class is especially relevant to DESI because an effective phantom-to-quintessence transition is much easier to realize in the Jordan-frame expansion history of modified gravity than in a single minimally coupled canonical scalar. The cost is that the model must pass Solar-System screening, gravitational-wave speed bounds, CMB lensing, RSD, and weak-lensing growth constraints, so the parameter space that changes distances enough to affect $H_0$ while leaving growth acceptable is narrow.

\subsubsection{Interacting dark energy-dark matter models}\label{subsubsec:IDEDM}

In the last section above, we have briefly mentioned the interacting type of dark sectors as combined early-late solutions to the Hubble tension and other cosmological tensions. We will discuss it more below. See Refs.~\cite{vanderWesthuizen:2025vcb,vanderWesthuizen:2025mnw,vanderWesthuizen:2025rip} for the most recent review and references therein.

The interacting dark energy-dark matter (IDE) scenarios~\cite{Wang:2016lxa} are motivated by the fact that the two dominant components of the late Universe, dark matter and dark energy, are inferred only gravitationally, so a non-gravitational exchange inside the dark sector is not excluded by laboratory fifth-force bounds on baryons. On the other hand, at the theoretical level, IDE was originally introduced to address the coincidence problem that if the dark sectors exchange energy, their ratio need not evolve as steeply as in uncoupled $\Lambda$CDM, and scaling or quasi-scaling solutions can become attractors~\cite{Amendola:1999er,Wetterich:1994bg,Khoury:2003aq,Khoury:2003rn,Upadhye:2012vh}. The interaction between dark energy and dark matter allows part of the dark matter to decay into dark energy. Since dark matter itself has spatial fluctuations, the dark energy interacting with it also acquires spatial dependence, hence an essentially inhomogeneous modification. The IDE increases the late-time dark-energy fraction, thereby directly raising the Hubble expansion rate. At the same time, the reduction of dark matter requires a larger Hubble constant to keep the physical dark-matter density fraction $\omega_\mathrm{cdm}=\Omega_\mathrm{cdm}h^2$ consistent with CMB constraints. Furthermore, allowing part of the dark matter to decay into dark energy also reduces the late-time growth of matter structures, thereby alleviating the $S_8$ tension. Moreover, as an inhomogeneous modification of the late Universe, it can avoid the strong inverse-distance-ladder and CDDR constraints on late-time homogeneous models discussed before. In contrast to purely late-time $w$CDM or $w_0w_a$CDM, IDE changes not only the dark-energy pressure but also the effective dilution of cold dark matter; this extra freedom can shift the CMB-inferred $\Omega_m$, $H_0$, and $\sigma_8$ in directions unavailable to uncoupled dark energy. In all these respects, the IDE can alter both the background distance ladder and the growth of structure, making it attractive for the simultaneous Hubble and $S_8$ tensions~\cite{DiValentino:2017iww,DiValentino:2019ffd,Pan:2019gop,Lucca:2020zjb,Pan:2023mie,Shah:2024rme}.

IDE is traditionally formulated at the level of coupled background equations of motions,
\begin{align}
\dot{\rho}_c+3H\rho_c=Q,\qquad
\dot{\rho}_x+3H(1+w_x)\rho_x=-Q ,
\end{align}
so that $Q>0$ corresponds to energy transfer from dark energy to dark matter and $Q<0$ to energy transfer from dark matter to dark energy. If the late-time dark matter density is modified relative to the standard $\rho_c\propto a^{-3}$ law, the CMB can infer a different present-day matter abundance and a different expansion rate while keeping the angular acoustic scale approximately fixed. This is why early Planck+SH0ES analyses found that IDE could raise the CMB-inferred $H_0$, especially when the dark-energy EoS was allowed to be phantom-like. The same interaction also modifies perturbation friction and source terms, so it can either suppress or enhance growth depending on the sign, time dependence, and momentum structure of the coupling~\cite{Yang:2018euj}. Thus IDE is not merely a late-time distance modification but a coupled background-plus-perturbation deformation. The traditional phenomenological choices of the interaction kernel $Q$ are
\begin{align}
Q=3H\xi\rho_c,\qquad Q=3H\xi\rho_x,\qquad
Q=3H\xi(\rho_c+\rho_x),\qquad
Q=3H\xi\frac{\rho_c\rho_x}{\rho_c+\rho_x},
\end{align}
together with variants where $H$ is replaced by a local expansion scalar, or where $\xi$ is allowed to vary with redshift or with the dark-sector density ratio~\cite{Wang:2006qw,Jesus:2008xi,Boehmer:2009tk,Costa:2016tpb,Yang:2018euj,Yao:2022kub,BeltranJimenez:2021wbq,Hoerning:2023hks,vanderWesthuizen:2023hcl}. Couplings proportional to $\rho_c$ are often more active at early times and therefore more tightly constrained by CMB and perturbation stability, while couplings proportional to $\rho_x$ are more naturally late-time deformations and have been widely used in Hubble-tension studies~\cite{Hoerning:2023hks,Benisty:2024lmj}. Nonlinear density kernels and sign-switching kernels are designed to reduce the tension between the $H_0$-preferred and $S_8$-preferred coupling branches by allowing the energy-transfer direction to change during cosmic evolution~\cite{Boehmer:2009tk,Sabogal:2025mkp,Silva:2025hxw,Wang:2026wrk}.

Another formulation of IDE is based on the conserved total energy-momentum tensor,
\begin{align}
\nabla_\mu T^{\mu\nu}_{(c)}=Q^\nu,\qquad
\nabla_\mu T^{\mu\nu}_{(x)}=-Q^\nu ,
\end{align}
where $Q^\nu$ can be parallel to the dark-matter four-velocity, parallel to the dark-energy four-velocity, or decomposed into energy transfer plus pure momentum transfer. Energy-transfer models change the homogeneous densities, and therefore the background distances directly; pure momentum-transfer models can leave the background close to uncoupled quintessence while modifying the perturbation equations, the dark-sector drag, and the growth rate~\cite{Skordis:2015yra,BeltranJimenez:2021wbq,Pourtsidou:2025sdd}. A more appealing formulation of IDE is by microphysical origin directly with coupled quintessence from conformal or disformal matter couplings~\cite{Amendola:1999er} or kinetically mixing~\cite{Alexander:2022own,Montani:2024pou}, interacting vacuum or decaying-$\Lambda(t)$CDM models~\cite{Brito:2024bhh}, running-vacuum models~\cite{SolaPeracaula:2026pgi}, scalar-mediated dark fifth forces, and effective phenomenological reconstructions. The more microphysical the model is, the more it must satisfy stability, screening, and particle-physics consistency conditions; the more phenomenological it is, the more it risks fitting tensions without a controlled ultraviolet or effective-field-theory interpretation.
Here, the perturbative stability issue is crucial for IDE model-building as the relativistic perturbations must be stable to render a consistent cosmological evolution. Simple constant-$w_x$ fluids with naive couplings can develop early-time large-scale non-adiabatic instabilities, with curvature perturbations blowing up on super-Hubble scales even for small coupling~\cite{Valiviita:2008iv}. This instability led to several stability prescriptions: choosing the sign of $Q/(1+w_x)$ to avoid the dangerous ``doom factor'', transforming the coupling by factors involving $(1+w_x)$, using parametrized post-Friedmann (PPF) closures instead of an ill-defined dark-energy pressure perturbation, or deriving the coupling from a well-posed scalar-field Lagrangian~\cite{Li:2014eha,Skordis:2015yra,Zhang:2017ize,Feng:2018yew,Dai:2019vif}. Any viable IDE solution of the Hubble tension must therefore specify not only $Q$ at the background level, but also the rest-frame sound speed, initial conditions, momentum-transfer prescription, and treatment of the $w_x=-1$ crossing if an effective phantom regime is invoked. See also Ref.~\cite{Harko:2022unn} (and references therein) for a thermodynamic treatment.

Observational constraints of IDE are severe because IDE affects multiple observables all at once. Planck temperature, polarization, and lensing fix the acoustic scale, matter-radiation equality, early ISW effect, CMB lensing amplitude, and the shape of the matter power spectrum; BAO fixes combinations of $D_M(z)/r_d$ and $H(z)r_d$; SNe fix the relative low-redshift luminosity-distance curve; RSD and weak lensing constrain $f\sigma_8$ and $S_8$; cosmic chronometers provide direct $H(z)$ information; and SH0ES provides an external local calibration of the distance ladder. Early positive results showed that Planck+SH0ES could favor nonzero coupling at more than $2\sigma$, with a phantom-like EoS improving the fit~\cite{DiValentino:2017iww}. Later analyses including BAO, SNe, RSD, and cosmic shear found a more mixed picture: coupling can reduce the nominal $H_0$ discrepancy, but full joint datasets often pull the model back toward $\Lambda$CDM or reveal internal dataset incompatibilities~\cite{Yang:2018euj,DiValentino:2019ffd,Hoerning:2023hks}. A particularly important recent constraint is the $H_0$-$S_8$ tradeoff. DESI DR2 work emphasizes that in traditional IDE a negative coupling branch can mitigate $H_0$ while worsening $S_8$, whereas a positive branch can reduce $S_8$ while exacerbating the Hubble tension, depending on the sign convention and precise kernel~\cite{Silva:2025hxw}. This explains why a single fixed-sign coupling often fails as a complete solution: the direction needed to change the distance ladder is not necessarily the direction needed to suppress growth. Late-time analyses independent of $H_0$, $r_d$, and $M_B$ find only mild, at most $\sim2\sigma$, evidence for interaction from low-redshift data alone~\cite{Benisty:2024lmj}. Therefore, robust support for IDE cannot come merely from adding a SH0ES prior; it should appear consistently in uncalibrated SNe, BAO, chronometers, RSD, lensing, and CMB lensing. DESI has further sharpened the situation. DESI DR1 and DR2 favor dynamical dark energy in the CPL plane when combined with CMB and SNe, typically in the $w_0>-1$, $w_a<0$ direction, but this DESI-preferred late-time evolution by itself tends not to solve the Hubble tension and can even worsen it in inverse-distance-ladder fits. IDE is therefore being reconsidered because an interaction can reproduce part of the effective dark-energy evolution while also changing the dark-matter dilution law~\cite{Petri:2025swg}. DESI DR2 IDE analyses find no strong evidence for interaction in the simplest model, with a mild preference around the $1$-$2\sigma$ level and mixed information-criterion results~\cite{Pan:2025qwy,Giare:2024smz} (see, however, Refs.~\cite{Wang:2025znm,Li:2026xaz} with high significance of detection for a particular IDE model). More flexible new interacting models, such as the $\widetilde{\Lambda}$CDM and $e\widetilde{\Lambda}$CDM constructions, can raise $H_0$ substantially and reduce the tension with SH0ES, but the conclusion depends on whether the added interaction degree of freedom is interpreted as genuine physics or as an effective parameterization of DESI-era distance residuals~\cite{Zhang:2025dwu}.

The limitations of current IDE constructions are obvious. Phenomenological couplings such as $Q=3H\xi\rho_x$ are useful for data analysis, but $H$ is a global background quantity and must be replaced by a covariant local scalar in a fundamental theory. Similarly, specifying $Q$ at the background level does not determine the momentum transfer, sound speed, rest frame, or nonlinear behavior. Scalar-field completions reduce this ambiguity but introduce their own constraints: fifth-force screening, radiative stability of a light scalar, compatibility with $w_x\simeq -1$ (see, however, Ref.~\cite{Perez:2020cwa} for unimodular gravity with $w=-1$), absence of ghosts or gradient instabilities, and consistency with structure formation. As a result, the most statistically flexible IDE parameterizations are often the least theoretically controlled, while the most controlled Lagrangian models are often too constrained to move $H_0$ enough. Another limitation is the nonlinearity. Most constraints are obtained using linear Boltzmann codes, but IDE changes halo growth, virialization, nonlinear clustering, and possibly halo bias. Nonlinear matter-power modeling is still less mature for interacting dark sectors than for $\Lambda$CDM, and this limits the use of DESI full-shape clustering, weak lensing, clusters, and small-scale probes. Since the difference between a genuine dark-sector interaction and an effective $w(z)$ can be small at the background level, nonlinear and perturbative observables are exactly where the model should be tested. Without validated nonlinear simulations and emulators for IDE, apparent agreement with $S_8$ or RSD data may remain approximate.

Future perspective of IDE is to replace background-only IDE with covariant, perturbatively stable, and falsifiable interaction models. A viable next-generation IDE model should specify: (i) the covariant transfer vector $Q^\nu$; (ii) the scalar, fluid, or vacuum degrees of freedom generating it; (iii) the stability conditions for scalar and metric perturbations; (iv) the nonlinear matter-power prescription; (v) the mapping to effective $w(z)$ and to DESI observables; and (vi) the predicted relation between $H_0$, $S_8$, neutrino mass, and CMB lensing. In this sense, the future of IDE is less likely to be a single constant coupling $\xi$, and more likely to involve sign-switching interactions, momentum-transfer dark drag, interacting vacuum, or early+late unified dark-sector dynamics. The conservative assessment is that IDE remains one of the few late-time frameworks capable of addressing both distances and growth, but current data do not yet provide decisive evidence that it is the correct resolution of the Hubble tension; its strongest role at present is as a structured phenomenological bridge between DESI dynamical dark energy, $S_8$ suppression, and possible early-Universe sound-horizon reductions.

\subsubsection{Effective vs. apparent dark-energy equation of state}\label{subsubsec:EffAppEoS}

In interacting dark energy-dark matter models, the EoS inferred from a phenomenological fit should not be immediately identified with the physical EoS of the dark-energy field. For a coupled quintessence realization, the physical degrees of freedom are the scalar field and the interacting dark matter, whose background equations may be written as
\begin{align}
\dot{\rho}_\varphi+3H(1+w_\varphi)\rho_\varphi&=-Q,\\
\dot{\rho}_{\rm DM}+3H\rho_{\rm DM}&=+Q,
\end{align}
with, for the nonminimally coupled quintessence (NMCQ) example~\cite{Wang:2025znm},
\begin{align}
Q=\frac{{\cal A}'(\varphi)}{{\cal A}(\varphi)}\dot{\varphi}\rho_{\rm DM}.
\end{align}
The important point is that $\rho_{\rm DM}$ no longer scales exactly as $a^{-3}$. Therefore, when observations are interpreted with a noninteracting CDM plus dark-energy template, part of the true interacting dark-matter evolution may be artificially reassigned to the dark-energy sector. This is the origin of the distinction between the physical scalar-field EoS $w_\varphi$, the usual effective dark-energy EoS $w_{\rm DE}^{\rm eff}$, and the apparent EoS $w_{\rm DE}^{\rm app}$ reconstructed by a mismatched phenomenological model such as $w_0w_a$CDM, as we elaborated below. This distinction is especially relevant after DESI, because the CPL fit suggests a phantom-crossing behavior around $z\sim0.5$, whereas a stable single canonical quintessence field cannot smoothly cross $w=-1$.

The usual effective dark-energy density is obtained by first subtracting from the interacting dark matter a fiducial cold dark matter part (with a fiducial constant $\rho_\mathrm{CD,fid}$),
\begin{align}
\rho_{\rm CDM}^{\rm fid}=\rho_{\rm DM,fid}a^{-3},\qquad
\Delta\rho_{\rm DM}\equiv\rho_{\rm DM}-\rho_{\rm CDM}^{\rm fid},
\end{align}
and then absorbing the residual non-cold dark matter part into the dark-energy sector,
\begin{align}
\rho_{\rm DE}^{\rm eff}=\rho_\varphi+\Delta\rho_{\rm DM}.
\end{align}
Since the subtracted CDM piece is pressureless, the corresponding effective EoS defined from $ \dot{\rho}_{\rm DE}^{\rm eff}+3H(1+w_{\rm DE}^{\rm eff})\rho_{\rm DE}^{\rm eff}=0$ is
\begin{align}
w_{\rm DE}^{\rm eff}
=\frac{w_\varphi\rho_\varphi}{\rho_{\rm DE}^{\rm eff}}
=\frac{w_\varphi}{1+(\rho_{\rm DM}-\rho_{\rm CDM}^{\rm fid})/\rho_\varphi}.
\end{align}
In the dilaton-coupled NMCQ example with ${\cal A}(\varphi)=\exp(-\beta\varphi/M_{\rm Pl})$, the best-fit evolution can keep $\Delta\rho_{\rm DM}>0$ when the fiducial component is chosen as $\rho_{\rm DM,0}a^{-3}$ (iff you expect all dark matter today is cold, that is, there is no noncold dark matter at present day, $\Delta\rho_{\rm DM,0}=\rho_{\rm DM,0}-\rho_{\rm CDM,0}^{\rm fid}=0$). Then, for a canonical quintessence field with $w_\varphi>-1$, the effective EoS remains nonphantom, $w_{\rm DE}^{\rm eff}\geq w_\varphi>-1$. Thus the underlying model contains neither a phantom scalar nor a physical ghost instability. This behavior differs from interacting constructions in which the coupling function is chosen so that $\Delta\rho_{\rm DM}$ becomes negative, allowing the effective EoS itself to cross the phantom divide.

The apparent EoS relevant for interpreting a CPL reconstruction is different. A $w_0w_a$CDM fit assumes its own noninteracting CDM density,
\begin{align}
\rho_{\rm CDM}^{\rm CPL}=\rho_{\rm CDM,0}^{\rm CPL}a^{-3},
\end{align}
and therefore decomposes the same total dark-sector density as
\begin{align}
\rho_\varphi+\rho_{\rm DM}
=\rho_{\rm DE}^{\rm app}+\rho_{\rm CDM}^{\rm CPL}.
\end{align}
This defines
\begin{align}
\rho_{\rm DE}^{\rm app}
=\rho_\varphi+\rho_{\rm DM}-\rho_{\rm CDM}^{\rm CPL}
=\rho_{\rm DE}^{\rm eff}
+\rho_{\rm CDM}^{\rm fid}-\rho_{\rm CDM}^{\rm CPL},
\end{align}
and hence
\begin{align}
w_{\rm DE}^{\rm app}
=\frac{w_{\rm DE}^{\rm eff}\rho_{\rm DE}^{\rm eff}}{\rho_{\rm DE}^{\rm app}}
=\frac{w_\varphi\rho_\varphi}{\rho_{\rm DE}^{\rm app}}
=\frac{w_\varphi}{1+(\rho_{\rm DM}-\rho_{\rm CDM}^{\rm CPL})/\rho_\varphi}.
\end{align}
The equality $w_\varphi\rho_\varphi=w_{\rm DE}^{\rm eff}\rho_{\rm DE}^{\rm eff}=w_{\rm DE}^{\rm app}\rho_{\rm DE}^{\rm app}$ only says that all the artificially subtracted matter pieces are pressureless. The denominator, however, depends on the comparison model. Therefore, if the CPL fit assigns a different matter fraction from a given IDE model, the mismatch $\rho_{\rm DM}^{\rm IDE}-\rho_{\rm CDM}^{\rm CPL}$ is misidentified as part of dark energy, and $w_{\rm DE}^{\rm app}$ can cross $-1$ even when $w_\varphi$ and $w_{\rm DE}^{\rm eff}$ never do~\cite{Wang:2025bkk}. This is a long-neglected subtlety in the IDE community.

This resolves the apparent contradiction between DESI-motivated phantom crossing and canonical interacting quintessence. In the physical NMCQ description, the scalar field remains nonphantom and the usual effective dark energy can remain above the phantom divide. In the phenomenological CPL description, however, the same background expansion is forced into a noninteracting-CDM plus dark-energy split; the wrong CDM subtraction changes the reconstructed dark-energy density and can make its derivative behave as if the EoS crossed $w=-1$. Around the transition from dark-matter domination to dark-energy domination, this mismatch becomes large enough to generate an apparent crossing near $z\simeq0.5$, reproducing the DESI-like phantom-crossing signature without introducing a fundamental phantom degree of freedom for both the bare scalar field and the effective DE component.

At still higher redshift, the apparent component may even pass through zero and its EoS may diverge. This is not a physical singularity, because $\rho_\varphi$ and $\rho_{\rm DM}$, the quantities actually evolved in the interacting model, remain regular. The divergence only reflects that $\rho_{\rm DE}^{\rm app}$ is an artificial residual obtained by subtracting the CDM density preferred by the wrong template. Consequently, an apparent negative dark-energy density or divergent apparent EoS should not be interpreted as a pathology of the interacting dark-sector model. It is instead a diagnostic of the dark degeneracy: background distance data constrain the total expansion history, while the separation into CDM and dark energy is model dependent~\cite{Petri:2025swg}.

The broader lesson is that the EoS measured by a phenomenological reconstruction is not necessarily the EoS of a physical dark-energy substance. For noninteracting dark energy, the CPL EoS can be read directly as a property of the dark-energy fluid, subject to the usual caveats about parametrization. For interacting dark sectors, the same CPL EoS is an apparent quantity that absorbs both pressure evolution and matter-sector misassignment. Therefore, DESI-like evidence for phantom crossing can be interpreted in two logically distinct ways: either as a genuine violation of the nonphantom bound, requiring quintom~\cite{Feng:2004ad,Feng:2004ff,Guo:2004fq}, modified gravity, or other beyond-canonical dynamics, or as an apparent crossing produced by dark-sector interaction while the intrinsic dark energy remains nonphantom~\cite{Wang:2025bkk,Wang:2026wrk}. This distinction is crucial for model selection: the relevant question is not merely whether a model reproduces a reconstructed $w(z)$ curve, but whether it gives a consistent decomposition of the dark sector, aligns the Planck-DESI-SNe matter fractions, and remains stable at the perturbation level.

\subsection{The Hubble tension as a phenomenon}\label{subsec:HubbleDiversity}

The observational results from the early Universe have small errors and are relatively concentrated. If the Hubble-constant tension truly originates from statistical~\cite{Hogas:2026urs} or observational systematic errors, it is generally thought that the source is more likely to lie in measurements of the late or local Universe. However, the distribution of late/local-Universe measurements is rather scattered. Although they are systematically higher than early-Universe results, it is difficult to attribute the discrepancy to a single systematic error. If they can be explained by a single systematic error, however, that systematic error must contain new physics that has not been theoretically modeled. For this reason, it remains necessary to examine systematic errors carefully, for example, direct tests of large-scale homogeneity and isotropy~\cite{Rameez:2019wdt,Luongo:2021nqh,Aluri:2022hzs,Cowell:2022ehf}.

The error budget of the SNe Ia distance-ladder measurement can be divided mainly into three parts: calibration errors among different rungs of the distance ladder, errors in the standardization of supernovae as standard candles, and cosmic variance of the supernova sample. The calibration error of the distance ladder has now been reduced to below $1\%$, so it will not be discussed further. The standardization error of supernovae as standard candles arises from the fact that, although supernovae are ideal standard candles in theory, in real observations their light curves are affected by various factors associated with the supernova progenitor, such as white-dwarf accretion or white-dwarf merger models, and by the local environment. Their light curves therefore exhibit a certain dispersion and require various corrections in order to be standardized.

The quantity used to test standardization is the Hubble residual, or distance-modulus residual, $\Delta\mu\equiv\mu_\mathrm{obs}-\mu_\mathrm{mod}$. The observed distance modulus is $\mu_\mathrm{obs}=m-M+\alpha x_1-\beta c+\Delta_\mathrm{bias}+\Delta_M$, while the theoretical distance modulus is $\mu_\mathrm{mod}=5\lg d_L(z)+5\lg( c/ H_0/\mathrm{Mpc})+25$, as defined above. Here $\alpha x_1$ and $\beta c$ are corrections for the shape and color of the light curve, respectively, $\Delta_\mathrm{bias}$ comes from simulations, and the most noteworthy correction is the host-galaxy stellar-mass step correction $\Delta_M$. Without this mass step correction, the observed Hubble residuals of supernovae exhibit a special correlation~\cite{Kelly:2009iy,SNLS:2010kps,SDSS:2010swx,Gupta:2011pa,Johansson:2012si,Childress:2013xna}: supernovae in more massive host galaxies have more negative Hubble residuals, namely $\mu_\mathrm{obs}$ is smaller relative to $\mu_\mathrm{mod}$. Equivalently, the effective estimate of the supernova absolute magnitude $M$ is larger, so the intrinsic luminosities of SNe Ia appear dimmer in more massive host galaxies. Standardization must therefore artificially make them appear brighter, and the mass correction takes the form of a step function $\Delta_M=\gamma\Theta(M_\mathrm{host}-M_\mathrm{step})-\frac{\gamma}{2}$ to effectively reduce $M$ to be brighter, where $\gamma>0$ and $M_\mathrm{step}$ is usually around $10^{10}$ solar masses.

It is worth noting that, according to galaxy-formation theory, larger galaxies form in larger dark-matter halos, and larger dark-matter halos are more likely to be distributed in denser environments~\cite{Sheth:2000ii}. It is therefore natural to conjecture that the Hubble residuals of SNe Ia may also correlate with the matter-density environment of their host galaxies. Before introducing this correlation, we first review in Sec.~\ref{subsubsec:deltaH0} the relation between the cosmic variance of the supernova sample in Hubble-constant measurements and the local density of the observer, as well as the sample variance with respect to the local density contrast of the sample volume when the sample volume is not centered at the observer (recall that this sample variance is actually the cosmic variance when the sample volume is centered at the observer). We then introduce in Sec.~\ref{subsubsec:Chameleon} a model that matches the observed physical picture of this non-local sample variance, which might be extracted in Sec.~\ref{subsubsec:LambdaLCDM} as a particular IDE model that admits a scale-dependent effective cosmological constant. In this new picture, the Hubble tension might as well be regarded as an existing phenomenon~\footnote{See also Refs.~\cite{Erdem:2024vsr,Erdem:2025xtr} with similar philosophy but totally different realization.} rather than a discrepancy to be eliminated.

\subsubsection{Sample-variance systematics}\label{subsubsec:deltaH0}

The Hubble-flow range $0.0233<z<0.15$ is taken as the cleanest representative of late-time SNe because its lower bound suppresses cosmic variance from local density fluctuations~\cite{Scrimgeour:2012wt,1992AJ....103.1427T,Shi:1995nq,Shi:1997aa,Wang:1997tp,Kenworthy:2019qwq,Sinclair:2010sb,Marra:2013rba,Ben-Dayan:2014swa,Camarena:2018nbr}, while its upper bound minimizes sensitivity to late-Universe cosmological evolution. It is this suppressed cosmic variance of Hubble-flow SNe Ia that essentially rules out the long-conjectured inhomogeneous modification of the late-time/local cosmological principle~\cite{Aluri:2022hzs} by placing us in a local cosmological void. In such a void, the galaxy distribution is sparse, so the matter-density fraction is lower than in other regions of the Universe. Correspondingly, the dark-energy distribution is relatively higher, and the local Hubble expansion rate is larger. It is worth noting that as early as the mid-1990s, Chinese scholars were among the first internationally to propose using a local cosmic void to explain the overestimation of the Hubble constant~\cite{Wu:1994ai,Wu:1996pt}. Recent galaxy-survey observations~\cite{Lavaux:2011zu} seem to support the possibility that we live in a local cosmological void of radius 300 Mpc and depth $-30\%$, namely the KBC (Keenan-Barger-Cowie) void~\cite{Keenan:2013mfa}. This led to the suggestion that such a void might be the cause of the Hubble-constant crisis~\cite{Hoscheit:2018nfl,Banik:2025dlo}. However, if SNe Ia are used to trace the Hubble expansion rate at different redshifts, the claimed underdense region is found to be incompatible with observations beyond its radius~\cite{Kenworthy:2019qwq,Lukovic:2019ryg}. Therefore, there is no sufficiently large and sufficiently deep local cosmological void capable of resolving the Hubble-constant problem~\cite{Cai:2020tpy}. In fact, the lower redshift bound of Hubble-flow SNe Ia is specifically chosen in such a way that the theoretically expected uncertainty in measuring $H_0$ from Hubble-flow SNe Ia can be pinned down below $1\%$, as we will see shortly below. In particular, including local structures can make the Hubble tension even worse in $\Lambda$CDM~\cite{Giani:2023aor}.

\paragraph{Local cosmic variance.}

For supernovae within the Hubble-flow range, the Hubble constant can be estimated through the simple relation between distance and recession velocity in the Hubble law, $H_0D_L=v_r=c z_\mathrm{cos}$. Here $z_\mathrm{cos}$ should be the redshift caused purely by background expansion, and the contribution from the peculiar velocity of the supernova should be subtracted in advance. However, measurements of supernova peculiar velocities are not always clean enough, so some residual peculiar-velocity contribution might always be left unaccounted for. The measured Hubble constant therefore contains not only the contribution from the pure background expansion, but also a contribution from unmodeled peculiar velocities. In principle, there is therefore always a systematic error, which gives the bias between the Hubble constant measured from a specific supernova sample and the true Hubble constant associated with the background expansion.

Theoretically, we can compute this Hubble bias $\delta_H(\mathrm{observer};\mathrm{sample})$ for different supernova sample distributions~\cite{Yu:2022wvg}. In particular, we recovered the well-known result that, for an ultra-local supernova sample, namely a sample distributed infinitesimally locally around the observer, the Hubble bias is negatively correlated with the observer's local matter-density contrast $\delta_\mathrm{m}$~\cite{1992AJ....103.1427T},
\begin{align}\label{eq:TCO}
\delta_H(r_\mathrm{obs};r_\mathrm{sam}\equiv r_\mathrm{obs})=-\frac{f}{3}\delta_\mathrm{m}(r_\mathrm{obs}),
\end{align}
where the proportionality coefficient is one third of the matter-structure growth factor $f(\Omega_\mathrm{m})\approx\Omega_\mathrm{m}^{-0.55}$, if the $\Lambda$CDM model is used for perturbative calculations. Thus, if an observer is located in a local void, the local density contrast is negative, and the observer will tend to overestimate the Hubble constant. This is why a local void has been considered as an explanation for the Hubble-constant tension. However, more detailed calculations show that, for a supernova sample distributed within a certain radial range around the observer, the standard deviation of the Hubble bias decreases with radius~\cite{Camarena:2018nbr}. Therefore, by selecting sufficiently distant supernova samples, for example above the lower Hubble-flow redshift limit $z\gtrsim0.023$, the standard deviation of the Hubble bias contributed by an observer in a local void can be controlled to below $1\%$. Hence a local void smaller than the Hubble-flow redshift range, also called a Hubble bubble, cannot solve the Hubble-constant crisis we face. This should be distinguished from cosmological-scale void scenarios excluded above~\cite{Ding:2019mmw,Haslbauer:2020xaa,Banik:2021woo,Mazurenko:2023sex,Mazurenko:2024gwj}.

\paragraph{Non-local sample variance.}

For the Hubble bias from arbitrary sample distributions~\cite{Yu:2022wvg}, there is a very special case. If the host galaxies of the selected supernova sample all reside in environments with the same local $R$-scale averaged matter-density contrast $\delta_\mathrm{m}^R$, then the Hubble bias corresponding to this supernova sample is also negatively correlated with the local $R$-scale averaged matter-density contrast of the supernova host galaxies,
\begin{align}
\delta_H(r_\mathrm{obs};r_\mathrm{sam})=-\frac{f}{3}\bigg\langle\frac{R^2}{d^2}\bigg\rangle\delta_\mathrm{m}^R,
\end{align}
but the proportionality coefficient is suppressed by the factor $R^2/d^2$. Here, $d=|r_\mathrm{sam}-r_\mathrm{obs}|\gg R$ is the distance from the sample supernova to the observer, which we assume to be larger than the homogeneity scale $R\simeq70h^{-1}\,\mathrm{Mpc}$, and the angle brackets denote the average over all sample supernovae. This differs from the local cosmic variance relation in Eq.~\eqref{eq:TCO}. The local density correlated with the Hubble bias is no longer the observer's local density, but the local density of the sample supernova host galaxies. We therefore call this correlation non-local sample variance.

Unexpectedly, when we directly tested the above non-local sample-variance relation using real observational data~\cite{Yu:2022wvg}, we found a non-negligible conflict between observations and theoretical predictions. Specifically, we used the matter-density field reconstructed from the BOSS DR12 (Baryon Oscillation Spectroscopic Survey Data Release 12) galaxy survey to estimate the arbitrary $ R$-scale-averaged matter-density contrast around the host galaxies of every PantheonPlus SNe Ia. We then grouped supernovae with roughly the same local matter density and fitted their Hubble constants. The study found that supernovae in denser environments give larger fitted Hubble constants, that is, a positive slope between the Hubble bias $\delta_H$ and sample density $\delta_m^R$. This is contrary to the negative correlation (negative slope) expected from the non-local sample-variance relation, and the discrepancy reaches nearly $3\sigma$ at the scale $R=60$ Mpc/h. We call this conflict the slope ($\delta H_0$) tension. It is different from both the Hubble tension and the $S_8$ tension, and is a new cosmological tension at the perturbative level that reveals, at a deeper level, the possible existence of new physics beyond the current standard cosmological model based on general relativity.

\subsubsection{Chameleon dark energy model}\label{subsubsec:Chameleon}

Since observations indicate that the Hubble constant we measure always contains a systematic error from the supernova sample distribution, and that this systematic error is correlated with the matter-density environment of the supernova host galaxies, a natural question is whether this systematic error is also present in other observations. One can see that almost all early-time global-background measurements of the Hubble constant, including those involving supernovae, are systematically smaller than late-time local measurements. This is because matter-density growth is larger in the late/local Universe, and various late-time luminosity-distance indicators or even their calibrators all reside in such high-density environments, namely galaxies or their dark host halos.

A theoretical model that naturally generates such a positive correlation between the Hubble constant and the local density contrast is the chameleon dark energy model~\cite{Cai:2021wgv}. The original motivation for the chameleon mechanism was to screen modified-gravity effects in high-density environments on small scales. The mechanism assumes that a scalar field couples to the local matter density in a specific way, so that the scalar field has a larger effective mass in high-density environments. The corresponding fifth/dark force~\cite{Hogas:2025mii,Archidiacono:2022iuu} then has a shorter range, thereby screening the fifth/dark force within the high-density region. However, the chameleon mechanism also has an accompanying effect: in a high-density environment, the potential energy of the scalar field at its vacuum expectation value is also higher. In other words, its effective cosmological constant is larger, and hence the local Hubble expansion rate is also larger than the background. In this model, the effective cosmological constant changes with matter-density fluctuations averaged on different scales, but on any fixed scale it is equivalent to the standard $\Lambda$CDM model. The simplest realization is a non-minimally coupled quintessence model,
\begin{align}
    S=S_{\rm GR}
    +S_{\rm SM}[\psi_{\rm SM};g_{\mu\nu}]
    +S_{\rm DM}[\psi_{\rm DM};\tilde{g}_{\mu\nu}\equiv{\cal A}^2(\varphi)g_{\mu\nu}]
    +S_\varphi ,
\end{align}
with
\begin{align}
    S_\varphi=\int d^4x\sqrt{-g}
    \left[
        -\frac12 g^{\mu\nu}\partial_\mu\varphi\partial_\nu\varphi
        -V(\varphi)
    \right],
\end{align}
where $V(\varphi)$ and $\mathcal{A}(\varphi)$ are decreasing and increasing functions of $\varphi$, respectively, such that $\varphi$ can track the effective potential minima $V_\mathrm{eff}(\varphi)=V(\varphi)+(\mathcal{A}/\mathcal{A}_0)\rho_\mathrm{DM,0}(a/a_0)^3\equiv V(\varphi)+\rho_\mathrm{DM}$.

Thus, when measuring our local Hubble constant using distance-indicator samples located in high-density environments, the result contains three contributions. The first comes from the matter-density environment in which we ourselves reside; on small scales this contribution tends to be positive. The second comes from the sum of matter-density fluctuations between us and the distance-indicator samples; if the distance indicators are sufficiently far away, this contribution should average to approximately zero. However, if the distance indicator is calibrated by an intermediate probe, then the matter-density environment around that calibrator should also be considered. The third comes from the matter-density environment of the distance-indicator samples themselves; this contribution is generally positive. Therefore, the measured Hubble constant is always larger than the contribution from the true background expansion. In this physical picture, early-time measurements give a smaller Hubble constant because matter-density fluctuations in the early Universe are always smaller, so they reflect the true background-expansion component.

In addition, the $S_8$ tension can also be explained in this picture. The larger the growth of matter perturbations, the faster the local Hubble expansion rate, which in turn dilutes the original growth of matter perturbations, and hence also explains the $\gamma$ tension. The equilibrium value of $S_8$ is therefore naturally smaller than expected in the standard cosmological model. Moreover, because a larger local matter-density fluctuation at late times is diluted by a larger local cosmological constant, it can allow larger matter-density fluctuations at high redshift relative to a spatially fixed cosmological constant, as in the standard cosmological model. This naturally explains the unexpectedly large number density of massive galaxies recently observed by JWST at high redshifts. In the future, we will study this model, as a special case of an interacting dark energy model, more carefully at the perturbative level and non-linear scales. Eventually, in this picture, the Hubble tension is not resolved but explained as a diverse phenomenon at local scales perturbatively, depending on local environments of distant distance indicators/calibrators.

\subsubsection{Scale-dependent cosmological constant}\label{subsubsec:LambdaLCDM}

Finally, we return to what may be called one of the ultimate problems in cosmology: the cosmological-constant problem, the other being the origin of the Universe. On the one hand, the Universe is described on small scales by local quantum field theory, whose vacuum energy density is given by an effective cosmological constant. Specifically, if gravity is still described by the classical gravitational field equation, $G_{ab}+\lambda_\mathrm{bare} g_{ab}=8\pi G\langle T_{ab}\rangle$, then by Lorentz covariance the energy-momentum tensor corresponding to vacuum energy should take the form $\langle T_{ab}\rangle=-\langle\rho\rangle g_{ab}$~\cite{Firouzjahi:2023wbe}. Its conservation equation $\nabla^aT_{ab}=0$ implies that the vacuum energy density $\langle\rho\rangle$ is a constant, which in effective field theory should be set by the ultraviolet cutoff scale, for example $\langle\rho\rangle\sim\Lambda_\mathrm{UV}^4\sim M_\mathrm{Pl}^4$. The final field equation $G_{ab}+\lambda_\mathrm{eff} g_{ab}=0$ describes vacuum energy through an effective cosmological constant that includes the bare cosmological constant and matter fluctuations, $\lambda_\mathrm{eff}=\lambda_\mathrm{bare}+8\pi G\langle\rho\rangle$, for example $\lambda_\mathrm{eff}\sim\lambda_\mathrm{bare}+M_\mathrm{Pl}^2$.

On the other hand, on large scales the Universe obeys the cosmological principle and can therefore be described by the FLRW metric $\mathrm{d}s^2=-\mathrm{d}t^2+a(t)^2\delta_{ij}\mathrm{d}x^i\mathrm{d}x^j$. For the current dark-energy-dominated Universe, the scale factor evolves approximately as $a(t)=a(0)e^{Ht}$, namely nearly exponential expansion, where the measured Hubble expansion rate $H=\pm\sqrt{\Lambda/3}$ defines a constant $\Lambda$. Current observations of cosmic acceleration constrain this constant to be $\Lambda=3H^2\leq 3H_0^2\sim10^{-120}M_\mathrm{Pl}^2$. It is worth noting that we encounter the so-called fine-tuning problem of the cosmological constant, $\lambda_\mathrm{bare}\sim-(1-10^{-122})M_\mathrm{Pl}^2$, only when we identify the effective cosmological constant $\lambda_\mathrm{eff}$ originating from extremely small scales with the $\Lambda$ inferred from extremely large-scale observations, namely $\lambda_\mathrm{eff}=\Lambda$. But there is no reason to think that spacetime remains homogeneous and isotropic on extremely small scales; indeed, spacetime is more likely to fluctuate violently on such scales~\cite{Firouzjahi:2023wbe,Firouzjahi:2024faf}. Therefore there is no reason to accept $\lambda_\mathrm{eff}=\Lambda$, and the fine-tuning problem of the cosmological constant is avoided from the beginning.

The very definition of the cosmological-constant problem may therefore point to its own way out: the effective cosmological constant may be a scale-dependent physical quantity. It can be very large on extremely small scales because spacetime fluctuates violently there, while on extremely large scales, because spacetime tends toward homogeneity and isotropy, it may be averaged by some mechanism, such as those in Refs.~\cite{Wang:2017oiy,Cree:2018mcx,Wang:2019mbh,Wang:2019mee,Wang:2019wwg,Wang:2023tzm}, to a very small value. Our chameleon dark-energy model also provides a similar picture to some extent. The coupling of the chameleon field to differently averaged matter densities on different scales gives effective cosmological constants of different sizes, and the Hubble-constant tension might as well be a reflection of this physical picture on two scales: the CMB scale and Hubble-flow scale~\cite{Wagner:2023rC}.

\section{Conclusions and discussions}\label{sec:conclusion}

Modern cosmology has gone through the historical stages of hot big-bang cosmology, inflationary cosmology, and precision cosmology, and has finally forged the standard cosmological model: the six-parameter $\Lambda$CDM model with inflation, dark matter, and dark energy as its key ingredients within general relativity. This model can fit almost all observational facts over more than ten billion years of cosmic history, from galactic to cosmological scales. However, as a phenomenological model, the theoretical origins of its ingredients remain unknown, and the persistent Hubble tension in recent years has posed serious challenges to it. Yet a crisis also contains opportunity. The Hubble tension~\cite{DiValentino:2024yew,Efstathiou:2024dvn} may be such a historical opportunity to glimpse the underlying new physics beneath the standard model of cosmology.

In this review, after recognizing the systematic shift between early-Universe constraints and late-Universe measurements of the Hubble constant, we briefly summarize various solutions from early and late/local Universe. However, they are strongly constrained by the so-called ``early-time no-go theorems''~\cite{Krishnan:2020obg,Jedamzik:2020zmd,Lin:2021sfs,Vagnozzi:2021gjh,Philcox:2022sgj,Vagnozzi:2023nrq,Pedrotti:2024kpn} and ``late-time no-go theorems''~\cite{Benevento:2020fev,Camarena:2021jlr,Efstathiou:2021ocp,Cai:2021weh,Cai:2022dkh,Keeley:2022ojz,Huang:2024erq,Huang:2024gfw,Pedrotti:2025ccw,Ling:2025lmw,Wang:2026kor,Tiwari:2026pzk}.
In view of this dilemma, we review in this paper two additional tensions of the $\Lambda$CDM model around the homogeneity scale at both background and perturbation levels: an intercept ($-5a_B$) tension~\cite{Wang:2026kor} in the magnitude-distance relation between local and late Universes and a slope ($\delta H_0$) tension~\cite{Yu:2022wvg} in the non-local sample variance of Hubble-constant measurements. 
Addressing the intercept tension suggests possible new physics between the second and third rungs of the distance ladder~\cite{Huang:2024erq,Huang:2024gfw,Huang:2025som}, while the slope tension hints at an opposite deviation from the $\Lambda$CDM model for a locally faster expansion rate in denser regions, reminiscent of the late-time chameleon-like dark energy model that admits a larger effective cosmological constant for a larger density contrast~\cite{Cai:2021wgv}. This model agrees with both early and late observations by reproducing the Hubble tension as an existing phenomenon, predicting a decreasing trend in measuring the Hubble constant~\cite{Krishnan:2020vaf,Dainotti:2022bzg,Jia:2022ycc,Hu:2023jqc,Jia:2024wix,Bousis:2024rnb,Jia:2026vdt} from higher redshift data due to decreasing density contrast, which also allows for a larger density perturbation and a slower growth rate to resolve the matter-perturbation amplitude tension and growth tension. Furthermore, the usual two cosmological constant problems can also be addressed in this picture, where the coincidence problem is explained for the chameleon field tracking the matter evolution at the late time, and the fine-tuning problem is evaded for a scale-dependent cosmological constant (dubbed $\Lambda_L$CDM model).

Currently, we might be at a similar edge of the ``Planck moment'' for extracting the ``quantum'' concept in the 1900s from interpreting the interpolation between two conflicting measurements of black-body radiation spectrum at low and high frequencies. Only in the present case, such conflicting measurements occur for the Hubble constant from low and high redshifts. This Hubble tension has plagued us for the past decade, from 2016 to 2026. Solving the Hubble tension might be the clue to extract a similar ``quantum'' concept of modern cosmology in the next decade beyond the current paradigm. 

\appendix

\acknowledgments
This review is partially based on a much shorter version~\cite{Cai:2023sli} in Chinese in 2023 and a talk given at the Frontier Forum on Science and Technology, Academic Division of Chinese Academy of Science, held in 2024 at Ningbo city, Zhejiang Province, China.
We acknowledge the support from the National Key Research and Development Program of China Grants No. 2021YFC2203004 and No. 2021YFA0718304, the National Natural Science Foundation of China Grants No. 12422502, No. 12547110, No.12588101, No. 12235019, and No. 12447101, and the China Manned Space Program Grant No. CMS-CSST-2021-B01 and No. CMS-CSST-2025-A01.

\bibliographystyle{JHEP}
\bibliography{refer}

\end{document}